\documentclass[aps,prd,twocolumn,groupedaddress,showpacs,showkeys,nofootinbib]{revtex4}

\usepackage{graphicx}
\usepackage{dcolumn}
\usepackage{bm}

\font\eightit=cmti8

\begin{document}


\title{Measurements of Bottom Anti-Bottom Azimuthal Production\\
 Correlations in Proton-Antiproton Collisions at $\sqrt{s}= 1.8~\rm{TeV}$ }

\author{
D.~Acosta,$^{14}$ T.~Affolder,$^{7}$ M.G.~Albrow,$^{13}$ D.~Ambrose,$^{36}$   
D.~Amidei,$^{27}$ K.~Anikeev,$^{26}$ J.~Antos,$^{1}$ 
G.~Apollinari,$^{13}$ T.~Arisawa,$^{50}$ A.~Artikov,$^{11}$
% ...
}


\affiliation{
$^{1}$  {\eightit Institute of Physics, Academia Sinica, Taipei, Taiwan 11529, Republic of China} \\
% ...
$^{7}$  {\eightit University of California at Santa Barbara, Santa Barbara, California 93106} \\ 
% ...
$^{11}$  {\eightit Joint Institute for Nuclear Research, RU-141980 Dubna, Russia} \\
% ...
$^{13}$ {\eightit Fermi National Accelerator Laboratory, Batavia, Illinois
60510} \\
$^{14}$ {\eightit University of Florida, Gainesville, Florida  32611} \\
% ...
$^{26}$ {\eightit Massachusetts Institute of Technology, Cambridge,
Massachusetts  02139} \\   
$^{27}$ {\eightit University of Michigan, Ann Arbor, Michigan 48109} \\
% ...
$^{36}$ {\eightit University of Pennsylvania, Philadelphia, 
        Pennsylvania 19104} \\   
% ...
$^{50}$ {\eightit Waseda University, Tokyo 169, Japan} \\
% ...
}

\date{\today}

\begin{abstract}

We have measured the azimuthal angular correlation of $b\overline{b}$ production, using $86.5~\rm{pb^{-1}}$ 
of data collected by Collider Detector at Fermilab (CDF) in $p\overline{p}$ collisions at $\sqrt{s}$=1.8 TeV
during 1994$-$1995. In high-energy $p\overline{p}$ collisions, such as at the
Tevatron, $b\overline{b}$ production can be schematically categorized into three mechanisms. The leading-order (LO) process is
``flavor creation,'' where both $b$ and $\overline{b}$ quarks substantially participate in the hard scattering
and result in a distinct back-to-back signal in final state. 
The ``flavor excitation'' and the ``gluon splitting'' processes, which  appear at next-leading-order (NLO), are 
known to make a comparable contribution to total $b\overline{b}$ cross section, while providing very different opening angle distributions from the 
LO process. 
An azimuthal opening angle between bottom and anti-bottom, $\Delta\phi$, has been 
used for the correlation measurement to probe the interaction creating $b\overline{b}$ pairs. 
The $\Delta\phi$ distribution has been obtained from two different methods. One method measures the  
$\Delta\phi$ between bottom hadrons using events with two reconstructed secondary vertex tags. The other method 
uses $b\overline{b}\rightarrow(J/\psi X)(\ell X^{\prime})$ events, where the charged lepton ($\ell$) is an
electron ($e$) or a muon ($\mu$), to measure $\Delta\phi$ between bottom quarks.  The $b\overline{b}$ purity is determined as a function of
$\Delta\phi$ by fitting the decay length of the $J/\psi$ and the impact parameter of the $\ell$. 
Both methods quantify the contribution from higher-order production mechanisms by the fraction of the $b\overline{b}$ 
pairs produced in the same azimuthal hemisphere, $f_{toward}$. The measured $f_{toward}$ values 
are consistent with both parton shower Monte Carlo and NLO QCD predictions. 

\end{abstract}

\pacs{13.85.-t,12.38.Qk,14.65.Fy}
\keywords{QCD, heavy quark production}

\maketitle

\section{Introduction}
\label{intro}
	The dominant $b$ quark production mechanism at the Tevatron is believed to be pair production through the strong interaction.  However, predictions from next-to-leading order (NLO) perturbative QCD~\cite{MNR} have so far failed to describe the observed $b$ single quark production cross section~\cite{ua1-91,cdf-93a,cdf-93b,cdf-95,cdf-02,d0-95,d0-00}.   Differential cross section measurements have also been systematically higher than theoretical predictions~\cite{ua1-94,d0-00b,cdf-96,cdf-97}.  Possible explanations for the disagreement between the measured and predicted cross sections involve improved $b$ fragmentation models~\cite{frag}, and non-perturbative $b\overline{b}$ production mechanisms~\cite{nonpert}, and supersymmetric production mechanisms~\cite{squark}.  

Studying $b\overline{b}$ correlations gives additional insight into the effective contributions from higher-order QCD processes to $b$ quark production at the Tevatron.
  For example, the lowest-order QCD $b\overline{b}$ production diagrams contain only the $b$ and $\overline{b}$ quarks in the final state.  Momentum conservation requires that these quarks be produced back-to-back in azimuthal opening angle, $\Delta \phi$, and with balanced momentum transverse to the beam direction, $p_{T}$.  However, when higher-order QCD processes are considered, the presence of additional light quarks and gluons in the final state allows the $\Delta\phi$ distribution to become more spread out and the $b$ transverse momenta to become more asymmetric.  Previous measurements of azimuthal correlation distributions have yielded varying levels of agreement with NLO predictions~\cite{ua1-94,d0-00b,cdf-96,cdf-97,cdf-04}.   Additional measurements related to $b\overline{b}$ production are needed to determine whether experimental measurements are consistent with the Standard Model picture of $b\overline{b}$ production.


	The NLO QCD calculation of $b\overline{b}$ production includes diagrams from each production mechanism up to $\mathcal{O}(\alpha_{S}^{3})$.  The NLO calculation is the lowest order approach that returns sensible results because certain classes of diagrams which first appear at $\mathcal{O}(\alpha_{S}^{3})$--often referred to as flavor excitation and gluon splitting diagrams (see below)--provide contributions of approximately the same magnitude as the lowest-order diagrams, which are $\mathcal{O}(\alpha_{S}^{2})$.  This contribution can be understood by considering the cross section for $gg\to gg$ which is approximately two orders of magnitude larger than the cross section for $gg\to b\overline{b}$.  Higher-order $b\overline{b}$ diagrams can be formed from the leading-order diagram $gg\to gg$ by adding a $g\to b\overline{b}$ vertex to either in the initial or final state, but even with the $\mathcal{O}(\alpha_{S})$ suppression, these higher-order diagrams still provide contributions that are numerically comparable to the leading-order terms~\cite{MNR,ua1-94}.  Therefore, higher-order corrections to $b\overline{b}$ production cannot be ignored, and a recent measurement indicates that the higher order diagrams contribute a factor of four above the leading order term~\cite{cdf-04}.

\begin{figure}[htbp]
\includegraphics[width=8.6 cm]{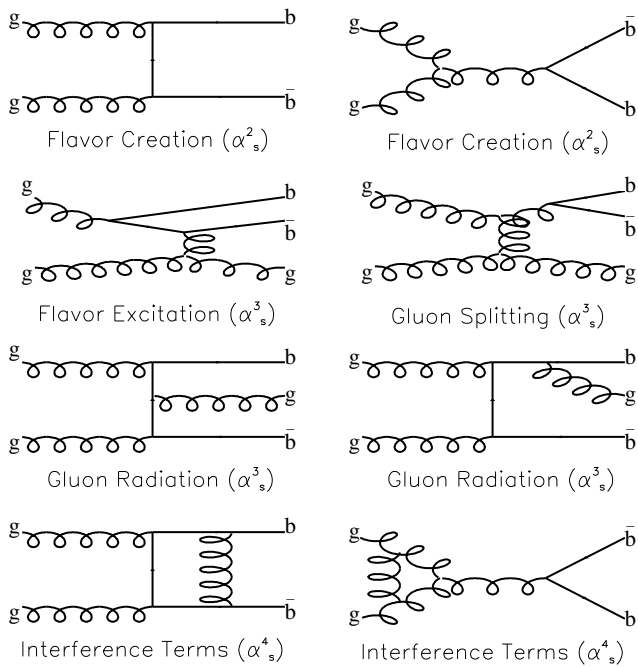}
\caption{\label{fig:feynman} Example Feynman diagrams that contribute bottom production.  The bottom two virtual exchange diagrams enter into the NLO calculation through interferences with leading-order terms.  Interferences between the flavor creation, flavor excitation, and gluon splitting diagrams, as well as the virtual exchange diagrams, are ignored in the parton shower approximation }
\end{figure}

An alternative approach to estimating the effects of higher-order corrections is the parton shower model implemented by the {\sc Pythia} ~\cite{pythia1,pythia2} and {\sc Herwig}~\cite{herwig} Monte Carlo programs\footnote{{\sc Herwig} and {\sc Pythia} use the exact matrix elements for all parton-parton two-to-two scatterings.  However, all two-to-$N$ ($N > 2$) processes are estimated using the ``leading-log'' approximation, which becomes exact in the limit of ``soft'' or ``collinear'' emissions.  As a result, such Monte Carlo programs are often said to use the ``leading-log approximation.''}.  The parton shower approach is not exact to any order in $\alpha_{S}$ but rather tries to approximate corrections to all orders by using leading-order matrix elements for the hard two-to-two QCD scatter and adding addition initial- and final-state radiation using a probabilistic approach.  	In this approximation, the diagrams for $b\overline{b}$ production can be divided into three categories:
\begin{description}
\item[Flavor Creation] refers to the lowest-order, two-to-two QCD $b\overline{b}$ production diagrams.  This process includes $b\overline{b}$ production through $q\overline{q}$ annihilation and gluon fusion, plus higher-order corrections to these processes.  Because this production is dominated by two-body final states, it tends to yield $b\overline{b}$ pairs that are back-to-back in $\Delta\phi$ and balanced in $p_{T}$.
\item[Flavor Excitation] refers to diagrams in which a $b\overline{b}$ pair from the quark sea of the proton or antiproton is excited into the final state because one of the quarks from the $b\overline{b}$ pair undergoes a hard QCD interaction with a parton from the other beam particle.  Because only one of the quarks in the $b\overline{b}$ pair undergoes the hard scatter, this production mechanism tends to produce $b$ quarks with asymmetric $p_{T}$.  Often, one of the $b$ quarks will be produced with high rapidity and not be detected in the central region of the detector.
\item[Gluon Splitting] refers to diagrams where the $b\overline{b}$ pair arises from a $g\to b\overline{b}$ splitting in the initial or final state.  Neither of the quarks from the $b\overline{b}$ pair participate in the hard QCD scatter.  Depending on the experimental range of $b$ quark $p_{T}$ sensitivity, gluon splitting production can yield a $b\overline{b}$ distribution with a peak at small $\Delta\phi$.
\end{description}

Figure~\ref{fig:feynman} illustrates some lowest-order examples of each type of diagram.  The general trend is that flavor creation diagrams, being dominated by two-body $b\overline{b}$ final states, tend to produce back-to-back $b\overline{b}$ pairs balanced in $p_T$, while flavor excitation and gluon splitting, which necessarily involve multiparticle final states including a $b\overline{b}$ pair and light quarks or gluons, produce $b\overline{b}$ pairs that are more smeared out in $\Delta\phi$ and $p_{T}$.  Categorizing $b\overline{b}$ diagrams in this scheme becomes ambiguous at higher order in perturbation theory.  In the parton shower approximation, flavor creation, flavor excitation, and gluon splitting processes can be separated exactly based on how many $b$ quarks participate in the hard two-to-two scatter.    Interference terms among the three production mechanisms, as well as virtual exchange diagrams, are neglected as higher-order effects in this approximation.  

 	Refs.~\cite{field} and \cite{cdf-04} show that parton shower Monte Carlo programs, which include sizeable contributions from the higher-order $b$ production mechanisms of flavor excitation and gluon splitting, are able to better reproduce the observed $b$ production cross section.  Studying $b\overline{b}$ correlations provides a way to tell whether such large contributions from these higher-order processes are supported by the data.

	In this paper, we present two new CDF measurements of the $\Delta\phi$ spectrum in $b\overline{b}$ production in $p\overline{p}$ collisions at $\sqrt{s} =1.8~\mathrm{TeV}$.  These measurements are made using approximately 90 pb$^{-1}$ of data collected during the 1994-1995 Tevatron run (known as Run Ib).  In addition to providing new information about the entire range of the $b\overline{b}$ $\Delta\phi$ spectrum, these analyses are more sensitive than previous measurements to the low $\Delta\phi$ region, where flavor excitation and gluon splitting make a larger contribution.  
	
	One analysis begins with a sample of events containing an 8 GeV electron or muon to enhance the $b$ quark content of the sample by taking advantage of the relatively high semileptonic $B$ branching ratio.  These events are then searched for the presence of displaced secondary vertices indicating the decay of a long lived $B$ hadron, using a vertexing algorithm similar to the SECVTX algorithm used for the top quark analyses~\cite{secvtx, cdf-01}.  This analysis requires that the decay vertices for both $B$ hadrons in the event be reconstructed and extracts the $B$ hadron $\Delta\phi$ distribution from the $\Delta\phi$ distribution measured between the reconstructed secondary vertices.  The direction of each $B$ hadron is inferred using the vector sum of the momenta from the secondary vertex tracks and $\Delta\phi$ is defined as the azimuthal angle between the inferred directions of the two $B$ hadrons.  This technique yields a high-statistics sample of double-tagged $b\overline{b}$ events and retains sensitivity to $b\overline{b}$ pairs with small opening angles.  The second analysis detects the presence of $b$ quark decays in the data entirely through leptonic signatures.  The decay of one $b$ is tagged by reconstructing the decay of a $J/\psi\to \mu^+\mu^-$, which provides the trigger signature that defines this sample.  Events are also required to contain an electron or muon consistent with the semileptonic decay of the second $b$.  This approach does not yield as many double-tagged events as the first, but it retains the highest sensitivity for $b\overline{b}$ production at small opening angles and has fewer backgrounds.  Both analyses produce consistent results indicating that roughly one fourth of the $b\overline{b}$ pairs produced in the momentum and rapidity range to which these analyses are sensitive have $\Delta\phi < 90 ^{\circ}$.  In addition, both analyses are at least qualitatively consistent with the contribution from higher order $b\overline{b}$ predicted by {\sc Pythia} and {\sc Herwig}, further supporting the significance of the flavor excitation and gluon splitting production mechanisms at the Tevatron.

\section{Detector}


The CDF detector has a cylindrical symmetry about the beam-line, making it convenient to use 
a cylindrical coordinate system with the $z$-axis along the proton beam direction. We define $r$ to be
the distance from the beam-line and $\phi$ to be the azimuthal angle measured from the direction 
pointing radially outward in the plane of the Tevatron ring. It is also useful to use the polar 
angle $\theta$ measured with respect to the $z$-axis, and pseudorapidity $\eta = -{\rm ln}\;({\rm tan}\;(\theta/2))$. In the approximation of massless particles, the pseudorapidity equals the rapidity $y = (1/2){\rm ln}\;((E + p_{z})/(E-p_{z}))$, which is  the invariant boost of the particle along the $z$-axis.  
The CDF detector is described in detail elsewhere~\cite{cdfdet}. In the following, we focus on the elements 
most relevant to these analyses.

\begin{figure}[htbp]
\includegraphics[width=8.6cm]{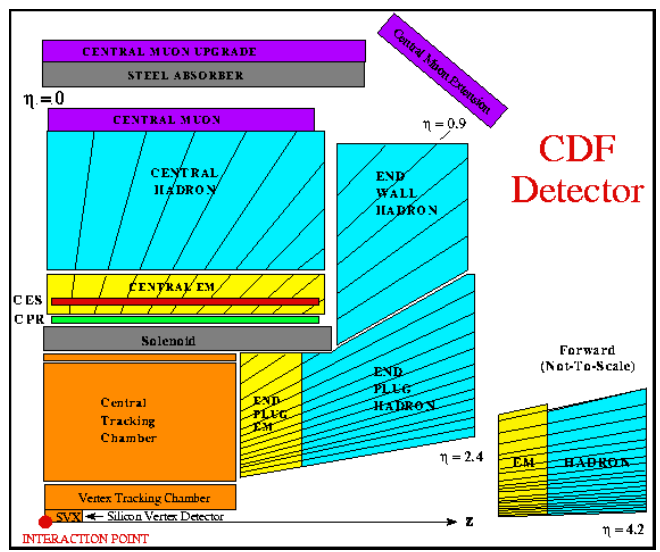}
\caption{\label{fig:cdfdet} Schematic of a quarter cross-section of the CDF Run 1b detector.}
\end{figure}


The tracking system, consisting of three different sub-detectors, the central tracking 
chamber (CTC)~\cite{CTC}, the vertex detector (VTX), and the silicon vertex 
detector (SVX$^{\prime}$)~\cite{SVX}, is immersed in a uniform 1.4 T solenoidal magnetic 
field in order to measure the charged particle momentum in a plane transverse to the 
$z$-axis, denoted as $p_{T}= p\;{\rm sin}\;\theta$. A charged track reconstruction begins with 
the measurements made in the CTC, which is a large cylindrical multi-wire drift chamber 
in 3.2 {\rm m} length along the $z$-axis and centered at the nominal interaction point of the CDF detector. 
It contains a total of 84 layers of wires positioned between $r=31$ and $133$ {\rm cm}. The
layers are arranged in the alternating groups of 12 with wires parallel to the $z$-axis, known as axial superlayers, 
and 6 with wires in $\pm 3^{\circ}$ stereo angles, known as stereo superlayers. The VTX, sitting 
inside the inner radius of the CTC, is a time projection chamber that provides a precise 
particle trajectory measurement in the $r$$-$$z$ plane and ultimately allows the determination of the $z$-location of the primary interaction point. The innermost system 
is the SVX$^\prime$ covering from $r=2.9$ to $8.1$ {\rm cm}. This four-layer detector allows high precision determination of 
particle trajectories in the $r$$-$$\phi$ plane. Combined, the whole tracking system provides 
a $p_{T}$ resolution of $\delta p_{T}/p_{T} = [(0.0009 \times p_{T})^{2} + (0.0066)^2)]^{1/2}$ 
and an impact parameter resolution of $\delta d_{0} = [13 + (40 \;{\rm GeV}/c)/p_{T}] \;\mu\mathrm{m}$.

	
The central electro-magnetic calorimeter (CEM)~\cite{cem}, located outside the radius of the CTC and segmented 
in a projective tower geometry, is designed to be deep enough to contain electro-magnetic showers initiated 
by electrons or photons. The CEM consists of alternating layers of lead absorber and polystyrene scintillator.
A set of wire and strip tracking chambers, known as CES,
are embedded in the CEM near the shower maximum or depth of greatest energy deposition, to measure the transverse 
shower profile. An electron is identified from a track reconstructed in the tracking system that points 
to an energy deposition in the CEM of appropriate size and matches to a cluster in the CES. 
For more penetrating particles, the central hadronic calorimeter (CHA)~\cite{cha} is 
located behind the CEM. The CHA is constructed from alternating layers of steel absorber
and scintillator, and also segmented in a projective tower geometry.  The CHA is used in 
these analyses primarily to reject hadrons that might fake an electron or a muon signature.


Muons are detected by their ability to penetrate the material in the calorimeter.  
Three sets of chambers are positioned outside the CHA to identify muons. The 
first set, known as the central muon (CMU)~\cite{cmu} is located at $r=3.47$ {\rm m}.
A particle traveling perpendicular to the $z$-axis from the primary interaction point 
must traverse 5.4 pion interaction lengths of material to reach the CMU. An additional set of chambers, 
the central muon upgrade (CMP)~\cite{cmp} is arranged in a rectangular array around the CMU behind 
an additional 60 {\rm cm} of steel shielding to provide further discriminating power between real muons and hadronic punch-through.  
To penetrate to the CMP, a particle traveling perpendicular to the $z$-axis from the primary 
interaction point has to pass through 8.4 pion interaction lengths of material. The CMU and the CMP 
detectors cover $|\eta| < 0.6$.  Another set of chambers, the central muon extension (CMX), 
consisting of four arches of drift chambers located behind 6.2 pion interaction lengths of material,
covers $0.6 < |\eta| < 1.0$. In addition, the CMX drift tubes are sandwiched between two layers 
of scintillator that provide fast timing information to the trigger.  Segments reconstructed from hits in the chambers are known as ``stubs''.


The CDF uses a three-level trigger system. The first two levels, named Level-1 and Level-2, 
are implemented in hardware and reduce the data rate from the full 300 kHz beam crossing rate 
to a more manageable 20 Hz. The third level, named as Level-3, consists of software 
algorithms that run a stream-lined version of the full CDF reconstruction software.  
The triggers used for these analyses rely on lepton identification through matching 
energy deposition in the CEM (for electron) or muon hits in the CMU, the CMP, and the CMX (for muon) 
with charged particle tracks reconstructed in the CTC.

\section{Secondary Vertex Tag $B$ Hadron Correlation Analysis}

\subsection{Overview}



The $\Delta\phi$ distribution of two reconstructed secondary vertex tags has been obtained from data
as a probe to investigate the $b\overline{b}$ production mechanisms and compared to the predictions 
based on {\sc Pythia} and {\sc Herwig} Monte Carlo (MC) programs. We correct our data for detector 
effects and background contributions using MC information in order to extract the $\Delta\phi$ 
distribution of $B$ hadrons that can be directly compared to the theoretical predictions. 
We choose to measure the $\Delta\phi$ distribution of $B$ hadrons rather than $b$ quarks, 
since our secondary vertex tags are more directly related to $B$ hadrons than $b$ quarks. 
Converting our measurement from the $B$ hadron level to the $b$ quark level would introduce 
a dependence on $b$ quark fragmentation models that we wish to avoid.

This analysis uses the largest sample of double-tagged $B$ hadron decays ever collected at a hadron collider, extracted from the data taken by CDF during the 1994$-$1995 run of the Tevatron (Run Ib).
To create a sample enhanced in $b$ quark content, we take advantage of the high purity of CDF lepton 
triggers as well as the significant impact parameters of $B$ decay daughters. Each candidate event 
is required to contain a lepton, either an electron or a muon, presumably coming from the 
semileptonic decay of one $B$ hadron, and the displaced secondary vertices of both $B$ hadrons. After 
background removal, we obtain a sample of approximately 17,000 events.

\subsection{Secondary Vertex Tagging}
\label{sec:bvtx}

Our secondary vertex tagging algorithm looks for tracks consistent with coming from a secondary vertex, 
significantly displaced from the primary vertex, using the precise tracking information. This algorithm is based on 
the BVTX algorithm used for the $B^{0}-\overline{B^{0}}$ mixing analysis~\cite{bdmixing}, which is a modified 
version of the SECVTX algorithm used for the top quark analysis~\cite{secvtx,cdf-01}. The main difference between the 
version of the BVTX used here and the version used for the previous CDF analyses is the ability to locate more 
than one secondary vertex per jet searched. For extensive details on the BVTX and the modifications made 
for this analysis, see Refs.~\cite{bdmixing} and ~\cite{lannon-thesis}. Below we summarize the secondary vertex finding approach.

The secondary vertex finding begins by first locating the primary interaction vertex for the event using the precise 
tracking information. Next the tracks in the event passing quality cuts are grouped into jets using a cone-based 
clustering algorithm with a cone size of $\Delta R =\sqrt{(\Delta \phi)^{2}+(\Delta \eta)^{2}}=1.0$. Each jet is 
then searched for the presence of one or more secondary vertices displaced from the primary.  Because the secondary vertex finding is done on a jet-by-jet basis, this algorithm is not able to handle the case where the $B$ decay products are contained in more than one jet.  However,
the relatively large cone-size used in this analysis was chosen to reduce the number of times the a $B$ decay would span more than one jet.   The secondary vertex finding 
is done in two steps for each jet. The first step finds secondary vertices containing at least three tracks. When the first 
step fails to find any more secondary vertices in a jet, the second step is attempted in which the individual track cuts 
are made more stringent and two-track secondary vertices are accepted. Each secondary vertex found is required to be 
significantly displaced from the primary and not to be consistent with the decay of a $K_{S}^{0}$ or $\Lambda$.

\subsection{Sample Selection} \label{sec:sample-selection}

This analysis starts with the data sample used for the measurement of time dependent $B^{0}-\overline{B^{0}}$ 
mixing~\cite{bdmixing}, which is a loosely selected sample that requires each event to have at least an electron 
or a muon with $p_{T} > 8 \;{\rm GeV}/c$ identified using the standard CDF lepton identification 
cuts~\cite{lannon-thesis}, and at least one reconstructed secondary vertex. This sample is known as the BVTX sample, 
after the name of the secondary vertex tagging algorithm used to create it. The BVTX sample consists of over 
480,000 electron-triggered events and over 430,000 muon-triggered events.

The strategy for extracting candidates from the BVTX sample is as follows: Because the BVTX sample was collected with a number of different lepton triggers, we impose specific trigger requirements to ensure the electron and muon subsamples have comparable kinematic properties. Next, the data sample is 
reprocessed by the modified version of the BVTX algorithm (see section~\ref{sec:bvtx} above) and each event is required to contain 
at least two secondary vertex tags. The separation $L_{xy}$ between each secondary vertex  and the primary vertex in the plane perpendicular 
to the beam-line divided by the uncertainty on the measurement ($\sigma_{L_{xy}}$) is required to be $L_{xy}/\sigma_{L_{xy}}\ge 2$.
To reduce the chance of tagging the same $B$ decay with two poorly measured tags, the 2-dimensional separation 
between the secondary vertex tags is also required to be $|\Delta L_{xy}|/\sigma_{\Delta L_{xy}}\ge 2$.  $\Delta L_{xy}$ is defined to be the distance between the two secondary vertex tags as measured in the plane perpendicular to the beam.
Each tag pair is required to have an invariant mass greater than 6 GeV/$c^{2}$ to reduce the chance that a tag pair results 
from a $B \to D \to X$ decay chain. For a tag pair failing either the $|\Delta L_{xy}|/\sigma_{\Delta L_{xy}}$ or the 
invariant mass cuts, only the tag with the longest 2-dimensional separation from the primary vertex is removed. 
Finally, since the trigger requirements for this sample assume 
at least one of the $B$ hadrons decay semileptonically, the trigger lepton is required to be within a cone of 
$\Delta R = 1.0$ of one of the vertices.

\subsection{Sample Composition}  \label{sec:sample-comp}


We have isolated a high purity $b\overline{b}$ sample in section~\ref{sec:sample-selection} with small contamination from other sources.  Table~\ref{table:sample-comp} 
shows the sample composition, including background sources that make a contribution to the sample. 
We briefly summarize each background contribution below.

\begin{table*}
\begin{ruledtabular}
\begin{tabular}{p{10cm}l}
Scenario & Classification \\
\hline
The tracks in the tag are from the same $B$ decay (including any tracks from a secondary $D$ decay) &	Good Tag (Signal) \\
The tag contains random prompt tracks not associated with the decay of any long-lived particle & Mistag (Background) \\
The tracks in the tag are from a $B$ decay (including secondary $D$ decay) that has already been tagged with other tracks. &	Sequential Double-Tag (Background) \\
The tag tracks are from a prompt $D$ decay-in other words, a $D$ not associated with the decay of a $B$. & Prompt Charm (Background) \\
\end{tabular}
\caption{\label{table:sample-comp} Different sources of tags and their classification as signal or background.}  
\end{ruledtabular}
\end{table*}

A mistag happens when the secondary vertex tagging algorithm tries to fit a vertex from a set of tracks that do not physically originate from a common vertex. Due to errors caused by the tracking performance, it is possible to find a set of prompt tracks that seem to intersect at a vertex displaced from the primary vertex.  These 
vertices distort the correlation spectrum and must be removed. One way to identify mistags 
is by looking at the distribution of $L_{xy}$, which is signed based on the inferred direction of the particle, 
namely the direction of the secondary vertex, relative to the primary vertex. A particle that seems to be 
moving out from the primary vertex at the time of decay obtains a positive $L_{xy}$, 
while a particle that seems to have been moving towards the primary vertex gets a negative $L_{xy}$.  A particle is deemed to be moving away from the primary vertex if the angle between the tag displacement vector (measured from the primary vertex to the secondary vertex tag) and the tag momentum vector is less than 90$^\circ$, and towards the primary vertex otherwise. 
A secondary vertex corresponding to the decay of real long-lived particle is expected to have 
a positive $L_{xy}$. However, the finite resolution of the tagging algorithm can yield a negative 
contribution. As a consequence, mistags make an $L_{xy}$ distribution that 
is symmetric about zero. We make use of this feature of mistags to subtract them statistically from the data.  To understand better how the $L_{xy}$ distribution is used for mistag subtraction, consider the case of an analysis involving only single tags.  Half of the total mistag background appears in the negative $L_{xy}$ region.  The positive portion of the $L_{xy}$ distribution contains the other half of the mistags, as well as real secondary vertex tags.  Therefore, by subtracting twice the number of negative $L_{xy}$ tags from the entire sample of tags, we are left with only the good secondary vertex tags.  For analyses such as this one, which considers pairs of secondary vertex tags, the calculation is given by 


\begin{equation}
N_{GG} = N_{++} - N_{+-} - N_{-+} + N_{--},
\label{eq:mistag-sub}
\end{equation}


\noindent
where $N_{GG}$ is the estimated number of tag pairs in which both tags are legitimate secondary vertex tags, while 
$N_{++}$ is the number of tag pairs in which both tags have $L_{xy} > 0$, $N_{+-}$ and $N_{-+}$ are the number of 
tag pairs in which one tag has a positive $L_{xy}$ and the other has a negative $L_{xy}$, and $N_{--}$ is the number 
of tag pairs in which both tags have $L_{xy} < 0$.  Conceptually, in this equation, we are using the second and third terms to subtract mistags from the tag pairs represented by the first term.  However, $N_{+-}$ and $N_{-+}$ each contain a contribution from the case where both tags are mistags and by subtracting them both, this contribution is double-counted.  The last term in the equation corrects this.  To obtain a mistag subtracted distribution, Eq.~\ref{eq:mistag-sub} is applied on a bin-by-bin basis.


Another possible source of background involves tagging more than one secondary vertex from a single $B$ decay.  
These tags, known as sequential tags, are most likely to occur when the $B$ decay involves the production of 
a $D$ hadron that travels a certain distance from the $B$ decay vertex before itself decaying. The invariant 
mass cut of 6 GeV/$c^2$ eliminates virtually all contribution from this source. Although this cut does reduce the tagging efficiency at low opening angle, it is necessary to keep the sequential tag background from overwhelming the signal in that region.  This efficiency reduction is accounted for in the Monte Carlo modeling of the data.  It is also possible that some 
sequential tag pairs arise from tracking errors that cause tracks originating from a common vertex 
to be reconstructed as coming from two vertices that are very close together. The cut on the significance of 
the 2-dimensional separation between the tags ($|\Delta L_{xy}|/\sigma_{\Delta L_{xy}}$) eliminates nearly 
all these tag pairs. The residual contribution from sequential double tags is estimated 
in section~\ref{sec:corr-and-syst}.


Finally, a background source of legitimate secondary vertices is direct $c\overline{c}$ production.
In general, most $D$ hadrons have a much smaller lifetime than $B$ hadrons. However, those $D$ hadrons 
that do live long enough to produce a secondary vertex capable of being tagged by BVTX will not be 
removed or accounted for by any of the methods mentioned above. In addition, it is possible to have 
events in which multiple heavy flavor pairs, such as $b\overline{b} + c\overline{c}$ and $b\overline{b} + b\overline{b}$, are produced. 
For example, in a flavor creation event, an additional $c\overline{c}$ pair may be produced through 
gluon splitting. In such events it is possible for the $b\overline{b}$ to contribute one tag and 
the $c\overline{c}$ to contribute another. Although the rate of multiple heavy flavor production 
is much lower than single $b\overline{b}$ production, the opportunity to tag more displaced vertices 
in a given event can provide an enhancement in tagging acceptance, meaning such processes cannot be 
discounted outright. Our MC studies indicate that the combined contribution to the tag pair sample from prompt 
charm and multiple heavy flavor production is not large, roughly 10\%.  
The subtraction of this contribution and the associated systematic error are described in section~\ref{sec:corr-and-syst}.

\subsection{Monte Carlo Samples}

\par The parton shower MC programs, {\sc Pythia}~\cite{pythia1} and {\sc Herwig}~\cite{herwig}, are used to generate large samples of 
$b\overline{b}$ events. Because flavor creation, flavor excitation, and gluon splitting mechanisms do not interfere with each other 
in the parton shower model, each mechanism is generated separately. For {\sc Pythia}, the flavor creation samples are 
generated as the heavy flavor production process using massive matrix elements for $q\overline{q} \to b\overline{b}$ 
and $gg \to b\overline{b}$ diagrams. Flavor excitation and gluon splitting samples are produced 
as the generic QCD 2-to-2 process using massless matrix elements, and then separated from 
other QCD processes by examining the partons that participate in the 2-to-2 hard scattering. Three 
{\sc Pythia} samples with different amounts of initial state radiation (ISR) are generated for each mechanism: The samples are dubbed low, medium, and high ISR, as explained in appendix~\ref{sec:bvtx-mc}.  The choice to investigate different ISR settings in {\sc Pythia} is motivated primarily because the ISR tuning of {\sc Pythia} was changed in the recent past based on studies of heavy flavor production~\cite{norb1,norb2}, and the new tuning produces a noticeably different $\Delta\phi$ spectrum from the previous version.  The low ISR sample corresponds to the most recent ISR settings in {\sc Pythia} while the high ISR sample reflects the previous default settings.  In principal, changes in the amount of final state radiation (FSR) would have a similar affect, but such an effect has not been studied here.
For all three {\sc Pythia} samples, the underlying event is tuned to match observations in CDF data~\cite{underlying-event}.  
On the other hand, because there are fewer parameters to tune, only one {\sc Herwig} sample is generated for each mechanism. 
The {\sc Herwig} flavor creation and flavor excitation samples are generated with heavy flavor production option 
including massive matrix element treatments of the LO flavor creation and flavor excitation diagrams.  
As in {\sc Pythia}, the {\sc Herwig} gluon splitting component results from generating all QCD 2-to-2 processes 
using massless matrix elements and retaining those events classified as gluon splitting based on the partons involved 
in the hard scattering. In addition, a small sample of $c\overline{c}$ events was generated using {\sc Pythia}, for the purpose 
of evaluating the possible effects of residual prompt charm as a background for this analysis. Both {\sc Pythia} and {\sc Herwig} generation used the CTEQ5L parton distribution functions.  See Appendix~\ref{sec:bvtx-mc} for more information about the {\sc Pythia} and {\sc Herwig} parameters used for this analysis.  For all samples, 
heavy flavor decays are handled by the CLEO {\sc QQ} MC program~\cite{cleo}.  
Finally, to make the MC data resemble the actual data as closely as possible, 
the MC events are passed through a CDF detector simulation, a CDF trigger simulation, 
and the same reconstruction and analysis code used for the actual data. Additional details 
regarding the generation of MC samples can be found in Ref.~\cite{lannon-thesis}.

\subsection{Data--Monte Carlo Comparison}  \label{sec:mc-data-comp}

After the Monte Carlo samples have been passed through the detector simulation as described above, the Monte Carlo predictions for the secondary vertex tag distributions can be compared directly with data.  Distributions involving individual tags have similar shapes for flavor creation, flavor excitation, and gluon splitting, and so these distributions can be used to check whether the detector simulation adequately models detector effects.  Distributions involving tag pairs, and therefore correlations, give information about how well the Monte Carlo models describe $b\overline{b}$ production.

\begin{figure}[htbp]
\includegraphics[width=8.6 cm]{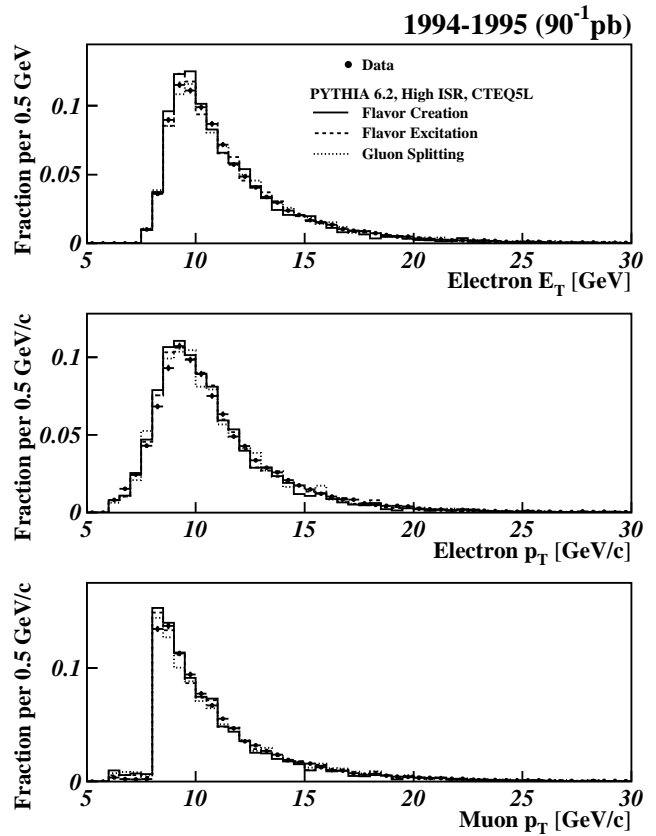} 
\caption{\label{fig:trig-lep-pt}  The trigger electron $p_{T}$ and $E_{T}$ and the trigger muon $p_{T}$ distributions from data compared to the high ISR {\sc Pythia} Monte Carlo sample.  The comparisons of data to other Monte Carlo samples is similar and can be found in Ref.~\cite{lannon-thesis}.  The muon trigger threshold is clearly visible in the lower plot.  The small number of muons below the 8 GeV trigger threshold come from events containing a second muon that passes the offline selection.}
\end{figure}

\begin{figure}[htbp]
\includegraphics[width=8.6 cm]{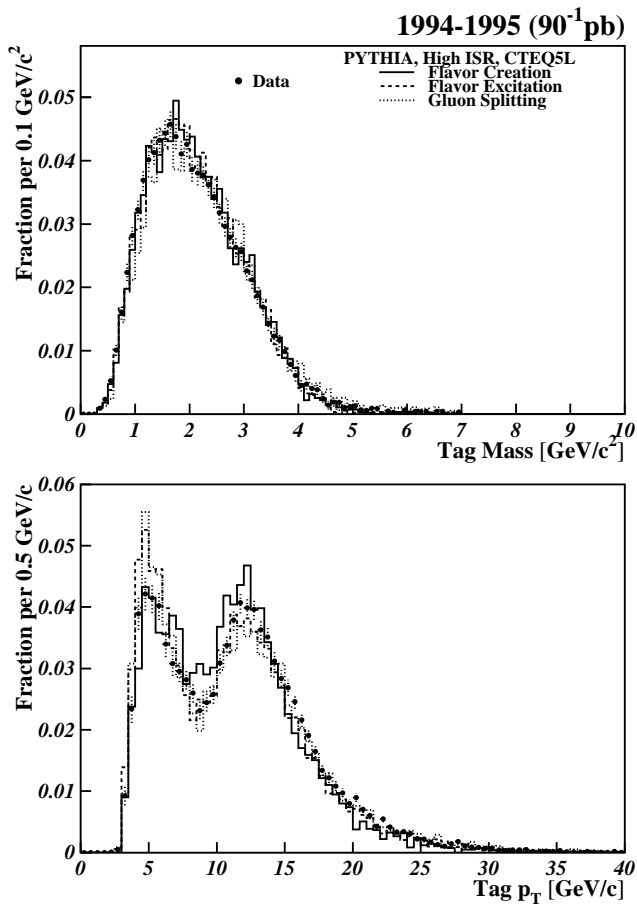}
\caption{\label{fig:tag-prop}  The secondary vertex tag distributions from electron data compared to the high ISR {\sc Pythia} Monte Carlo sample.  Comparisons involving muon data and comparisons to other Monte Carlo samples are similar and can be found in Ref.~\cite{lannon-thesis}.}
\end{figure}

	Figure~\ref{fig:trig-lep-pt} shows the comparison of the trigger electron $p_{T}$ and $E_{T}$ and the trigger muon $p_{T}$ between the data and each Monte Carlo sample.  Figure~\ref{fig:tag-prop} shows the comparison of the secondary vertex tag properties in the Monte Carlo and data.  In each case, the agreement between the measured spectrum from the data and the predicted spectra for each Monte Carlo sample indicates that the effects of trigger and reconstruction thresholds are adequately modeled in the simulation.  In addition, examining the Monte Carlo truth information for the $B$ hadrons tagged by the analysis code allows a determination of the effective minimum $B$ $p_{T}$ sensitivity for this analysis.  For the $B$ producing the 8 GeV/c trigger lepton, this measurement is sensitive only to $B$ hadrons with a minimum $p_{T}$ of 14 GeV/c, while the requirement that the other $B$ be tagged by the BVTX algorithm sets a minimum $p_{T}$ acceptance of 7.5 GeV/c $p_{T}$.

	Comparing tag pair correlations between the Monte Carlo samples and the data reveals whether {\sc Pythia} or {\sc Herwig} provide an adequate model of the higher-order contributions to $b\overline{b}$ production.  This analysis focuses on the transverse opening angle, $\Delta\phi$.  For tag pairs, $\Delta\phi$ is defined as the angle between the $p_{T}$ vectors determined by taking the vector sum of the $p_{T}$ from all the tracks involved in the tag.  The $\Delta\phi$ distribution is interesting to study because it is sensitive to contributions from flavor excitation and gluon splitting.  Also, the "broadness" of the back-to-back peak in $\Delta\phi$ is sensitive to the amount of initial-state radiation present in the Monte Carlo.  It should be noted that the shape of the $\Delta\phi$ distribution and the relative contributions from the three production mechanisms depend on the $p_{T}$ cuts placed on each of the $B$ hadrons.

	There are two possible approaches to normalizing the relative contributions in Monte Carlo from flavor creation, flavor excitation, and gluon splitting.  {\sc Pythia} and {\sc Herwig} each provide predictions for the cross section of each production mechanism, and these cross sections can be used to normalize their contributions relative to one another.  Alternatively, one could take the position that {\sc Pythia} and {\sc Herwig} may not correctly model the amount of each contribution, and the relative contributions should be determined to provide the best match to data.  In this analysis, both approaches are examined.  As described in the sections below, the data is compared to the Monte Carlo predictions in two ways.  First, the Monte Carlo prediction for the cross section of each production mechanism is used to normalize the flavor excitation and gluon splitting components relative to the flavor creation contribution.  In this ``fixed normalization'' scheme, the data is compared to the Monte Carlo using one arbitrary global normalization parameter.  The arbitrary global normalization is included because this analysis attempts only a shape comparison, not an absolute cross section measurement.  In addition, the Monte Carlo and data are compared using a ``floating normalization'' scheme.  In this comparison, each production mechanism is given an independent arbitrary normalization constant and the three normalizations are varied to yield the best match to data.

\begin{figure}[htbp]
\includegraphics[width=8.6 cm]{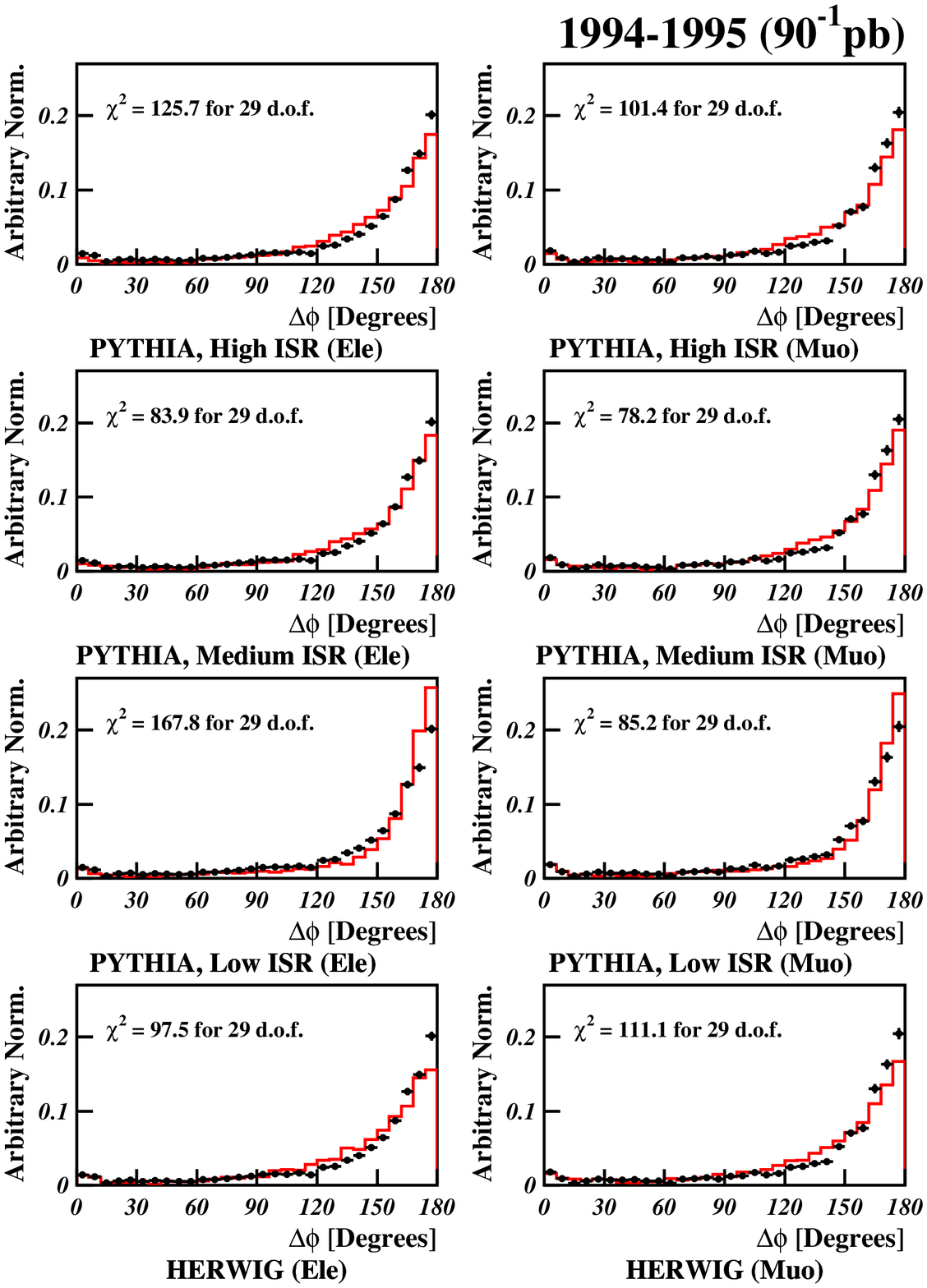}
\caption{\label{fig:1comp-fits}  Comparisons between the shape of the tag $\Delta\phi$ distribution for data (points) and the tag $\Delta\phi$ distribution predicted by the various Monte Carlo samples (line).  The contributions from flavor creation, flavor excitation and gluon splitting are added together according to the individual Monte Carlo cross section predictions for these contributions and only a single common normalization is varied to get the best match to data.  The $\chi^{2}$ values shown in the plots account only for the data and Monte Carlo statistical errors.}
\end{figure}

\begin{figure}[htbp]
\includegraphics[width=8.6 cm]{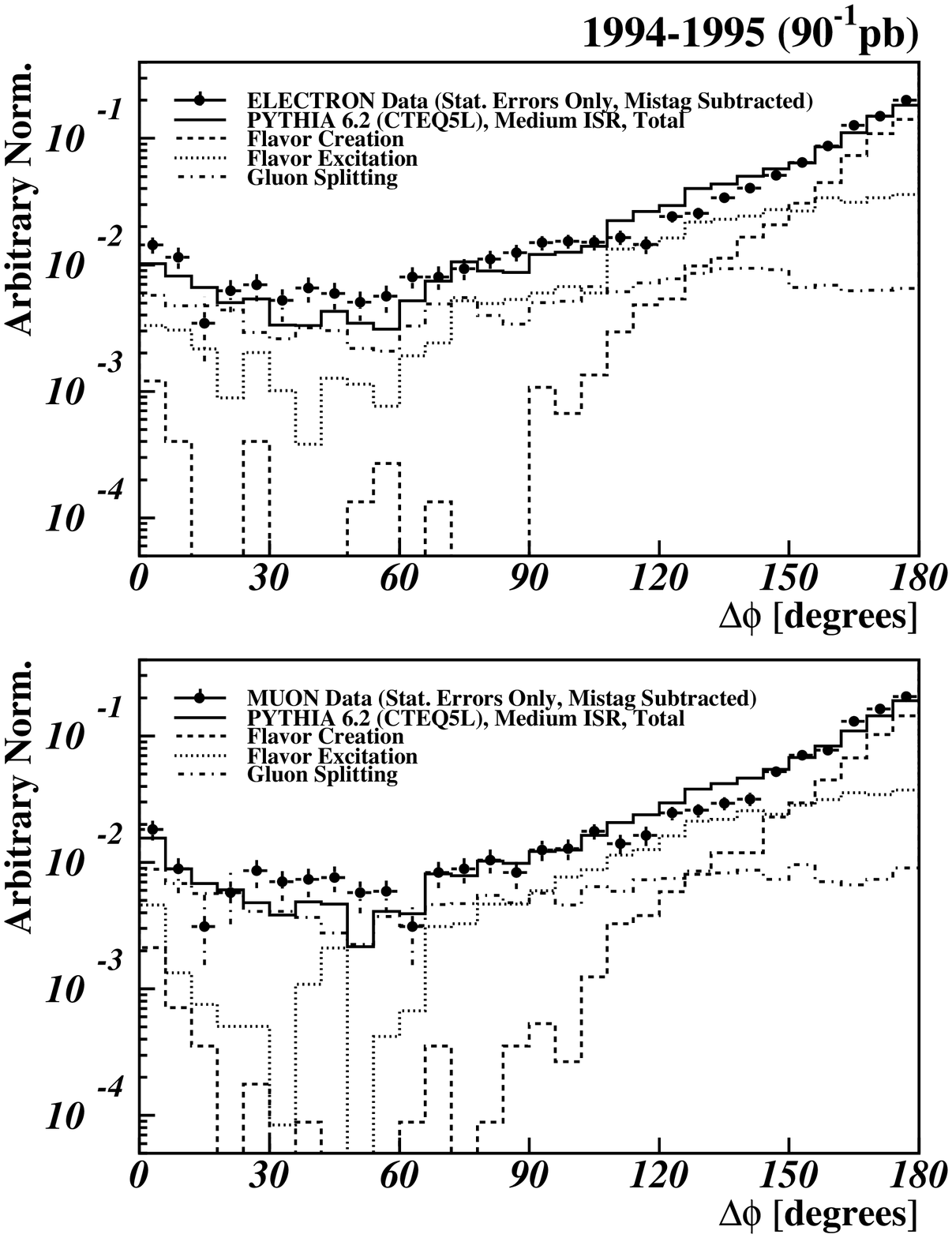}
\caption{\label{fig:1comp-midisr} A detailed comparison between the $\Delta\phi$ distribution from data (points, statistical errors only) and the $\Delta\phi$ distribution from {\sc Pythia} with medium ISR (solid line).  In addition, the contributions from flavor creation (dashes), flavor excitation (dots), and gluon splitting (dash-dots) are shown.  The contributions are normalized according to {\sc Pythia}'s cross section predictions and an arbitrary global normalization is used to give the best shape fit between data and Monte Carlo.}  Note that mistag subtraction applied to the individual { \sc Pythia} contributions can result in negative values for bins with few entries.  Consequently, the total {\sc Pythia} distribution can be less than one of the components in some bins. 
\end{figure}

	Figure~\ref{fig:1comp-fits} shows the comparison of the BVTX tag $\Delta\phi$ distribution from data to the distribution predicted by each Monte Carlo sample when the relative normalization of each production mechanism is based on the Monte Carlo prediction for the cross sections of the different production mechanisms.  From these $\Delta\phi$ comparisons it can be seen that each Monte Carlo model matches the qualitative features of the data, although there are definite differences in shape, as reflected by the poor $\chi^{2}$ values.  For the {\sc Pythia} sample with low initial-state radiation (ISR), the peak in the back-to-back region is too narrow, while for the medium and high ISR samples, the back-to-back peak is too broad.  Similarly, the {\sc Herwig} Monte Carlo sample also has a peak that is too broad at high $\Delta\phi$, perhaps even more so than in {\sc Pythia}.  However, aside from these discrepancies at high $\Delta\phi$, the rest of the $\Delta\phi$ distribution matches reasonably well between Monte Carlo and data using the normalizations predicted by the Monte Carlo generators for the different production mechanisms.  The $\chi^{2}$ values between the $\Delta\phi$ curves from Monte Carlo and data are listed in Table~\ref{table:prodfractions}.  On the basis of these $\chi^{2}$ values, it appears that {\sc Pythia} with the medium ISR value provides the best match to data when using the Monte Carlo's default normalization for the three production mechanisms.  Figure~\ref{fig:1comp-midisr} shows the breakdown of the contributions from the individual production mechanisms to the overall $\Delta\phi$ shape for this {\sc Pythia} sample.

\begin{figure}[htbp]
\includegraphics[width=8.6 cm]{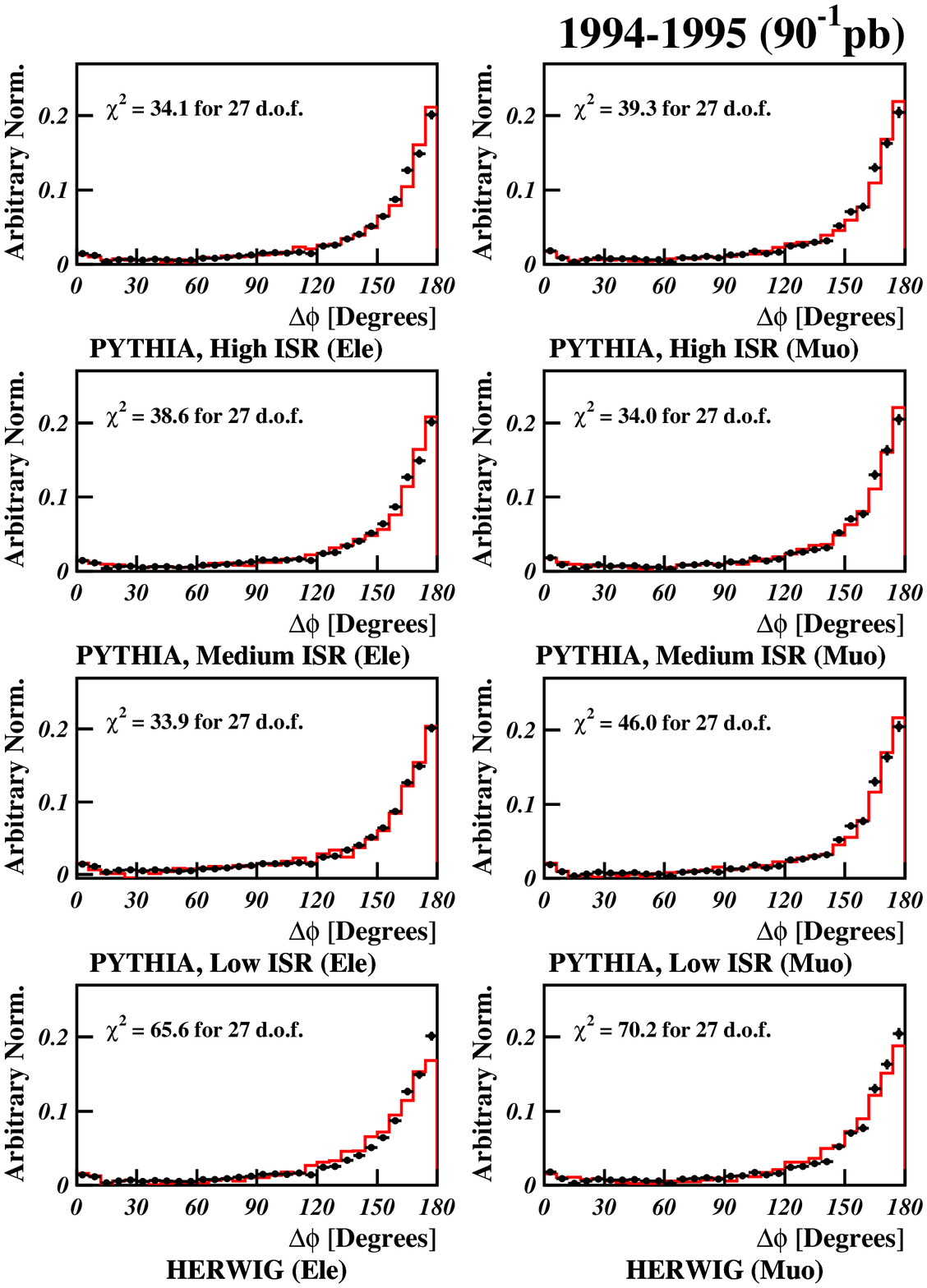}
\caption{\label{fig:3comp-fits}  Comparisons between the shape of the tag $\Delta\phi$ distribution for data (points) and the tag $\Delta\phi$ distribution predicted by the various Monte Carlo samples (line).  In these comparisons, the normalizations of each production mechanism were allowed to vary independently and were chosen to give the best fit between the Monte Carlo and the data.  Again, the fit  $\chi^{2}$ takes into account Monte Carlo statistics in addition to errors on the data.}
\end{figure}

\begin{figure}[htbp]
\includegraphics[width=8.6 cm]{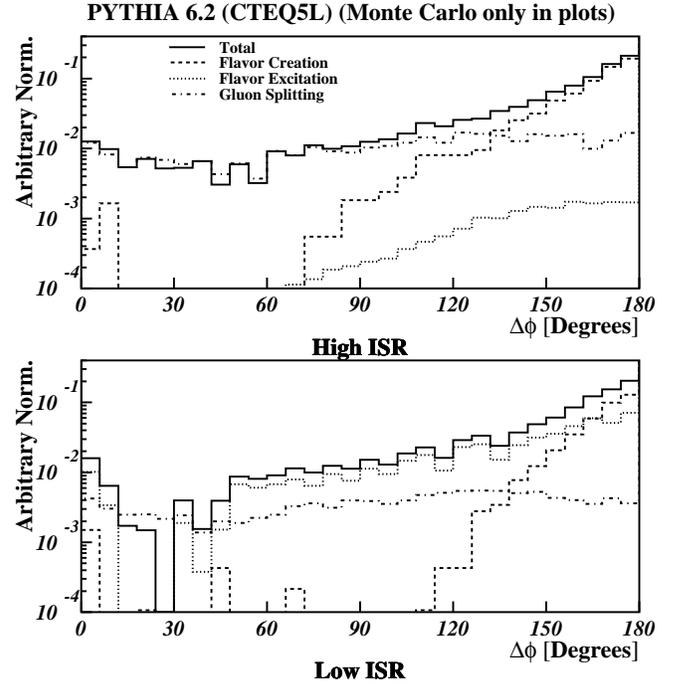}
\caption{\label{fig:3comp-pythia} A comparison of the contributions from flavor creation (dashes), flavor excitation (dots), and gluon splitting (dash-dots) to the total $\Delta\phi$ shapes (solid) for {\sc Pythia} with high ISR (top) and low ISR (bottom). Electron Monte Carlo is shown in the plots.  The muon plots can be found in Ref.~\cite{lannon-thesis}. The normalization of each component is set by the best fit of the three components to the    spectrum from data.  Note that mistag subtraction applied to the individual { \sc Pythia} contributions can result in negative values for bins with few entries.  Consequently, the total {\sc Pythia} distribution can be less than one of the components in some bins. }
\end{figure}

	However, since, in the parton shower approximation, the contributions from flavor creation, flavor excitation, and gluon splitting may be generated separately, each component can have a separate, arbitrary normalization and the three components can be fit for the combination of normalizations that gives the best match to the shape of the $\Delta\phi$ spectrum from data.  These fits are shown in Figure~\ref{fig:3comp-fits}.  When the normalizations of the individual components are allowed to float with respect to one another, one can obtain rather good agreement in shape between data and both the low ISR and high ISR {\sc Pythia} samples.  The fit of the low ISR {\sc Pythia} Monte Carlo to the data increases the broader contribution from flavor excitation to compensate for the narrowness of the back-to-back peak from flavor creation.  For the high ISR {\sc Pythia} samples, the peak at high $\Delta\phi$ is made narrower to match the data by all but eliminating the contribution from flavor excitation.  A comparison of the relative fractions of each production mechanism in the two {\sc Pythia} fits is shown in Figure~\ref{fig:3comp-pythia}.  The fit of the {\sc Herwig} sample to the data also tries to compensate for the excessive broadness of the {\sc Herwig} flavor creation peak at high $\Delta\phi$, but even after completely eliminating the flavor excitation contribution, the remaining contribution from flavor creation at high $\Delta\phi$ is too broad to model the data.  Table~\ref{table:prodfractions} compares the fit quality and effective contribution from flavor creation, flavor excitation, and gluon splitting in the fits of the various Monte Carlo samples to the data.  Both low ISR and high ISR {\sc Pythia} samples can be made to fit the data with approximately the same fit quality, which is unexpected, especially since the low ISR sample accomplishes this fit with a high flavor excitation content while the high ISR sample fits with almost no flavor excitation contribution.  In the end, there seems to be an ambiguity in {\sc Pythia} that allows a trade-off between initial state-radiation and the amount of flavor excitation.

\begin{table*}
\begin{tabular}{|c|rd|rd|rd|rd|}
\hline
\hline
 & \multicolumn{4}{c|}{Electrons}&\multicolumn{4}{c|}{Muons}\\
\cline{2-9}
 & \multicolumn{2}{c|}{Fixed Normalization}&\multicolumn{2}{c|}{Floating Normalization}& \multicolumn{2}{c|}{Fixed Normalization}&\multicolumn{2}{c|}{Floating Normalization}\\
\hline
{\sc Pythia}         & FC: & 43.7\% & FC: & 66.1\% & FC: & 41.4\% & FC: & 64.5\% \\
High ISR & FE: & 40.7\% & FE: &  1.7\% & FE: & 41.5\% & FE: &  8.1\% \\
               & GS: & 15.6\% & GS: & 32.2\% & GS: & 17.1\% & GS: & 27.4\% \\
$\chi^{2}/d.o.f.$&\multicolumn{2}{c|}{125.7/29}&\multicolumn{2}{c|}{34.1/27}&\multicolumn{2}{c|}{101.4/29}&\multicolumn{2}{c|}{39.3/27}\\
$\chi^{2}$ probability & \multicolumn{2}{c|}{$5.26\times10^{-14}$} & \multicolumn{2}{c|}{0.136} & \multicolumn{2}{c|}{$5.83\times10^{-10}$} & \multicolumn{2}{c|}{0.0595}\\
\hline
{\sc Pythia}         & FC: & 47.7\% & FC: & 65.3\% & FC: & 46.5\% & FC: & 63.2\% \\
Medium ISR & FE: & 35.8\% & FE: &  0.2\% & FE: & 35.3\% & FE: &  8.6\% \\
               & GS: & 16.5\% & GS: & 34.5\% & GS: & 18.2\% & GS: & 28.2\% \\
$\chi^{2}/d.o.f.$&\multicolumn{2}{c|}{83.9/29}&\multicolumn{2}{c|}{38.6/27}&\multicolumn{2}{c|}{78.2/29}&\multicolumn{2}{c|}{34.0/27}\\
$\chi^{2}$ probability & \multicolumn{2}{c|}{$3.07\times10^{-7}$} & \multicolumn{2}{c|}{0.0688} & \multicolumn{2}{c|}{$4.11\times10^{-6}$} & \multicolumn{2}{c|}{0.166}\\
\hline
{\sc Pythia}         & FC: & 68.3\% & FC: & 37.6\% & FC: & 63.9\% & FC: & 48.5\% \\
Low ISR & FE: & 12.0\% & FE: & 51.4\% & FE: & 13.9\% & FE: & 34.6\% \\
               & GS: & 19.7\% & GS: & 11.0\% & GS: & 22.2\% & GS: & 16.9\% \\
$\chi^{2}/d.o.f.$&\multicolumn{2}{c|}{167.8/29}&\multicolumn{2}{c|}{33.9/27}&\multicolumn{2}{c|}{85.2/29}&\multicolumn{2}{c|}{46.0/27}\\
$\chi^{2}$ probability & \multicolumn{2}{c|}{$1.76\times10^{-21}$} & \multicolumn{2}{c|}{0.169} & \multicolumn{2}{c|}{$1.96\times10^{-7}$} & \multicolumn{2}{c|}{0.0127}\\
\hline
{\sc Herwig}         & FC: & 57.6\% & FC: & 70.9\% & FC: & 55.7\% & FC: & 74.8\% \\
               & FE: & 24.0\% & FE: &  0.0\% & FE: & 23.1\% & FE: &  0.0\% \\
               & GS: & 18.4\% & GS: & 29.1\% & GS: & 21.2\% & GS: & 25.2\% \\
$\chi^{2}/d.o.f.$&\multicolumn{2}{c|}{97.5/29}&\multicolumn{2}{c|}{65.6/27}&\multicolumn{2}{c|}{111.1/29}&\multicolumn{2}{c|}{70.2/27}\\
$\chi^{2}$ probability & \multicolumn{2}{c|}{$2.45\times10^{-9}$} & \multicolumn{2}{c|}{$4.65\times10^{-5}$} & \multicolumn{2}{c|}{$1.52\times10^{-11}$} & \multicolumn{2}{c|}{$1.05\times10^{-5}$}\\
\hline
\hline
\end{tabular}
\caption{\label{table:prodfractions} Comparison of the effective contributions from flavor creation (FC), flavor excitation (FE), and gluon splitting (GS) to fits of the Monte Carlo $\Delta\phi$ to the data.  The fit $\chi^{2}$ takes into account Monte Carlo statistics in addition to errors on the data.  The ``$\chi^{2}$ probability'' entry refers to the probability of getting a worse fit, according to the $\chi^{2}$ distribution.}  
\end{table*}

\begin{figure}[htbp]
\includegraphics[width=8.6 cm]{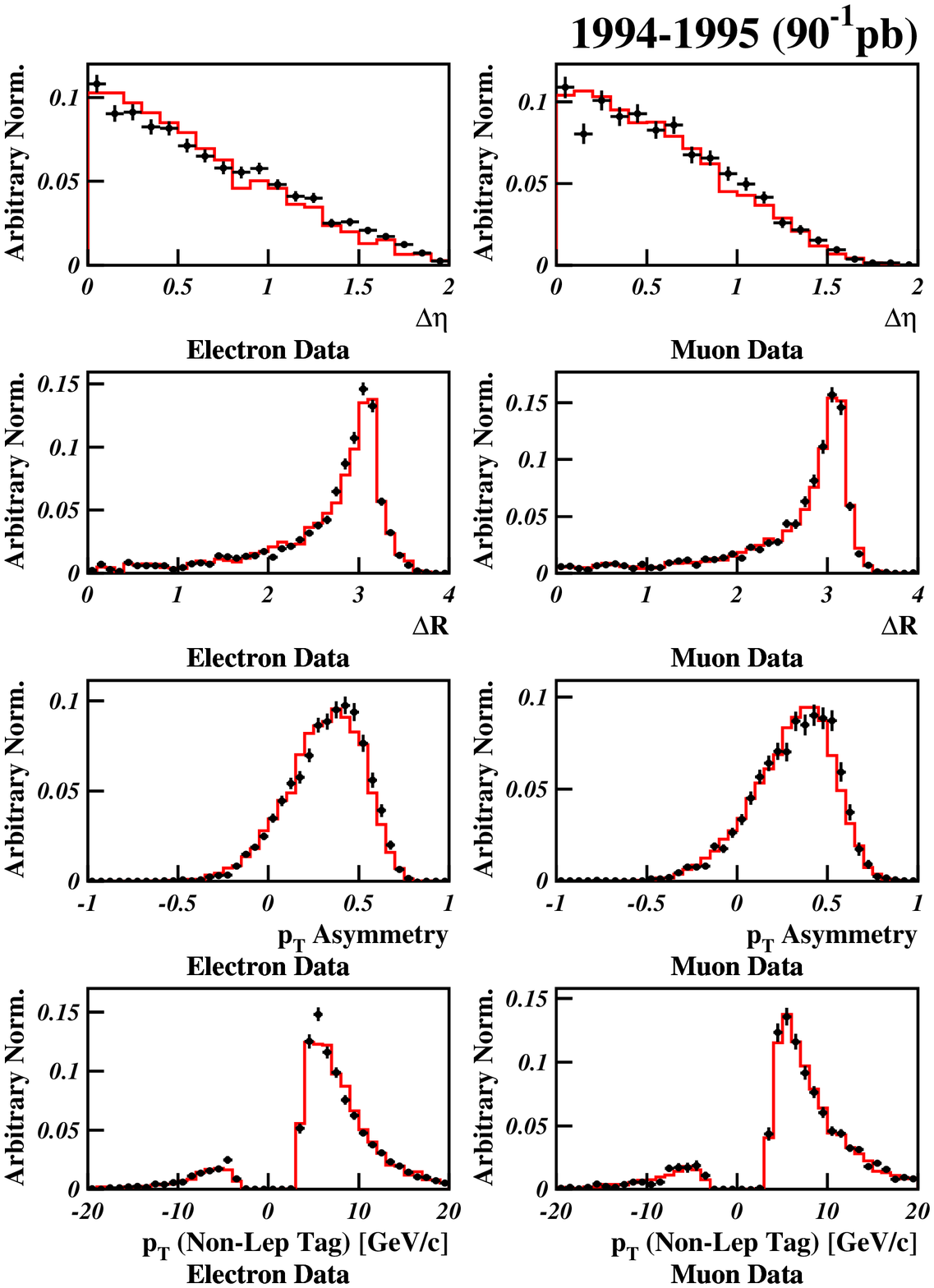}
\caption{\label{fig:other-corr} A comparison of {\sc Pythia} with high ISR to the data for several different correlation quantities.  The normalizations for the three production mechanisms in {\sc Pythia} have been determined by the fit of the {\sc Pythia} $\Delta\phi$ distributions to data. The $p_{T}$ asymmetry is given by $A_{p_{T}}=(p_{T}(lep)-p_{T}(non-lep)/(p_{T}(lep)+p_{T}(non-lep)$ .  In the $p_{T}(Non-Lep Tag)$ plot, the sign of the $p_{T}$ is determined by the opening angle between the lepton-tag and the non-lepton tag: negative for tag pairs with $\Delta\phi < 90^{\circ}$, positive otherwise.  The data are shown as points with statistical error bars only.  The solid line is {\sc Pythia} with high ISR}
\end{figure}

	In general, any of the Monte Carlo samples compared to the data shows reasonable qualitative agreement.  The Monte Carlo sample that best matches the data is {\sc Pythia} with medium or high ISR settings, when the individual normalizations of the flavor creation, flavor excitation, and gluon splitting are allowed to float separately to best fit the data.  Although the fit using {\sc Pythia} with low ISR is not so poor as to rule this model out completely, studies indicate that {\sc Pythia} with high initial state radiation does a better job of matching both the underlying event and minimum bias data at CDF~\cite{underlying-event}.  Therefore, we select the {\sc Pythia} sample, with high ISR and the relative normalizations of flavor creation, flavor excitation, and gluon splitting fixed by our fit to the $\Delta\phi$ distribution of the data, as the best Monte Carlo model of the data.  Comparisons indicate that the differences between {\sc Pythia} with medium or high ISR settings are minor.  Figure~\ref{fig:other-corr} shows a comparison of other correlations between the data and {\sc Pythia} with high ISR.  Although these plots show good agreement between the data and {\sc Pythia} for the overall shapes of the distributions, the shapes of the individual contributions from flavor creation, flavor excitation, and gluon splitting are not distinct enough to allow a separation of the components as was done for the $\Delta\phi$ distribution.

	It is interesting to note that before allowing the normalizations of each production mechanism to float in the fits, the agreement between {\sc Herwig} and the data is no worse than the agreement between the low ISR {\sc Pythia} sample and the data.  However, because the disparity between the data and the low ISR {\sc Pythia} sample comes from the narrowness in the flavor creation peak at high $\Delta\phi$, when the normalizations are allowed to float, the fit can alleviate the disagreement by increasing the peak width through a higher contribution from flavor excitation.  In contrast, for {\sc Herwig}, once the contribution from flavor excitation has been reduced to zero, the fit has no way to make the width of the back-to-back flavor creation peak smaller, short of the unphysical situation of setting the flavor excitation normalization negative.  If there were some other parameter for {\sc Herwig}, like {\sc Pythia}'s initial state radiation parameter, PARP(67), that could be used to tune the width of the back-to-back flavor creation peak, it may be possible to achieve good agreement between {\sc Herwig} and the data as well.

The results presented here can be compared to another analysis of lepton tags in  heavy-flavor events  presented in Ref.~\cite{cdf-04}.  That analysis compares {\sc Herwig} to double-tagged events using higher $E_T$ jet samples.  In addition to using a sample of double-tagged events at higher momentum, Ref.~\cite{cdf-04} also differs from this analysis in that it uses calorimeter based jets as opposed to the tracking jets utilized here, and the quantity measured is the azimuthal opening angle between the tagged jets rather than the angle between the tags themselves.  In agreement with this analysis, that one clearly shows the importance of the higher-order contributions in heavy flavor production, and also shows an agreement with {\sc Herwig} that is better than the agreement seen in this analysis.  Perhaps this suggests that the disagreement shown here between {\sc Herwig} and the data is related to {\sc Herwig}'s ability to model low $p_{T}$ $b$ production or $b$ fragmentation.

\subsection{Corrections and Systematics} \label{sec:corr-and-syst}

The correlations examined so far in the data involve pairs of BVTX tags, rather than pairs of $B$ hadrons.  
There are detector effects, such as the tagging efficiency for pairs of $B$ hadrons as a function of $\Delta\phi$, 
that distort the shape of the measured tag pair correlations from the true $B$ hadron distribution.  In addition, 
residual contributions from background can affect the shape of the tag pair distribution. For the comparison 
between MC and data, the detector effects are accounted for by using a detector and trigger simulation to 
adjust the MC to match the conditions in the data, while the backgrounds are assumed to be negligible.  
However, since the MC models examined in section~\ref{sec:mc-data-comp} match the data reasonably well, 
MC events can be used to determine the relationship between the measured tag pair distribution and 
the actual $B$ hadron distribution. In the sections below, two kinds of corrections to the tag pair 
$\Delta\phi$ distribution are considered: a correction for the relative tagging efficiency, which is 
a detector effect, and a correction for the contributions from mistags, prompt charm, and sequential decays 
that remain in the data after the steps taken in section~\ref{sec:sample-comp} to remove backgrounds.  
In addition, the MC is used to estimate the systematic uncertainties associated with correcting for 
the relative tagging efficiency and removing background events. These corrections and systematic 
errors are evaluated using several different MC samples to account for uncertainties involved in the MC model itself.


The BVTX tagging algorithm is not equally effective for all topologies of $b\overline{b}$ production.  In particular, it becomes more difficult for the BVTX algorithm to reconstruct both displaced secondary vertices as the opening angle between the two $B$ hadrons decreases.  This effect becomes especially 
severe when the two $B$ hadrons are both contained within the cone of a single jet for track clustering 
purposes. Furthermore, correlations between opening angle and $p_{T} (B)$ for the various production 
mechanisms can lead to differences in the relative efficiency for reconstructing tag pairs at different 
opening angles. These effects distort the shape of the $\Delta\phi$ distribution measured for tags from 
the true $B\overline{B}$ $\Delta\phi$ distribution.

We correct for these relative efficiency effects using the MC that best matches the data, as determined 
in section~\ref{sec:mc-data-comp}. Because we are only examining the shape of the $\Delta\phi$ distribution, 
our goal in making this correction is only to account for differences in the relative efficiency of the 
tagging algorithm, as a function of $\Delta\phi$. We do not attempt to correct for effects that impact 
all parts of the $\Delta\phi$ spectrum equally. For example, an overall shift in the muon trigger 
efficiency would not affect this correction. To determine the correction for each bin we take the ratio 
of the number of tag pairs reconstructed in the MC to the number of pairs that could have been reconstructed 
if the tagging algorithm had perfect efficiency. The number of tag pairs that would have been reconstructed
 assuming perfect efficiency is determined by looking at the generator level $B$ hadron $\Delta\phi$ distribution.  
For electron MC, to simulate the electron trigger, we require one $B$ hadron in the event to have a 
$p_{T} > 14.0 \;{\rm GeV}/c$ and $|\eta| < 1.0$. For the muon MC, we demand one $B$ hadron with 
$p_{T} > 14.0 \;{\rm GeV}/c$ and $|\eta| < 0.6$. For both cases, we require a second $B$ hadron with 
$p_{T} > 7.5 \; {\rm GeV}/c$ and $|\eta| < 1.0$. The cuts placed on the generator-level MC were determined 
by examining the $p_{T}$ and $\eta$ distributions for $B$ hadrons from MC events in which two BVTX tags were 
reconstructed. The $p_{T}$ and $\eta$ values were chosen by determining the cuts for which 90\% of the $B$ 
hadrons in the double-tagged MC events would pass. We take the $\Delta\phi$ distribution resulting from the 
event selection above and convolute it with a Gaussian resolution function with a width of 0.1086 radians, 
characteristic of the $\Delta\phi$ resolution of the BVTX tagging algorithm as measured in MC.  

\begin{figure}[htbp]
\includegraphics[width=8.6 cm]{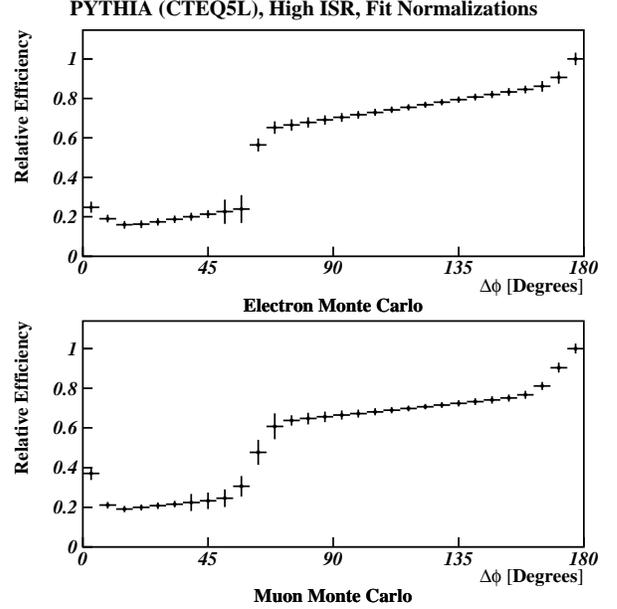}
\caption{\label{fig:releff} The bin-by-bin values for the relative efficiency returned by the fit.  The curves have been normalized so that the last $\Delta\phi$ bin has a value of one by definition.  The error bars on these curves indicate the statistical error on the bin values returned from the fit.  The statistical errors for the fit are correlated from bin to bin.}
\end{figure}

In order to minimize the effect of statistical fluctuations in the tagging efficiency determined from MC, 
we fit the tagging efficiency to an empirical function of the following form:

\begin{eqnarray}
\epsilon(\Delta\phi) &=& P_{1}\exp\left\{-\frac{1}{2}\left(\frac{\Delta\phi^{2}}{P_{2}^{2}}\right)\right\} + \nonumber\\
& &P_{3}\exp\left\{-\frac{1}{2}\left[\frac{(\Delta\phi - \pi)^{2}}{P_{4}^{2}}\right]\right\} + \nonumber\\
& &P_{5}\mathrm{freq}\left(\frac{\Delta\phi-P_{6}}{P_{7}}\right)+P_{8}(\Delta\phi)+P_{9},
\end{eqnarray}
where $\mathrm{freq}$ is the error function.  The relative efficiency curve resulting from this fit is shown in Fig.~\ref{fig:releff}.  The sharp step around $\phi = 60^{\circ}$, which is modeled by the error function term, comes from the transition from the case of finding secondary vertex tags in two separate jets to finding secondary vertex tags in the same jet. 
Since we are only interested in the effect of the efficiency on the shape of the $\Delta\phi$ distribution, 
and not on its absolute normalization, we have rescaled the curve in Fig.~\ref{fig:releff} 
so that the relative efficiency in the last $\Delta\phi$ bin is defined to be unity.  
Thus this curve shows the effect of the BVTX tagging efficiency for a given bin relative to the last $\Delta\phi$ bin.

There are two main contributions to the systematic uncertainty associated with the relative tagging efficiency correction. 
First, the statistical errors on the fit value for the relative efficiency correction factor should be propagated into 
systematic uncertainties on the corrected $\Delta\phi$ distribution. There is an additional systematic uncertainty that 
comes from the model used to calculate the relative efficiency correction. The {\sc Pythia} MC sample, with high ISR
and with the normalization of the different production mechanisms taken from the fit to the $\Delta\phi$ distribution 
in the data, is used as our baseline for the relative efficiency correction. However, other models, like the lower 
ISR {\sc Pythia} sample or {\sc Herwig} also match the data to varying degrees and so could also have been used.  
To account for this ambiguity, we compare the relative efficiency corrections from other MC models to our baseline model.  
In the worst case, the difference for the bin-by-bin relative efficiency correction factor is approximately equal in 
magnitude to the statistical error from the fit. Therefore, to account to modeling uncertainties in the relative 
efficiency correction, we increase the systematic error associated with the correction by a factor of $\sqrt{2}$ .
	

The mistag subtraction scheme used for this analysis relies on the assumption that 100\% of 
legitimate tags and 50\% of mistags have positive $L_{xy}$. The true fraction may be somewhat different.  
For example, if most of the events contain at least one $B$ hadron, then the $L_{xy}$ distribution of 
mistags may be biased towards positive values by the presence of actual displaced tracks in the events.  
Furthermore, the bias in $L_{xy}$ may depend on the topology of the event. To investigate any possible 
bias in the $L_{xy}$ distribution of mistags, we examined MC events containing mistags identified by 
matching tracking information to MC truth information. From MC sample to MC sample, the fraction of 
legitimate secondary vertex tags that have positive $L_{xy}$ varies from 0.97 to 1.0. For mistags, 
the positive $L_{xy}$ fraction varies from 0.45 to 0.55.  To estimate the possible effect of using 
the wrong fractions when performing mistag subtraction, we redo the mistag subtraction in the data 
using different assumptions about the positive $L_{xy}$ fraction for good tags and mistags.  
The mistag subtraction formula (Eq.~\ref{eq:mistag-sub}), generalized for an arbitrary fraction 
$p$ of good tags with positive $L_{xy}$ and an arbitrary fraction $q$ of mistags with positive $L_{xy}$, 
is given by  

\begin{eqnarray}
N_{GG} = \frac{(q-1)^{2}}{(q-p)^{2}}N_{++} + \frac{q(q-1)}{(q-p)^{2}}N_{+-} \nonumber\\
+ \frac{q^{2}}{(q-p)^2}N_{--} + \frac{q(q-1)}{(q-p)^2}N_{-+}
\label{eq:gen-mistag-sub}
\end{eqnarray}


Changing the positive $L_{xy}$ fractions from mistag subtraction affects both the normalization 
and the shape of the $\Delta\phi$ distribution. However, we are only concerned about the shape 
for this analysis. Therefore, before we compare the shape of the $\Delta\phi$ distribution using 
the standard mistag subtraction scheme to the shape obtained using alternative values for 
the positive $L_{xy}$ fractions, we normalize the distributions to unit area. To estimate the 
systematic error from mistag subtraction, we take the $\Delta\phi$ distributions calculated varying the $p$ 
and $q$ values in Eq.~\ref{eq:gen-mistag-sub} within their allowed ranges and fit them to the 
functional form for the $\Delta\phi$ distribution, given below, in order to minimize the effect of statistical fluctuations:

\begin{eqnarray}
f(\Delta\phi) & = &P_{1}\exp\Big[-\frac{1}{2}\Big(\frac{\Delta\phi}{P_{2}}\Big)^{2}\Big] + \nonumber\\ 
& & P_{3}\exp\Big[P_{4}(\Delta\phi-\pi)^{2} + P_{5}(\Delta\phi)\Big] + \nonumber\\
& & P_{6}\Delta\phi+P_{7}
\end{eqnarray}


We then calculate the maximum deviation between the result from the default mistag subtraction 
scheme and the results obtained from varying the positive $L_{xy}$ fractions. This maximum 
deviation is assigned as the systematic error on the $\Delta\phi$ shape from mistag subtraction.

\begin{figure}[htbp]
\includegraphics[width=8.6 cm]{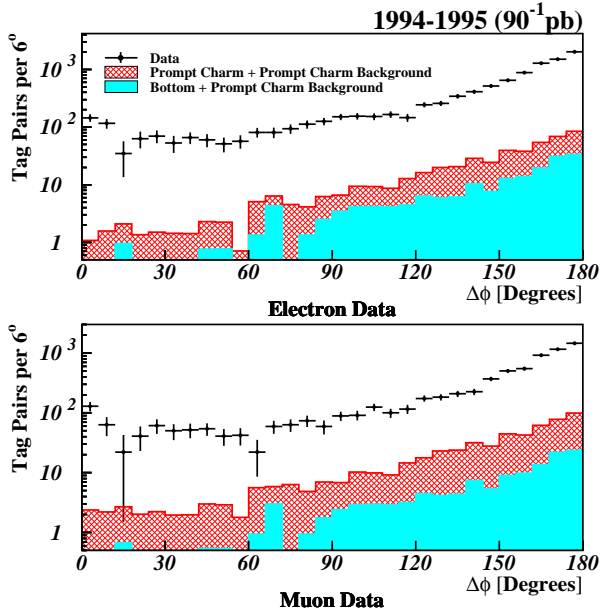}
\caption{\label{fig:charm-bkg} The estimated shape of the background from direct $c\overline{c}$ production (the hatched area) and multiple-heavy flavor ($b\overline{b}+b\overline{b}$ and $b\overline{b}+c\overline{c}$) production (the solid area).  The points with error bars show the data.}
\end{figure}

The bin-by-bin contribution to the double-tag $\Delta\phi$ distribution from prompt charm 
is estimated primarily using MC. The overall amounts of prompt charm and $b\overline{b} + c\overline{c}$ 
double tags are estimated by comparing the relative rate of obtaining a double-tagged $b\overline{b}$ 
MC event to the rates for double-tagging $c\overline{c}$ and $b\overline{b} + c\overline{c}$ MC events.  
This approach estimates that 2.9\% (6.0\%) of the tag pairs in this sample come from $c\overline{c}$ 
production for electron (muon) data, and 1.8\% of the tag pairs in both the electron and the muon samples  
comes from $b\overline{b} + c\overline{c}$ production. 
The $\Delta\phi$ shape for the $c\overline{c}$ and $b\overline{b} + c\overline{c}$ contributions is 
estimated by applying the measured relative tagging efficiency as a function of $\Delta\phi$ to 
the generator level $c\overline{c}$ and $b\overline{b} + c\overline{c}$ $\Delta\phi$ distributions.  
The resulting estimated contamination from prompt charm to the double-tag $\Delta\phi$ distribution 
is shown in Fig.~\ref{fig:charm-bkg}. The systematic error on this correction is estimated by 
performing several checks on the data. One check involves comparing the $\Delta\phi$ spectrum 
for double-tagged events in which the invariant mass of the tracks for each tag is greater than 2 GeV/$c^{2}$ 
to the spectrum when both tags have an invariant mass less than 2 GeV/$c^{2}$. The former sample is enhanced 
in $b\overline{b}$ content relative to prompt charm, while the latter sample has a greater contribution from 
prompt charm. Both subsamples have far fewer statistics than the main sample.  The $\Delta\phi$ shapes of 
these two subsamples agree within the statistics of the samples, suggesting a negligible contribution from prompt charm.  
An alternative estimation of the prompt charm contribution can be obtained by fitting the tag mass distribution to 
template shapes derived from $b\overline{b}$ events (including tags of secondary charmed mesons) and $c\overline{c}$ events. 
The results of these fits suggest a prompt charm contamination roughly a factor of two larger than the MC estimates, 
although still a relatively small contribution at 7.1\% for the electron data and 13.3\% for the muon data.  
As a result of the differences between these two alternate estimates of the prompt charm contribution and 
the MC method used to set the normalization of our prompt charm correction, we set the systematic error 
on the prompt charm correction equal to the size of the correction in each bin.

The MC is also used to determine the residual contribution from sequential double tags.  
Based on examining MC events in which two tags are identified to come from the same $B$ decay, 
we determine that after mistag subtraction, 25.9\% of the tags removed by the 6 GeV/$c^{2}$ 
mass cut were from sequential tag pairs. Furthermore, using Monte Carlo it was also determined that for every 100 mistags removed by the 6 GeV/c$^{2}$ mass cut, roughly 2.41 events remained in this sample, yielding an efficiency for this cut of 97.6\%.  In the data, after mistag subtraction, the 6 GeV/$c^{2}$ 
invariant mass cut removes 471 tags from the electron sample and 598 tags from the muon sample.  
Using the numbers derived from the MC above, this means that of the tags removed by the invariant 
mass cut, 122.1 electron and 155.0 muon tags come from sequential double tag pairs, 
and an estimated 2.9 electron sequential tag pairs and 3.7 muon sequential tag pairs remain in the data after this cut.  
The $\Delta\phi$ distribution of the sequential tag pairs is also determined using MC to be well described 
by the positive half of a Gaussian distribution with a mean of zero and a width of 0.122 radians.  To correct for the sequential 
double tag contribution in the data, we take the estimated number of sequential double tags, with a half-Gaussian 
distribution as described above, and subtract them from the $\Delta\phi$ bins in the data.  
The systematic error on this correction is set equal to the size of the correction.

\subsection{Final Distribution and Comments}

\begin{figure}[htbp]
\includegraphics[width=6.6 cm]{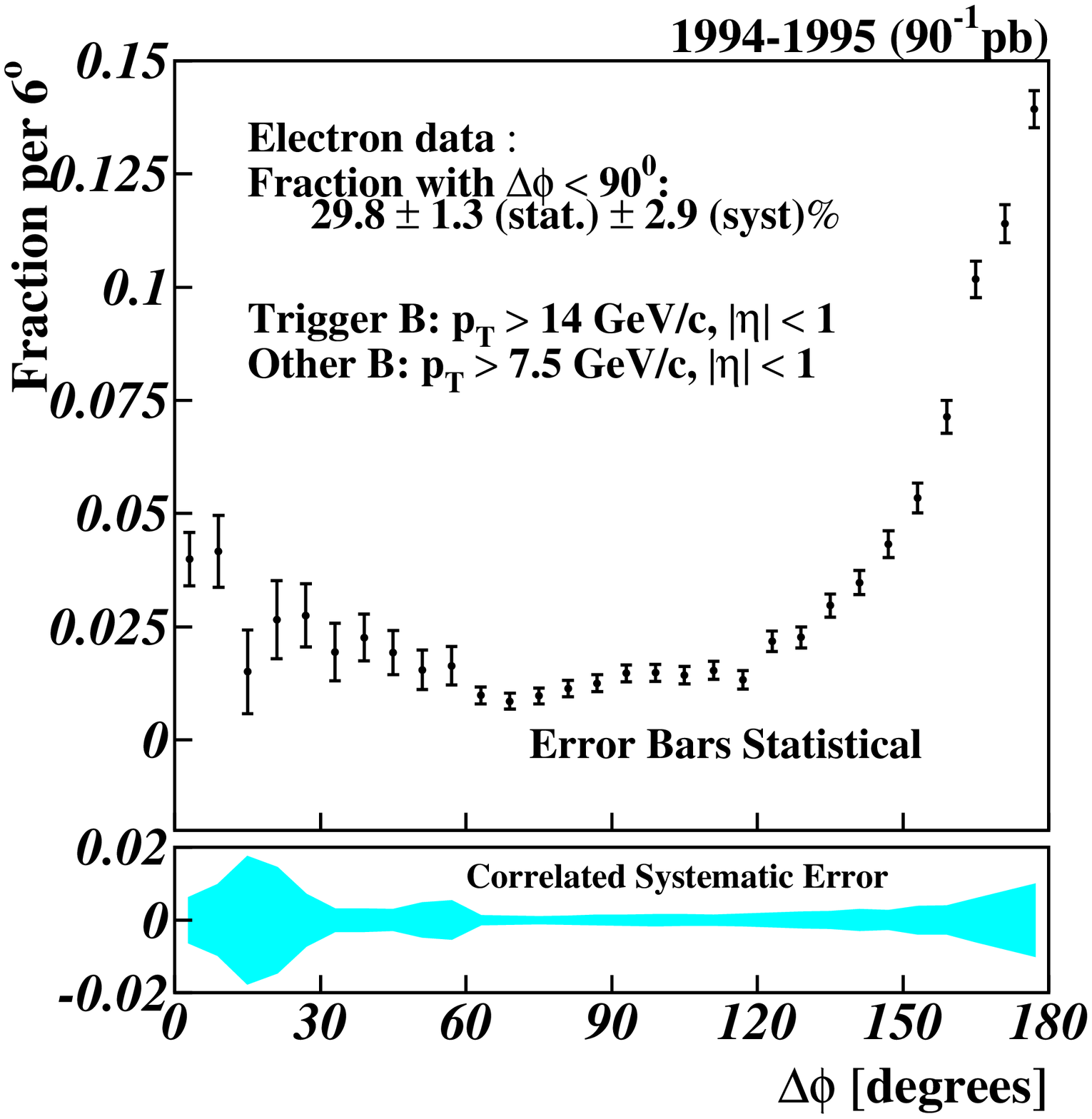}
\includegraphics[width=6.6 cm]{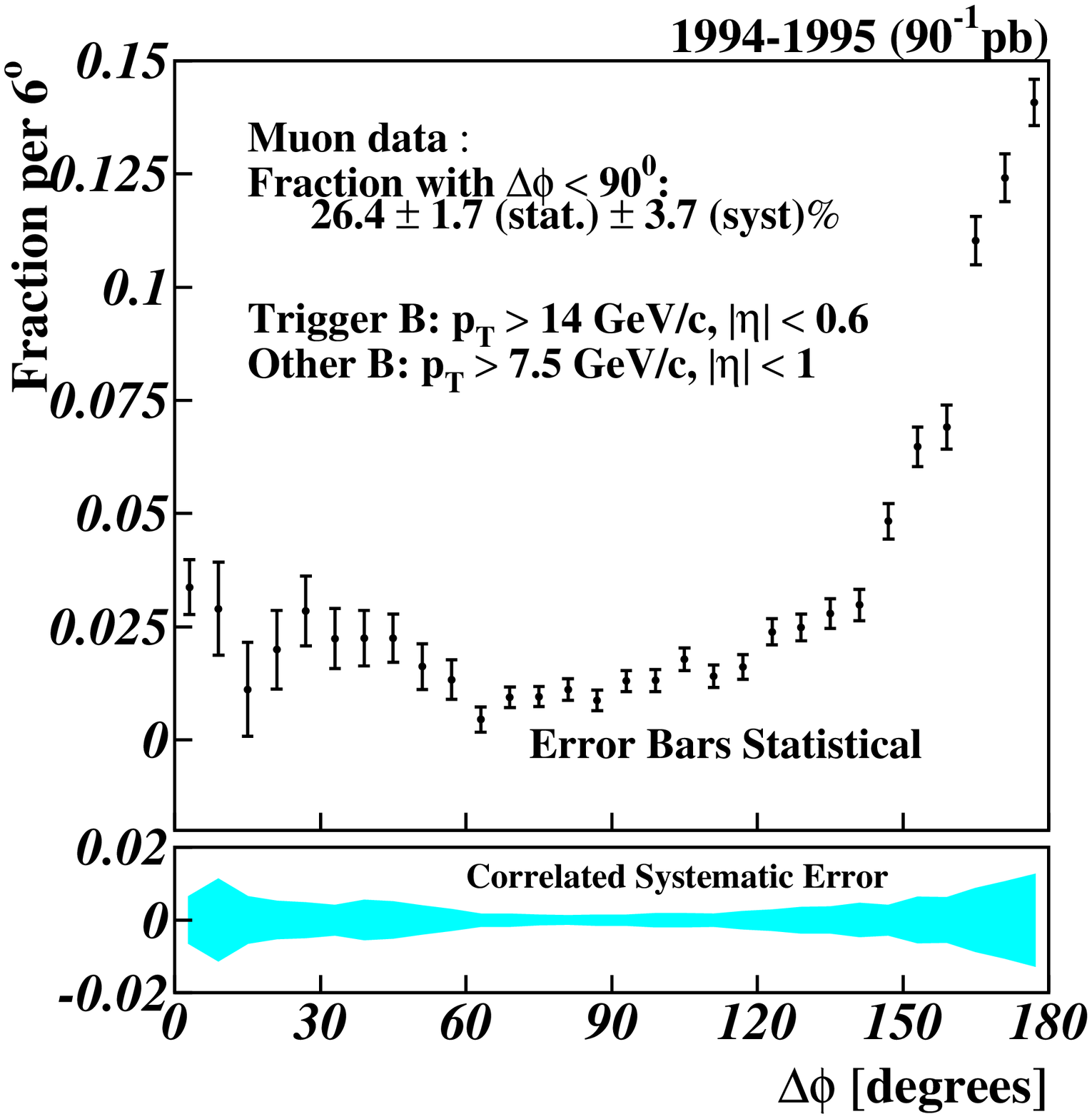}
\caption{\label{fig:dphi-corrected-emu} The final, corrected $\Delta\phi$ distribution for the double-tagged electron (left) and muon (right) data.  The corrections made to the data include mistag subtraction, sequential removal, prompt charm subtraction, and the relative tagging efficiency correction.  The error bars display the statistical error on the points.  The filled region at the bottom indicates the systematic errors.  The systematic errors are correlated from bin to bin.  Mistag subtraction provides the dominant contribution to the systematic errors.}
\end{figure}

\begin{figure}[htbp]
\includegraphics[width=8.6 cm]{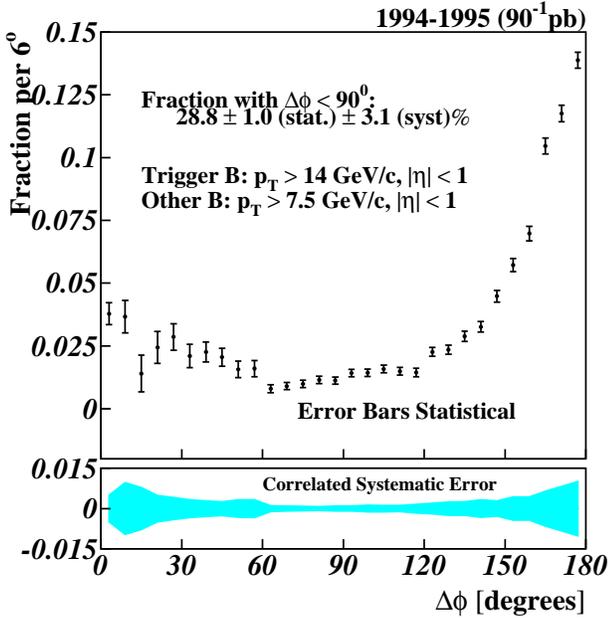}
\caption{\label{fig:dphi-corrected-comb}  The combined, corrected electron and muon $\Delta\phi$ distribution from the double-tagged analysis.  In making this plot, we ignored the difference in $\eta$ acceptance between the electron and muon triggers.  The corrections made to the data include mistag subtraction, sequential removal, prompt charm subtraction, and the relative tagging efficiency correction.  The error bars display the statistical error on the points.  The filled region at the bottom indicates the systematic errors.  The systematic errors are correlated from bin to bin.  Mistag subtraction provides the dominant contribution to the systematic errors.}
\end{figure}

	Figure~\ref{fig:dphi-corrected-emu} shows the final, corrected tag $\Delta\phi$ distribution, including systematic errors.  To obtain this distribution, the contributions from residual sequentials and prompt charm are removed from the mistag-subtracted distributions.  Then the relative efficiency corrections derived in Section~\ref{sec:corr-and-syst} are applied.  Systematic errors from the various corrections are combined in quadrature to give the total systematic error.  Mistag subtraction gives the largest contribution to the systematic error.  The final corrected tag $\Delta\phi$ distribution provides a measurement of the $B - \overline{B}$ $\Delta\phi$ distribution where the $B$ providing the trigger electron (muon) has $p_{T} > 14.0$ GeV/c and $|\eta| < 1.0 (0.6)$, and the other $B$ has $p_{T} > 7.5$ GeV/c and $|\eta| < 1.0$, with a  $\Delta\phi$  resolution of $6.22^{\circ}$.  This distribution can be compared to generator-level $\Delta\phi$ distributions from Monte Carlo that have been convoluted with a Gaussian resolution function to account for our $\Delta\phi$ resolution.  Finally, ignoring the small difference in $\eta$ acceptance between the electron and muon samples, these two distributions can be combined to give the overall $B$ hadron $\Delta\phi$ distribution, shown in Figure~\ref{fig:dphi-corrected-comb}.  Table~\ref{table:final-results} specifies the corrected fraction in each $\Delta\phi$ bin as well as the breakdown of the systematic errors for each bin.

\begin{table*}
\begin{ruledtabular}
\begin{tabular}{cccc|cccc}
    &          &                    &                  &\multicolumn{4}{c}{Systematic Error Components} \\
Bin	& Fraction &	Statistical Error &	Systematic Error & Sequential & Prompt Charm & Mistag Subtraction & Relative Efficiency \\
\hline
$  0^{\circ}-  6^{\circ}$ & 0.03901 &	0.00462 & 0.00593 & 0.00060             & 0.00051 & 0.00421 & 0.00411 \\
$  6^{\circ}- 12^{\circ}$ &	0.03765 & 0.00684 & 0.01042 & 0.00044             & 0.00082 & 0.00982 & 0.00336 \\
$ 12^{\circ}- 18^{\circ}$ &	0.01347 & 0.00774 & 0.00833 & 0.00013             & 0.00125 & 0.00810 & 0.00149 \\
$ 18^{\circ}- 24^{\circ}$ & 0.02498 & 0.00674 & 0.00544 & 0                   & 0.00084	& 0.00472 & 0.00257 \\
$ 24^{\circ}- 30^{\circ}$ &	0.02942 & 0.00561 & 0.00472	& 0                   & 0.00087 & 0.00370 & 0.00279 \\
$ 30^{\circ}- 36^{\circ}$ &	0.02152 & 0.00493 & 0.00372	& 0                   & 0.00074 & 0.00309 & 0.00194 \\
$ 36^{\circ}- 42^{\circ}$ &	0.02323 & 0.00420 & 0.00336	& 0                   & 0.00069 & 0.00256 & 0.00206 \\
$ 42^{\circ}- 48^{\circ}$ &	0.02077 & 0.00379 & 0.00298	& 0                   & 0.00101 & 0.00211 & 0.00185 \\
$ 48^{\circ}- 54^{\circ}$ &	0.01568 & 0.00349 & 0.00472	& 0                   & 0.00093 & 0.00171 & 0.00380 \\
$ 54^{\circ}- 60^{\circ}$ &	0.01651 & 0.00344 & 0.00461	& 0                   & 0.00043 & 0.00137 & 0.00438 \\
$ 60^{\circ}- 66^{\circ}$ &	0.00751 & 0.00167 & 0.00114 & 0                   & 0.00087 & 0.00054 & 0.00049 \\
$ 66^{\circ}- 72^{\circ}$ &	0.00869 & 0.00151 & 0.00102 & 0                   & 0.00084 & 0.00037 & 0.00044 \\
$ 72^{\circ}- 78^{\circ}$ & 0.00973 & 0.00153 & 0.00090 & 0                   & 0.00073 & 0.00030 & 0.00045 \\
$ 78^{\circ}- 84^{\circ}$ &	0.01156 & 0.00156 & 0.00079 & 0                   & 0.00059 & 0.00025 & 0.00047 \\
$ 84^{\circ}- 90^{\circ}$ &	0.01100 & 0.00155 & 0.00097 & 0                   & 0.00085 & 0.00022	& 0.00040 \\
$ 90^{\circ}- 96^{\circ}$ &	0.01423 & 0.00157 & 0.00130 & 0                   & 0.00084 & 0.00023 & 0.00046 \\
$ 96^{\circ}-102^{\circ}$ &	0.01395 & 0.00160 & 0.00128 & 0                   & 0.00121 & 0.00026 & 0.00040 \\
$102^{\circ}-108^{\circ}$ &	0.01559 & 0.00162 & 0.00120 & 0                   & 0.00117 & 0.00033 & 0.00040 \\
$108^{\circ}-114^{\circ}$ &	0.01474 & 0.00163 & 0.00172 & 0                   & 0.00106 & 0.00044 & 0.00034 \\
$114^{\circ}-120^{\circ}$ &	0.01370 & 0.00177 & 0.00212 & 0                   & 0.00159 & 0.00057 & 0.00029 \\
$120^{\circ}-126^{\circ}$ &	0.02203 & 0.00187 & 0.00259 & 0                   & 0.00195 & 0.00071 & 0.00045 \\
$126^{\circ}-132^{\circ}$ &	0.02244 & 0.00193 & 0.00268 & 0                   & 0.00242 & 0.00082 & 0.00045 \\
$132^{\circ}-138^{\circ}$ &	0.02813 & 0.00213 & 0.00344 & 0                   & 0.00246 & 0.00088 & 0.00059 \\
$138^{\circ}-144^{\circ}$ &	0.03128 & 0.00223 & 0.00303 & 0                   & 0.00328 & 0.00080 & 0.00069 \\
$144^{\circ}-150^{\circ}$ &	0.04471 & 0.00249 & 0.00465 & 0                   & 0.00279 & 0.00051 & 0.00106 \\
$150^{\circ}-156^{\circ}$ &	0.05622 & 0.00275 & 0.00466 & 0                   & 0.00444 & 0.00018 & 0.00137 \\
$156^{\circ}-162^{\circ}$ &	0.06983 & 0.00306 & 0.00716 & 0                   & 0.00419 & 0.00113 & 0.00169 \\
$162^{\circ}-168^{\circ}$ &	0.10516 & 0.00341 & 0.00914 & 0                   & 0.00583 & 0.00267 & 0.00319 \\
$168^{\circ}-174^{\circ}$ &	0.11783 & 0.00346 & 0.01105 & 0                   & 0.00688 & 0.00457 & 0.00391 \\
$174^{\circ}-180^{\circ}$ &	0.13944 & 0.00336 & 0.00419 & 0                   & 0.00779 & 0.00662 & 0.00419 \\
\end{tabular}
\caption{\label{table:final-results} The corrected fraction of combined electron and muon data in each bin as well as a breakdown of the components of the systematic errors on each bin.  The total systematic error is the sum in quadrature of the individual components.}  
\end{ruledtabular}
\end{table*}

	From the corrected data, we can also calculate the fraction of tag pairs in the ``towards'' region, defined by $\Delta\phi < 90^{\circ}$ .  This fraction is of interest because $\Delta\phi$  production in the ``towards'' region is dominated by the higher order production diagrams.  The towards fraction provides a single figure of merit to indicate the relative sizes of the contributions from flavor excitation and gluon splitting.  To account for correlated systematic errors, we calculate the towards fraction for our data by essentially repeating the analysis with two $\Delta\phi$ bins instead of thirty, and then taking the ratio of the ``towards'' bin over the total.  For the electron data, we obtain a towards fraction of $29.8 \pm 1.3 (stat.) \pm  2.9 (syst.)\%$.  For muon data, we obtain a towards fraction of $26.4 \pm  1.7 (stat.) \pm  3.7 (syst.)\%$.  The electron and muon samples are combined to give a towards fraction of $28.8 \pm  1.0 (stat) \pm  3.1 (syst)\%$.  Table~\ref{table:toward-away} shows the uncorrected number of tag pairs in the ``towards'' and ``away'' bins in the data and gives the corrections applied to obtain the final number.  Table~\ref{table:systematic-breakdown} breaks down the contributions to the systematic uncertainty on the towards fraction.

\begin{table*}
\begin{ruledtabular}
\begin{tabular}{ccccc}
	                               &\multicolumn{2}{c}{Electrons}&\multicolumn{2}{c}{Muons} \\
                                      & Towards & Away   & Towards & Away \\
\hline
Mistag-Subtracted Data                & 1210    & 8887   & 832     & 6260 \\
\hline
Charm Contamination                   & 42.1    & 442.6  & 52.9    & 500.3 \\
Sequential Contamination              & 2.9     & 0.0    & 3.7     & 0.0 \\ 
Relative Efficiency Correction Factor & 0.326   & 1.0    & 0.376   & 1.0 \\
\hline
Corrected Data                        & 3573.6  & 8444.4 & 2062.2  & 5759.7 \\
\end{tabular}
\caption{\label{table:toward-away} The number of events in the ``towards'' and ``away'' regions before and after applying corrections to the data.}  
\end{ruledtabular}
\end{table*}

\begin{table*}
\begin{ruledtabular}
\begin{tabular}{ccc}
	                                                & Electrons      & Muons \\
\hline
Towards Fraction                                  & 29.8\%         & 26.4\% \\
\hline
Statistical Error                                 & $\pm1.3\%$     & $\pm1.7\%$ \\
\hline
Mistag Subtraction Systematic Error               & $\pm2.0\%$	   & $\pm2.8\%$ \\
Sequential Removal Systematic Error               & $\pm0.05\%$    & $\pm0.09\%$ \\
Charm Subtraction Systematic Error                & $\pm1.3\%$     & $\pm2.2\%$ \\
Relative Efficiency Correction Systematic Error   & $\pm1.6\%$     & $\pm1.1\%$ \\
\hline
Total Systematic Error                            & $\pm2.9\%$     & $\pm3.7\%$ \\
\end{tabular}
\caption{\label{table:systematic-breakdown} The break-down of the systematic errors by contribution.  The total systematic error is the quadrature sum of the individual components.}  
\end{ruledtabular}
\end{table*}

\section{ ${\boldmath J/\psi}$--lepton ${\boldmath b}$ Quark Correlation Measurement}
\subsection{Overview}

\begin{figure}[htbp]
               \includegraphics[width=8.6 cm]{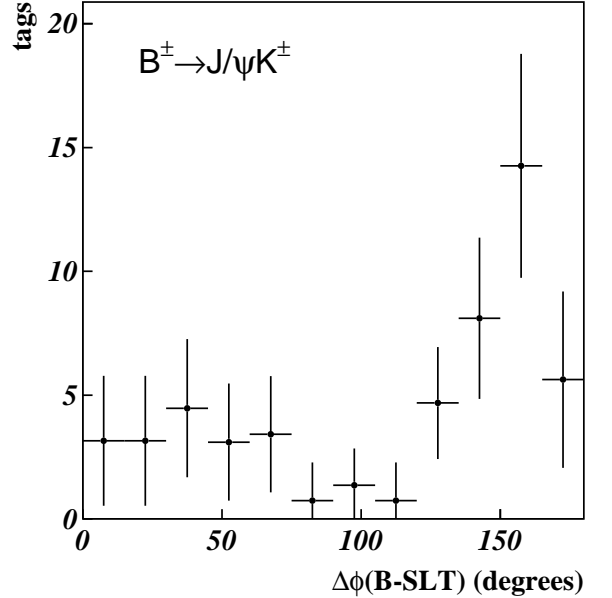}
\caption{\label{fig:sin2betastudy}
Sideband-subtracted $\Delta\phi$ distribution between fully
reconstructed $B^+ \rightarrow J/\psi K^+$ and soft leptons.}
\end{figure}

This measurement is optimized to measure the region in phase space least
understood in experimental measurements and theoretical predictions:
small $\Delta\phi$ where both bottom quarks point in the same
azimuthal direction.   As stated previously, earlier bottom quark angular
production measurements had little sensitivity to this region.  A study of opposite
side flavor tags using soft leptons for the CDF $\sin{2\beta}$
measurement~\cite{sin2betaprd} showed a significant number
of tags at small opening angles between fully reconstructed bottom decays and the soft leptons.
Figure~\ref{fig:sin2betastudy} shows the sideband-subtracted $\Delta\phi$
distribution between $B^+\rightarrow J/\psi K^+$ candidates and the soft leptons.  About $30\%$ of
the soft leptons are in the same azimuthal hemisphere, a fraction much
larger than expected from parton shower flavor
creation Monte Carlo ($\approx 5\%$ for {\sc Pythia} flavor creation).

This analysis uses the bottom pair decay signature of  $b \rightarrow
J/\psi X, \overline{b} \rightarrow \ell^+ X$.  The impact parameter of the
 additional lepton and the pseudo-$c\tau$ of
the $J/\psi$ are fit simultaneously in order to determine the
$b\overline{b}$ fraction of the two $\Delta\phi$ regions. 
Angular requirements
that were necessary in previous di-lepton measurements because of double sequential
semi-leptonic decay backgrounds
($b\rightarrow c \ell^- X ;c\rightarrow \ell^+ X'$) are
avoided by the chosen signal.  $B_c$ is the only particle that decays
directly into $J/\psi$ and an addition lepton.  The only other source of
candidates where the additional lepton and $J/\psi$ candidates originate from 
the same displaced decay are hadrons that fake leptons or decay-in-flight of 
kaons and pions.  The number
of events from $B_c$ or from `fake' leptons can be estimated well
by using techniques from Ref.~\cite{bcprd}.  Thus, no angle requirement
between the two candidate bottom decay products are necessary,
yielding uniform efficiency over the entire $\Delta\phi^{b\overline{b}}$
range.  Due to the limited size of the data sample, only $f_{toward}$, the fraction of $b\overline{b}$ pairs in the same azimuthal hemisphere, can be measured.

  The selection
criteria used in this analysis have similar bottom momenta and rapidity
acceptances to CDF's Run II displaced track(SVT)~\cite{SVT} and $J/\psi$
triggers, and the addition leptons have momenta very similar to the
opposite side taggers planned for Run II (opposite kaon,
opposite lepton and jet charge flavor taggers).  Therefore, this
measurement aids in the development and understanding of flavor taggers
for such Run II measurements as the $B_s$ mass difference.

\subsection{Sample Selection}
\label{jpsisel}

  The signal searched for in this analysis is
$b\rightarrow J/\psi X, \overline{b}\rightarrow \ell^+ X'$ where {$\ell$} can be an
electron or muon.  In this section, the Run IB $J/\psi$ data set is
described.  The offline selection criteria for both the $J/\psi$ and the additional lepton are also described.   

\subsubsection{$J/\psi \rightarrow \mu^+ \mu^-$ Selection}
\label{sec:jpsisel}

This analysis uses the CDF Run Ib $J/\psi$ data set obtained between
January 1994 and July 1995.  The CDF $J/\psi$ triggers and offline $J/\psi$ selection criteria utilized are the same as the $B_c$ discovery analysis at CDF~\cite{bcprd}.  In order to understand the
trigger efficiencies, we confirm that the $J/\psi$ candidate's muons are the two muon candidates which triggered the event.

After confirming the trigger, the position of the 
extrapolated track at the muon chamber is compared to the
position of the muon stub 
using a $\chi^2$ matching test, taking into account the effects of multiple scattering and energy loss in material.  The positions are required to match within 3 standard deviations in the r--$\phi$ projection and within 3.5 standard deviations in the r--z projection. 
       
Next, we require a high quality track for both muon
candidates.  The pseudo-lifetime ($c\tau$) of the
$J/\psi$ candidates is used in this analysis to
determine the bottom purity.  Therefore, 
SVX$^\prime$ information is required in order to improve the precision of the $c\tau$ measurement. 

In order to reject $J/\psi$ candidates with muons originating from
different primary interactions, the z position difference between the
two tracks is 
required to be less than 5 cm at the beam-line.  A vertex constrained fit of the two muon candidates is performed~\cite{CTVMFT}.   The $\chi^2$ probability of the
vertex fit is required to be better than $1\%$.  The vertex
constrained mass of the $J/\psi$ candidate is required to be
$2.9~{\rm GeV}<{\rm M}_{J/\psi}<3.3~{\rm GeV}$.  

\begin{figure}[htbp]
\includegraphics[width=8.6cm]{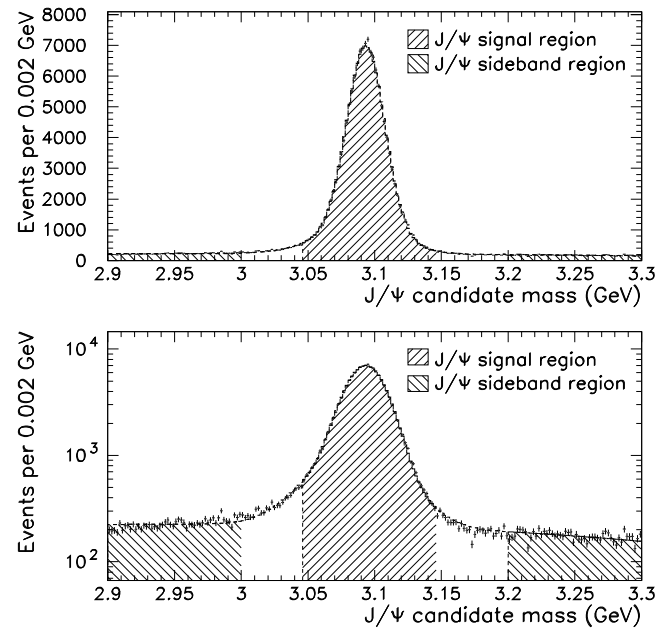}
\caption{Di-muon invariant mass distribution from events passing
selection criteria.  Top: Linear scale.  Bottom: Logarithmic scale.}
\label{fig:jpsimass}
\end{figure}

A total of 177,650 events pass the above selection cuts.   Figure~\ref{fig:jpsimass} shows
the $J/\psi$ mass distribution for these events.  In order to estimate the
number of $J/\psi \rightarrow \mu^+ \mu-$ candidates,  the mass has been
fit with
two Gaussians (used to model the $J/\psi$ signal) and a linear
background term.  A linear background has been assumed in many
previous CDF $J/\psi$ analyses~\cite{bcprd,blifeprd,jpsipolprl}.
The background under the mass peak is caused by irreducible 
decay-in-flight and punch-though backgrounds, Drell-Yan muons and
double sequential semi-leptonic decays where $b \rightarrow c \mu^- X, c
\rightarrow s \mu^+ X'$.  From the fit, $137780\pm440$ $J/\psi$
candidates are in the sample.  For this measurement, the  $J/\psi$
mass signal region is defined to be within $\pm 50~\rm{MeV}$ of the Particle
Data Group~\cite{pdg} world average value (3096.87 MeV). 
 The sideband regions are
chosen to be $2.900~\rm{GeV}\leq M_{J/\psi} \leq 3.000~\rm{GeV}$ and $3.200
~\rm{GeV} \leq M_{J/\psi} \leq 3.300~\rm{GeV}$.  The sideband regions contain 20,180
events.   The events in these regions are used later in the analysis to
describe the $c\tau$ shape of $J/\psi$ background
in the mass signal region.  

The ratio between the number of background 
$J/\psi$ events in the mass signal region to the background in the mass sideband region ($R_{side}$)
was determined to be $R_{side}=0.501\pm0.000043$ by the mass fit.  To
estimate the systematic uncertainty of this ratio,  a 2nd order polynomial is used to describe the
background term.  The resulting fit value is $R_{side}=0.545\pm0.008$.   The
difference between the two fits is taken to be the systematic
uncertainty yielding
$R_{side}=0.501\pm0.044.$

\subsubsection{CMUP $\mu$ Selection Requirements}
\label{sec:sltcmup}
The additional (non-$J/\psi$) muon is
required to have muon stubs 
in both the CMU and CMP (a CMUP muon).  Requiring both CMU and CMP muon stubs maximizes 
the amount of material traversed by the candidate, reducing the background due to hadronic
punch-though of the calorimeter.  The $\chi^2$ matching
requirements are the same as for the $J/\psi$ muons.  The muon candidates
are required to have a $p_T > 3~\rm{GeV}$; muons with lower $p_T$
will typically range out prior to the CMP due to energy loss in the
calorimeter and the CMP steel.

As the impact parameter is used to estimate the bottom purity of the muons, the
same track quality is required as for the $J/\psi$ muons.  Additionally,
the muon
candidate's track projection is required to be in the fiducial volume of the CMU and CMP.   The z positions of the $J/\psi$ candidate and the CMUP muon are
required to be within 5 cm of each other at the beam-line. 

In total, 247 CMUP candidate muons are found in the $J/\psi$ sample, out of which 51(142) CMUP candidates are in events where the $J/\psi$ candidate is in the mass sideband (signal) region.  

Of the 142 CMUP candidates with $J/\psi$ candidates in the $J/\psi$ mass signal region, 64 events
have the CMUP muon and the $J/\psi$ candidate in the same hemisphere in
the azimuthal angle (which will be known as toward); the other 78
events have the CMUP muon and $J/\psi$ candidate in the opposite hemisphere in the azimuthal angle (which is
denoted away).  

\subsubsection{Electron Selection Criteria}
\label{sec:sltel}
A method of the finding soft (relatively low momenta) electrons was developed for bottom
flavoring tagging in CDF's $B_d$ mixing and $sin(2\beta)$
measurements~\cite{bdmixing,sin2betaprd}.  These electrons have a
relatively high purity, a transverse momenta greater than 2 GeV, and an
understood efficiency.  The rate of hadrons faking an electron was
studied extensively in Ref.~\cite{bcprd}, making it possible to estimate the
background due to hadrons faking electrons.  The selection criteria is identical to Ref.~\cite{bcprd} in order to use the same fake rate estimates. 

 The selection criteria requires a high quality track which is consistent with an electron in the various detector systems.   Information on the energy (charge) deposited, the cluster location, and track matching $\chi^2$ variables are all used in order to reduce the rate of a hadron in the detectors' fiducial volume faking an electron to $\left( 6.4 \pm 0.6 \right) \times 10^{-4}$.

One source of electron background is photon
conversions, where a photon interacts with detector material and
converts into a $e^+ e^-$ pair.  In addition, $1.2\%$ of all neutral
pions decay into $\gamma  e^+ e^-$ directly (Dalitz decay).  To reduce
this background, conversions are searched for and vetoed by looking for
a conversion partner track.

The conversion requirements are the same as Ref.~\cite{bdmixing}.  Unfortunately, the conversion removal is not totally efficient.
Therefore, some of the soft electron candidates are residual conversion electrons,
where either the conversion pair track is not found due to tracking
inefficiencies at low $p_T$ or the conversion
electron selection is not fully efficient.
The rate of residual conversions is studied more in section~\ref{conversions}.

In total, 514 candidate electrons are found after conversion removal; 92 events have the $J/\psi$ candidate in the  mass sidebands
and 312 events have the $J/\psi$ candidate in the mass signal region.  In the $J/\psi$ mass signal region, 107(205) of the events are in the toward(away) regions in $\Delta\phi$.  In the $J/\psi$ mass signal
region, 6(9) events were vetoed as conversions in the toward(away)
$\Delta\phi$ bin.  In the $J/\psi$ mass sideband region, 5(4) events were
vetoed as conversions.

\subsection{Signal and Background Description}

The signal and backgrounds for both the $J/\psi + \mu$ and $J/\psi + e $
samples are very similar.  The basic technique to determine the
amounts of the various signal
and background components is with a simultaneous fit of the
pseudo-$c\tau$ (defined in section~\ref{jpsifit}) of the $J/\psi$ and the signed impact parameter of the non-$J/\psi$
lepton.  The impact parameter is signed to
distinguish between residual electron conversions and electrons from
bottom decay, as described in
section~\ref{conversions_shape}.   The impact parameter is signed
positive if the primary vertex lies outside the r--$\phi$ projection of the
particle's helix fit.

  As the $J/\psi$ and additional
lepton originate from separate bottom hadron decays, the impact parameter
of the additional lepton and the $c\tau$ of the $J/\psi$ are not
strongly correlated for the signal.  The backgrounds in this analysis
have two 
categories: one in which the
impact parameter and $c\tau$ are uncorrelated, and other where the impact parameter
and $c\tau$ are strongly correlated.   The impact parameter and the
$c\tau$ become strongly correlated when both the
$J/\psi$ and the additional lepton candidate originate from the same
displaced vertex.

In uncorrelated sources, the impact parameter and $c\tau$
shapes describing the background are determined independently.
$J/\psi$ candidates are assumed to originate from three
sources: direct $J/\psi$ production (including feed-down from $\chi_{c1}$,
$\chi_{c2}$, and $\psi(2s)$) where the $J/\psi$ decays at the
primary vertex, $J/\psi$ from bottom decay (including the feed-down from
higher $c\overline{c}$ resonances), and the non-$J/\psi$ background described by
the events in the $J/\psi$ mass sidebands.  
Lepton candidates are assumed to originate from the following
sources:  directly produced fake or real leptons from the primary
vertex, leptons from bottom decay (including $b\rightarrow cX \rightarrow
\ell X'$), lepton candidates with the fake $J/\psi$ candidate, and
residual conversion electrons.

In addition, two correlated sources of backgrounds exist.   The first
source is  $B_c\rightarrow J/\psi \ell^+ X$, which is a small but irreducible
background.  The impact parameter of the additional lepton and the
$c\tau$ of the $J/\psi$ is described by 
Monte Carlo techniques and
the overall size of the background is also estimated (see
section~\ref{bctemp}).   

The other correlated source of background occurs when a bottom hadron
decays into a $J/\psi$ and a hadron which is misidentified
as a lepton.  For electrons, this background is due to hadrons
(mostly $\pi^{\pm}$ and $K^{\pm}$) showering early in the calorimeter
and passing the electron identification selection.
For muons, there are two sources of this background.   The largest
source of 
correlated background is due to 
decay-in-flight of charged pions and kaons, which result in a real muon.  These real
muons are denoted as `fakes' in this analysis.  The other, smaller
correlated fake muon
background is caused by hadrons 
punching through the calorimeter and muon
steel shielding.  These background sources are more fully described
in section~\ref{bfaketemp}.
 
The following sections provide a description of the techniques used to
determine the impact parameter and $c\tau$ shapes of the various sources
and to estimate of the
number of residual conversions, $B_c \rightarrow J/\psi \ell X$ events, and $b \rightarrow J/\psi
\ell_{\rm{fake}} X $ events in the sample.

\subsubsection{$J/\psi$ $c\tau$ Signal and Background Distributions}
\label{jpsifit}

The direct $J/\psi$ and bottom decay $J/\psi~c\tau$ shapes are determined from a fit to the data, using a technique previously establish in Ref.~\cite{blifeprd,jpsipolprl}.  The relatively long average lifetime of the bottom hadron
($1.564\pm 0.014~\rm{ps}$~\cite{pdg}) allows one to distinguish between these two sources.  First, the signed transverse decay length $L_{xy}$ is determined in the r--$\phi$ plane. 

\begin{equation}
 L_{xy}\equiv\frac{\left(\vec{X}_{SV}-\vec{X}_{PV}\right) \cdot \vec{p}_T^{~J/\psi}}{p_T^{J/\psi}},
 \end{equation}
\noindent 
where $\vec{X}_{SV}$ and $\vec{X}_{PV}$ are the locations of the $J/\psi$
vertex and the primary vertex in the transverse plane, and
$\vec{p}_T^{~J/\psi}$ is the vector transverse momentum of the $J/\psi$.
Directly produced $J/\psi$ have a symmetric $L_{xy}$ distribution around
zero and bottom $J/\psi$ events will predominately have a
positive sign.  If the bottom decay was fully reconstructed, one could
 determine the proper decay length exactly ($c\tau_{proper}$) from the measured
$L_{xy}$ and $p_T$.  Because the bottom hadron is not fully reconstructed, a
`pseudo-proper decay length' ($c\tau$) is constructed using the kinematics of the
$J/\psi$ only:
\begin{equation}
c\tau \equiv \frac{L_{xy} \cdot m_{J/\psi}}{p_T^{J/\psi} \cdot F_{corr}(p_T^{J/\psi})}.
\end{equation}

\noindent 
$F_{corr}(p_T^{J/\psi})$, determined in Ref.~\cite{blifeprd}, is
the average correction factor for the partial reconstruction of the
bottom hadron. 

The events in the $J/\psi$ mass sidebands are used to model the
fake $J/\psi$ background under the $J/\psi$ mass signal peak.   Two
components are fit using an unbinned log-likelihood technique:
events from the primary vertex (direct) and events with lifetime from
heavy flavor (predominantly from $b\rightarrow c \mu^- X \rightarrow \mu^+ X'$). The direct events are
described by a symmetric resolution function chosen to be a Gaussian plus
two symmetric exponentials.  The events with lifetime are fit with a
positive only exponential.

\begin{figure}[htbp]
\includegraphics[width=8.6cm]{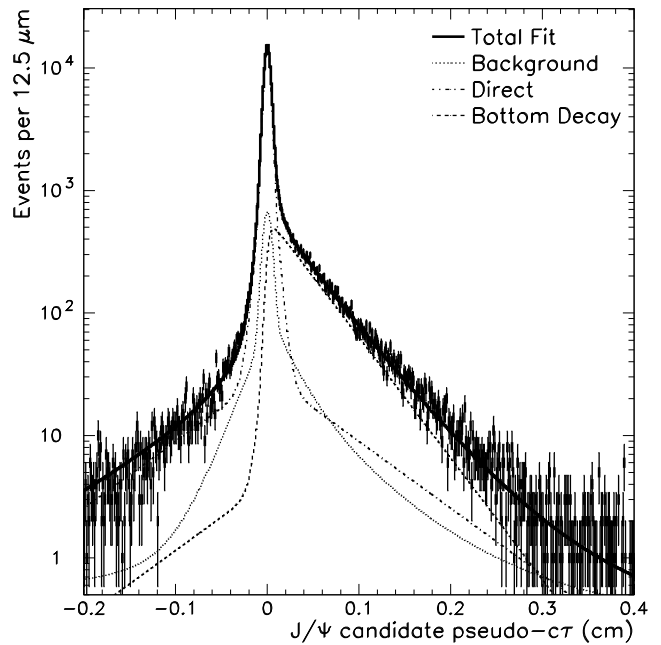}
\caption{Fit of $J/\psi$ signal region.} 
\label{fig:signalfit}
\end{figure}

Once the $c\tau$ shape of the mass sideband events is found, the
$c\tau$ shapes of directly produced and bottom decay $J/\psi$ can be
determined.  The shape of the directly produced $J/\psi$
($F_{direct}^{c\tau}(x)$) is 
parameterized by a Gaussian with two symmetric exponential tails; this
shape is the assumed 
resolution function of the $c\tau$ measurement.  The shape of $J/\psi$
events from 
bottom decay ($F_b^{c\tau}(x)$) is therefore
described as a positive exponential convoluted with the $c\tau$ resolution
function.  The background shape ($g_{back}$) is fixed to the value
obtained in the fit of the sideband region and the background fraction
($f_{back}$) is fixed to the value predicted by the $J/\psi$ candidate
mass fit.
Figure~\ref{fig:signalfit} shows the fit result of the signal region.   The fit average bottom proper decay length of $442\pm5~\mu\rm{m}$ is consistent with previous measurements at the Tevatron~\cite{blifeprd,jpsipolprl}.
The fit yields a bottom fraction of $16.6\%\pm 0.2\%$ or equivalently
$22150\pm270$ $J/\psi$ from bottom decay.

\subsubsection{Lepton Impact Parameter Signal Distributions}
\label{ipfit}

The impact parameter shape of bottom decay leptons is
determined by Monte Carlo simulation, using the prescription
from Ref.~\cite{field} for parton shower Monte Carlo programs.  In
this prescription, separate samples of flavor creation, flavor excitation, and
gluon splitting events are generated and then combined with the
relative rates predicted by the Monte Carlo programs.  In appendix~\ref{monte carlo}, the generation of the simulated events is described.

\begin{figure}[htbp]
\includegraphics[width=8.6cm]{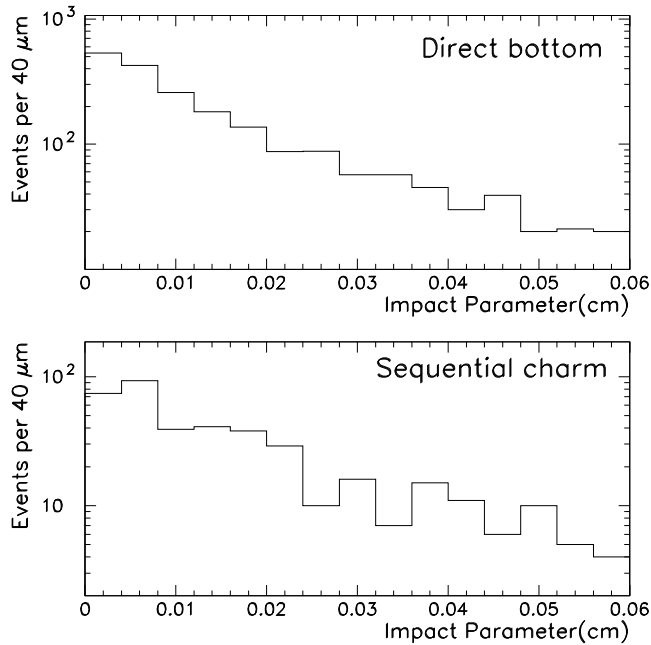}
\caption{The impact parameter of events with electrons passing
requirement in the flavor creation Monte Carlo sample.  Top: Direct bottom.  Bottom:
Sequential charm.}
\label{fig:seqc_vs_directb}
\end{figure}

\begin{figure}[htbp]
\includegraphics[width=8.6cm]{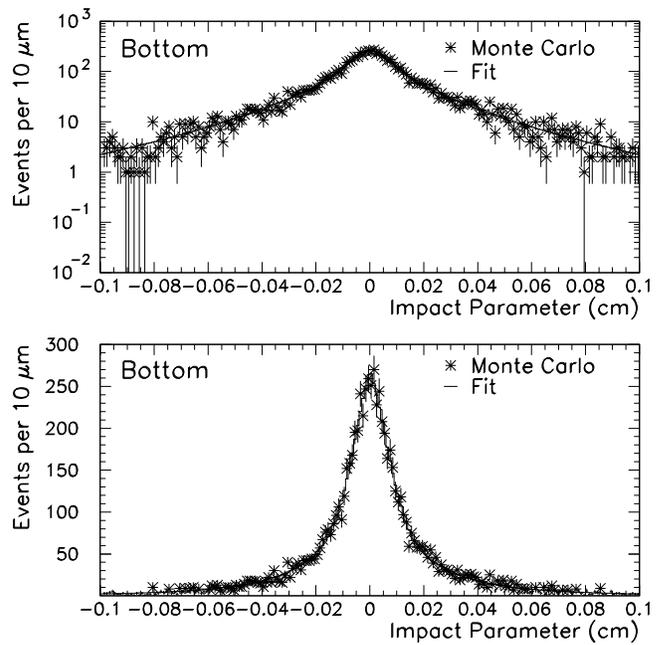}
\caption{The signed impact parameter distribution for the combined
electron bottom
Monte
Carlo.  Top: Logarithmic scale.  Bottom:
Linear scale.}
\label{fig:btempele}
\end{figure}

Figure~\ref{fig:seqc_vs_directb} shows the unsigned impact parameter for
direct bottom electrons ($b\rightarrow c e^-$) and sequential charm
electrons ($b\rightarrow c \rightarrow s e^+$) in the flavor creation
sample.  The impact parameter distributions are
very similar and cannot be fit for separately.  
The uncertainty on the relative rate of these two sources is one of the systematic uncertainties treated in
section~\ref{sec:systematics}.  The bottom decay impact parameter shape is fit with two symmetric exponentials and a Gaussian.  The fit to the simulated bottom impact parameter distribution for electrons is shown in fig.~\ref{fig:btempele}.

\subsubsection{Lepton Impact Parameter Background Distributions}
\label{ipfit_back}

In previous analyses~\cite{cdf-96,cdf-97}, the impact parameter
distributions for particles originating at the primary vertex were determined using jet
data.  Unfortunately, any data sample will have low level contamination
of heavy flavor (charm or bottom) at the $\approx0.1-1\%$ level, which
is larger than the non-Gaussian effects in the impact parameter
resolution.  In this analysis, the impact parameter shapes of directly produced
particles are determined with a 
Monte Carlo technique.   

{\sc Pythia} is used to generate the light quark and gluon subprocesses,
which are then passed through a detector simulation.  To be included in
the electron sample, the candidates must be a quality track
with a $p_T>2~\rm{GeV}$ and extrapolate into the electron fiducial region.   
For the muon direct sample, the candidates must have a quality track
with a $p_T>3~\rm{GeV}$ and extrapolate into the CMUP muon fiducial region.  
The simulation events which pass the selection criteria are fit with two symmetric exponentials and a Gaussian.  
The fit shapes are then used as input to the likelihood fit for particles originating from the primary vertex (direct tracks).

\subsubsection{Size of Residual Conversion Background}
\label{conversions}
One obvious source of electron background is residual conversions left in the sample
due to the inefficiency of finding the conversion pair.  In order to
estimate the number of residual conversions, a technique similar
to Ref.~\cite{sltelectron} is used.  
It is assumed 
that there are two independent causes for the lack of removal of a conversion
electron: the track pair is lost due to tracking inefficiencies at low
momenta or the selection requirements are not fully efficient.   By
measuring these two efficiencies and the rate of conversion
removal with the chosen conversion selection requirements, one determines the
residual electrons ($N_{resid}$).  The conversion electron that passes
the electron identification criteria is denoted as the conversion
candidate, and other electron that did not pass the electron
identification criteria is denoted as the pair candidate.

The number of residual electrons is equal to:
\begin{equation}
N_{resid}=N_{tag}\cdot \left(
\frac{1}{\epsilon_{cnv}(cut)}\cdot\frac{1}{\epsilon_{cnv}(p_T)}-1
\right)
\end{equation}

\noindent where  
$N_{tag}$ is the number of the conversions removed,
$\epsilon_{cnv}(cut)$ is the conversion finding efficiency, and
$\epsilon_{cnv}(p_T)$ is the tracking efficiency of the conversion pair.

The efficiency $\epsilon_{cnv}(cut)$ is measured using different sets of conversion
requirements, the tight (standard) and a loose set of cuts.
Assuming that the loose cuts are fully efficient, the ratio of
conversion pairs fit with tight and loose cuts yields
$\epsilon_{cnv}(cut)=72.3\pm6.5\%$.  In order to test this assumption, 2 additional
wider sets of cuts are used which yield no extra conversion candidates.  

The tracking efficiency of the pair candidates ($\epsilon_{cnv}(p_T)$) is
estimated with a Monte Carlo
technique similar to Ref.~\cite{bcprd}. The generation of the simulated conversions is detailed in 
Appendix~\ref{monte carlo}.  
$\epsilon_{cnv}(p_T)$ is estimated by comparing the $p_T$ distribution of 
the conversion partner in the simulation sample to the $p_T$ distribution of the conversion
partners in the data sample.  The simulation is normalized to the data in
the $p_T$ range where the tracking is fully efficient ($p_T>0.5~\rm{GeV}$).  The ratio of
the number of events seen in data versus the number of normalized
conversion candidates in simulation gives 
$\epsilon_{cnv}(p_T)=69\pm5(\rm{stat})\pm9(\rm{syst})\%$.   The systematic error includes the uncertainty in the conversion's $p_T$ spectra. 

  The ratio of the number of
residual to found conversions was found to be $R_{conv}=1.00\pm0.38$.  The
conversion veto removes 6(9) electron candidates in the
toward(away) $\Delta\phi$ region.  Thus, approximately 6.0 (9.0) residual conversion are
in toward(away) $\Delta\phi$ region.  About $5\%$ of the electron candidates are residual
conversions and have to be included in the $c\tau$--impact
parameter fit.  A total of 9 conversions (4 toward, 5 away) are found in
events with the $J/\psi$ candidate in the mass sideband region.

\subsubsection{Impact Parameter Shape of Residual Conversion Background}
\label{conversions_shape}

For conversion electrons from a primary
photon, the primary vertex alway lies outside
of the helix projection with perfect tracking.  To distinguish between conversions and bottom
decay electrons, the impact parameter is signed such that the impact
parameter is positive if the primary vertex is outside the r-$\phi$
projection of the track's helix and is
negative otherwise.  Conversion electrons are positively signed, and
Dalitz decay electrons and bottom decay electrons are equally negatively
and positively signed.

\begin{figure}[htbp]
               \includegraphics[width=8.5cm]{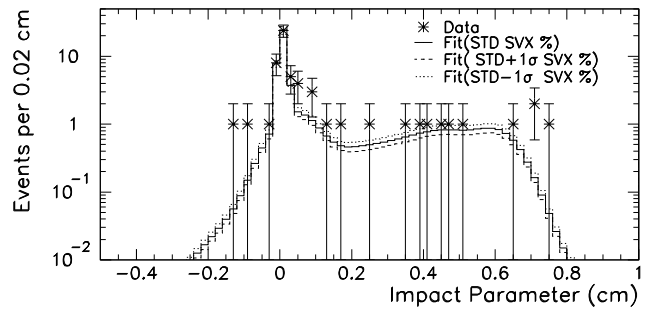}
\caption{The signed impact parameter distribution for conversions
found in data.  Solid line: Monte Carlo fit using the central value of
$f_{conv}^{GoodSVX}$.  Dashed line: Monte Carlo fit increasing $f_{conv}^{GoodSVX}$
by one sigma.  Dotted line: Monte Carlo fit decreasing $f_{conv}^{GoodSVX}$
by one sigma. }
\label{fig:convcomb}
\end{figure}

The vast majority of the conversion candidates are signed positive as
predicted (shown in fig.~\ref{fig:convcomb}).  Unfortunately, there is a large positive tail which is not
consistent with the number of silicon hits assigned to the track.  Since at
least 3 SVX$^\prime$ hits are required, one would expect the conversion candidates
either to originate from the first two silicon layers or the beam pipe, or
to be a $\pi^0$ Dalitz decay from the primary vertex.  These sources would produce
conversions with a impact parameter less than 0.04 cm.  Therefore, a
fraction of the conversion candidates must have mis-assigned silicon hits
and originate outside of the SVX$^\prime$.  The measured conversion radius for 25 of the 62 conversion candidates is greater than 6 cm, which is outside of the second silicon layer.

To construct an impact parameter shape for residual conversions, 
two Monte Carlo samples are generated: a sample in which the silicons hits are correctly assigned (low conversion radius, $R_{conv}<6~\rm{cm}$) 
and a sample in which silicon hits are falsely added (mostly at
high conversion radius, $R_{conv}>6~\rm{cm}$).  The relative fractions of
the two components are determined by data.  The fraction of conversions
with $R_{conv}<6~\rm{cm}$ ($f_{conv}^{GoodSVX}$) in data and 
simulation are matched, with the
uncertainty in the fraction in data included as a systematic uncertainty
in the shape.

Figure~\ref{fig:convcomb} shows the impact parameter of the
candidates found in data and the combined conversion impact parameter
shape normalized to data.  The combined impact parameter shape describes
the data adequately, including both the negative tail and large
positive tail.

\subsubsection{Additional Lepton Impact Parameter Distributions in Events with Fake $J/\psi$}
\label{sideband}
Events in which the $J/\psi$ candidate is in the mass sideband regions are used to describe the impact parameter shape of leptons in events with a fake $J/\psi$.  The composition of events in the $J/\psi$ mass sidebands are unknown; therefore, the shapes have to be  fit in a similar manner as section~\ref{jpsifit}.

For the muon sample, there is no knowledge of the contributions to the
additional muon's impact parameter distribution.  Therefore, the shape is
parameterized with a Gaussian and symmetric exponential.

The impact parameter shape in the electron sample for events in the $J/\psi$ mass sideband has an additional
complication; the sample contains residual
conversions.  The number of found conversions in the
signal region is used as a constraint on the number of residual conversion
events in the signal region.  Thus, the number of residual conversion events
fit in the sideband region has to be known.   In the sidebands, the
residual conversion fraction is fit for $f_{conv}=\frac{r_{conv}\cdot
n_{convside}}{N_{sideband}}$, where $r_{conv}$ and $n_{convside}$ are the fit
ratio of residual to found conversions and fit number of ``found''
conversions, respectively.  These quantities are constrained by the estimate of
$R_{conv}$ and the number of found conversions in the sidebands,
$N_{convside}$.  Since $r_{conv}$ is a component of the signal region
fit, the signal and sideband regions have to be fit simultaneously.

\subsubsection{$B_c \rightarrow J/\psi\ell^+X$ Background}
\label{bctemp}

$B_c$ decay is the only known process that yields a lepton and a $J/\psi$
from the same displaced vertex.   CDF's measurement~\cite{bcprd} of
the  $\frac{\sigma(B_c) \cdot BR(B_c
\rightarrow J/\psi \ell \nu)}{\sigma(B^+) \cdot BR(B^+ \rightarrow J/\psi
K)}$ and the $B_c$ lifetime allow one to estimate both the number and
impact parameter--$c\tau$ shape of this background.   The estimated number
of  $B_c \rightarrow J/\psi \ell X$ events in the samples is used as
a constraint in the fit.

Taking into account correlated uncertainties, the number of $B_c
\rightarrow J/\psi \ell X$ events estimated in the sample are
$N_{B_c}^e=10.0^{+3.5}_{-3.3}$ and $N_{B_c}^{\mu}=7.2 ^{+2.6}_{-2.4}$.  According to
Monte Carlo, over $99\%$ of the $B_c$ passing the selection requirements
have $\Delta\phi<\frac{\pi}{2}$ between the lepton and the
$J/\psi$.  Thus, this background is included in the fit only for the toward $\Delta\phi$ region.

Using a $B_c \rightarrow J/\psi \ell \nu$ Monte Carlo sample described in appendix~\ref{monte carlo}, the impact parameter--$c\tau$ shape is determined.  The shape takes into account the correlation in the impact parameter of the addition lepton and the c$\tau$ of the $J/\psi$.  As the c$\tau$ of the $J/\psi$ increases, the impact parameter of the additional lepton can be larger.

\subsubsection{$b \rightarrow J/\psi\ell_{fake}$ Background}
\label{bfaketemp}

The other source of background where the impact parameter and $c\tau$ are
correlated is bottom hadrons decaying to a real $J/\psi$ with a hadron
from the same decay faking an lepton.  The sources and rates for faking
leptons were studied extensively in Ref.~\cite{bcprd} and are used in
this analysis.  In Monte Carlo, more than $99\%$ of the events have
the $J/\psi$ and the fake lepton candidate in the same azimuthal hemisphere, and therefore this background is included only in the toward $\Delta\phi$ region.

The estimates of the amount and shapes of these backgrounds are made
using a Monte Carlo sample of $b \rightarrow J/\psi X$ events, described in appendix~\ref{monte carlo}.  Hadrons can ``fake'' an electron by showering early in the calorimeter.   ``Fake muons'' can be caused by  decay-in-flight of charged pions and kaons, and by hadrons not being completely absorbed in the calorimeter and leaving hits in the muon chamber.

  The rate that a hadron will fake an electron
was studied in Ref.~\cite{bcprd}.  The number of $b \rightarrow J/\psi e_{\rm{fake}}
X$ events is determined by using these fake rates and the Monte Carlo
sample.  To be included in the fake rate calculation, an event must pass
the $J/\psi$ selection and have a charged hadron in the electron identification fiducial volume with a track with SVX$^\prime$ hits
and a $p_T > 2~\rm{GeV}$.  Ideally, one would then apply the appropriate
fake rate for the particle's $p_T$ and isolation for
each track passing the selection, yielding the total fake rate.  The isolation is defined to be the scalar sum of the momenta of particles within a cone $\Delta R<0.2$, divided by the momentum of the particle in consideration.  Unfortunately,
the Monte Carlo used does not include particles from the underlying
event, fragmentation, gluon radiation or the other bottom hadron in the
event. Thus, the isolation in the simulation does not represent the
data.  A large fraction ($\sim 70\%$) of the events have a small isolation ($I<0.2$); this value of the isolation is used as a central value of the estimate.  The fake rates using the other isolation value estimates are used to quantify the systematic uncertainty of the background estimate.

Normalizing the Monte Carlo to the estimated number of $b \rightarrow
J/\psi X$ events in data yields a estimate of $2.85\pm0.03(\rm{stat.})\pm0.75(\rm{syst.})$ events, where the second error is the systematic uncertainty due
to modeling (or lack thereof) of the track isolation.

The impact parameter-$c\tau$ shape of this background is determined by a
fit to the Monte Carlo.  The Monte Carlo sample used in the fit consists of events which can included in the electron fake rate calculation. 

Decay-in-flight(DIF) of charged pions and kaons to muons is also a source of
correlated background, as long as the track is reconstructible.  The
probability of a decay-in-flight to have a reconstructible track is greatly
reduced by the SVX$^\prime$ requirements, as shown in Ref.~\cite{bcprd}.

The number of correlated background events from decay-in-flight is
determined in a manner similar to the fake electron estimate.   The
Monte Carlo is normalized in the same manner as the fake electron
calculation.  The $J/\psi$
candidate is required to pass the selection criteria in section~\ref{sec:jpsisel},
and the decay-in-flight candidates are required to have a SVX$^\prime$ track with
$p_T>3~\rm{GeV}$ and project into the CMU and CMP fiducial volumes.  The probability of decaying-in-flight is
determined for the given $p_T$ and particle species.  In Monte Carlo, $64.2\pm0.3\%$ of the
particles passing the requirements are kaons.  The decay-in-flight
background is estimated to average $9.9\pm2.1$ events.  The error
includes the
$12\%$ Monte Carlo calculation systematic uncertainty and a $17\%$
reconstruction efficiency
systematic uncertainty quoted in Ref.~\cite{bcprd}.

In Ref.~\cite{bcprd}, the decay-in-flight estimate was done using data.  In that analysis, the kaon fraction was measured to be
$(44\pm4.4)\%$.  The difference between the kaon fraction in Ref.~\cite{bcprd} and
the simulation could lead to a large systematic difference, because of the
difference in the kaon and pion decay-in-flight probabilities.   To estimate
this uncertainty, the Monte Carlo events are re-weighted in order to
match the kaon fraction measured by~\cite{bcprd}.  With the
re-weighting, the estimated number of correlated decay-in-flight
background is $8.7\pm2.0$.  The difference between the
two estimates is conservatively used as the systematic error, yielding a
final decay-in-flight estimate of $N_{B_{fake}}^{\mu,DIF}=9.9\pm2.4$.

The impact parameter--$c\tau$ shape of the decay-in-flight background is
determined by the same Monte Carlo sample.  In Ref.~\cite{cdf-97,bcprd}, it is
shown that the impact parameter distribution of reconstructible
decay-in-flight particles with SVX$^\prime$ information have the same impact
parameter distribution as the parent particle.  Similar to the fake
electron shape, the Monte Carlo events which
could be used in the DIF rate calculation
are fit in order to determine the DIF impact parameter--$c\tau$ shape.

Hadrons can also mimic muons by not being completely absorbed by the
calorimeter and leaving hits the muon chambers.  The probability of a
track punching-through the calorimeter was determined in Ref.~\cite{bcprd}.
The selection criteria of the punch-through estimate is the same as the
the decay-in-flight estimate.  The punch-though probability
of the tracks passing the requirements is calculated from its particle
type and momentum, yielding the final estimate.   An average
$1.76\pm0.70$ punch though events are expected in the data, including a $40\%$
systematic error used in ~\cite{bcprd}.

  As the punch-though rate is much larger for $K^+$ than
$K^-$ or $\pi^{\pm}$, the large difference in kaon fraction
between~\cite{bcprd} and Monte Carlo  (shown in the decay-in-flight
estimate) is a significant systematic shift in the punch-through estimate.   To be conservative, we
re-weight the data with the kaon 
fraction measured in Ref.~\cite{bcprd}; $1.23\pm0.46$ events are expected.
The difference between the two predictions is used as the estimate of the
systematic uncertainty in the prediction, yielding a final estimate of the
average number of correlated backgrounds from punch-through of
$N_{B_{fake}}^{\mu,PT}=1.76\pm0.88$ events.  In Ref.~\cite{suzuki}, the decay-in-flight
and punch-through backgrounds are shown to have the same lifetime shape
in the $B_c$ lifetime fit.  The decay-in-flight and punch-through
backgrounds are assumed to have the same impact parameter--$c\tau$
shape.

\subsection{Unbinned Likelihood Fit Results}

\label{sec:fitresults}
An unbinned log-likelihood fit is used to determine the estimated number of $b\overline{b}$ pairs and backgrounds in the two $\Delta \phi$ regions.   Inputs to the fit are the $J/\psi$'s $c\tau$, the additional lepton's impact parameter, and the shapes and the background estimates described in the previous sections.  The shapes are used to determine the sample compositions, with the background estimates used as constants.  In appendix~\ref{fit}, the complete details of the log-likelihood function is given. 

\begin{figure}[htbp]
\includegraphics[width=8.6cm]{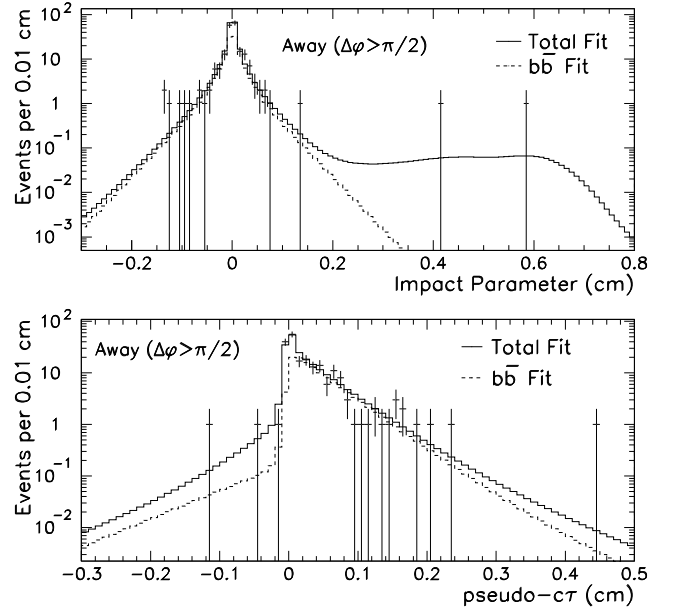}
\caption{Result of the $c\tau$--impact parameter fit for the
electron sample in the away bin.  Top:
Projection onto impact parameter axis.  Bottom: Projection onto $c\tau$ axis.}
\label{fig:eleaway}
\end{figure}

\begin{figure}[htbp]
 \includegraphics[width=8.6cm]{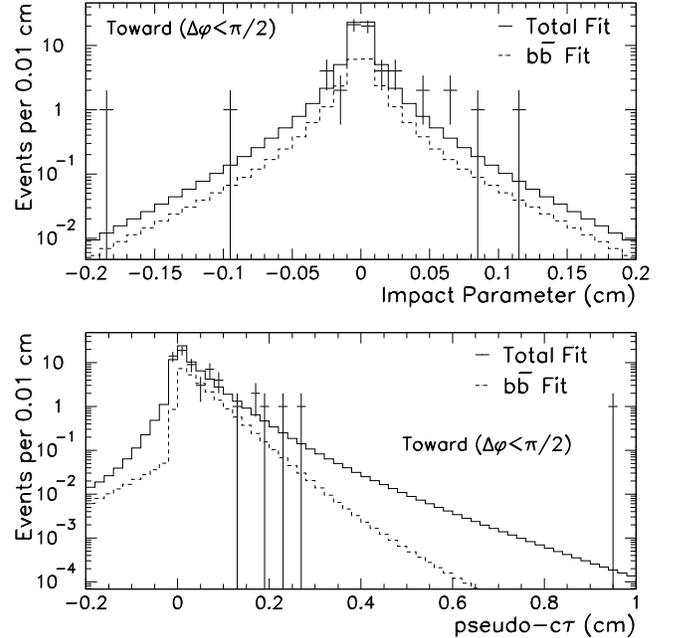}
\caption{Result of the $c\tau$--impact parameter fit for the
muon sample in
the toward bin.  Top:
Projection onto impact parameter axis.  Bottom: Projection onto $c\tau$ axis.}
\label{fig:muontoward}
\end{figure}

The log-likelihood ($-2\ln{\mathcal{L}}$) is minimized for both data
sets using MINUIT~\cite{minuit}.  The fit parameter errors are defined
by $\pm 1\sigma~(\Delta\mathcal{L}=1)$  contours of the likelihood
function using the MINOS option.   The results of the fit are shown in
table~\ref{table:fitresults}.   In order to display the fit result, the
log-likelihood function has been integrated in regions of impact
parameter--$c\tau$ space.  As examples of the fit quality, figures~\ref{fig:eleaway} and~\ref{fig:muontoward} show the fit results projected onto the impact
parameter and $c\tau$ axis for the electron sample in the away  $\Delta\phi$ region and muon sample in the toward $\Delta\phi$ regions, respectively.

The toward fraction measured in the two samples are:
\begin{eqnarray}
f_{toward}^\mu&=&34.5^{+9.2}_{-8.2}\%\\
f_{toward}^e&=&19.2^{+6.5}_{-5.9}\%
\end{eqnarray}
The measurement error includes both the statistical error as well as systematic
uncertainties due to the constraints.

As a test of the fitting technique, a set of 1000
toy Monte Carlo `experiments' have been generated.  The
results of the study, given in appendix~\ref{toymc}, show
that the fit results are unbiased and have proper errors. 

\begin{table*}
\begin{ruledtabular}
\renewcommand{\baselinestretch}{1.2}
\small\normalsize
 \begin{tabular}{ccccc}
Fit parameter & Electron & Electron Constraint & Muon & Muon Constraint\\ 
$n_{b\overline{b}}^{t}$ &$29.6^{+11.7}_{-10.4}$&&$23.0^{+ 7.6}_{-6.9}$& \\
$n_{bd}^{t}$ &$ 1.5^{+ 8.5}_{-8.1}$&&$ 1.6^{+ 4.6}_{-2.9}$ & \\
$n_{bconv}^{t}$ &0.6(fixed)&& N/A&  \\
$n_{db}^{t}$&0 (fixed)&&0 (fixed)&   \\
$n_{dd}^{t}$&$37.0^{+ 8.0}_{-7.3}$&&$11.3^{+ 5.1}_{-4.5}$& \\
$n_{dconv}^{t}$&$ 2.8^{+ 2.1}_{-1.7}$&& N/A&   \\
$n_{side}^{t}$&$45.4^{+ 6.9}_{-6.2}$&45&$32.9^{+ 5.7}_{-5.1}$&34  \\
$n_{B_{fake}}^{t}$ & $2.8^{+ 0.7}_{-0.7}$&$2.85\pm0.75$& $10.7^{+ 2.5}_{-2.5}$&$11.7\pm2.6$ \\
$n_{B_c}^{t} $& $10.0^{+ 3.2}_{-3.3}$&$10.0^{+3.5}_{-3.3}$&$5.1^{+ 2.5}_{-2.5}$&$7.2^{+2.6}_{-2.4}$\\  \hline
$n_{signal}^{t}$&107.1&107&68.2 &64\\ 
$n_{conv}^{t}$& 5.6 &6&N/A&\\ \hline\hline
$n_{b\overline{b}}^{a}$& $124.7^{+17.9}_{-16.7}$&&$43.6^{+10.2}_{-9.0}$&\\
$n_{bd}^{a}$& $-1.4^{+12.5}_{-12.2}$&&$8.1^{+8.0}_{-7.5}$&  \\
$n_{bconv}^{a}$&1.2(fixed)&&N/A&  \\
$n_{db}^{a}$& 0 (fixed)&& 0 (fixed)& \\
$n_{dd}^{a}$& $49.5^{+ 9.2}_{- 8.5}$&&$16.0^{+5.5}_{-5.2}$&\\
$n_{dconv}^{a}$&$ 6.0^{+ 2.6}_{-2.2}$&&N/A&  \\
$n_{side}^{a}$& $47.6^{+ 7.1}_{-6.5}$&47&$18.2^{+4.5}_{-3.9}$&17\\  \hline
$n_{signal}^{a}$&204.9&205 &76.8&78\\ 
$n_{conv}^{a}$& 9.5&9&N/A&\\ \hline\hline
$r_{side}$& $0.505^{+0.043}_{-0.043}$&$0.501\pm0.044$&$0.501^{+0.043}_{-0.043}$&$0.501\pm0.044$\\
$r_{conv}$ & $0.99 ^{+0.31}_{-0.28}$&$1.00\pm0.37$&N/A&\\ 

\end{tabular}
\renewcommand{\baselinestretch}{1.0}
\small\normalsize
\caption{Fit results and constraints for the electron and muon samples.  Appendix~\ref{fit} describes all the fit parameters in detail.
Variables $n_{signal}$ and $n_{conv}$ are not fit parameters but are functions of
fit parameters.}  
\label{table:fitresults}
\end{ruledtabular}
\end{table*}

\subsubsection{Fit Systematics}
\label{sec:systematics}
By measuring the fraction of bottom quark pairs produced in the
same hemisphere $f_{toward}$, the systematic uncertainties are minimized.
The selection in both the $\Delta\phi<\pi/2$ and  $\Delta\phi>\pi/2$ regions
are the same; therefore, the uncertainties in the lepton selection efficiency, tracking efficiency, luminosity, etc. mostly cancel in the fraction measurement.  In this section, the systematic uncertainty in the log-likelihood not already included in the fit is estimated.   The estimated size of these systematic uncertainties are collected in table~\ref{table:systematics}.

The sequential charm fraction ($f_{seq}$) that is used in the
bottom impact parameter shape ($F_{b}^{d_0}$) is derived from the simulation.
The uncertainty in the sequential charm fraction leads to a systematic
uncertainty in the determination of $F_{b}^{d_0}$, as sequential charm
leptons have a larger impact parameter than direct bottom leptons.  In
Ref.~\cite{cdf-97}, the relative
systematic uncertainty in $f_{seq}$ was studied.  The
relative uncertainty in $f_{seq}$ was $\pm19\%$, which is used in this analysis.  The value of $f_{seq}$ is varied by $\pm 1\sigma$, $F_{b}^{d_0}$ is re-fit, and then the new $F_b^{d_0}$ shapes are used to
re-fit $f_{toward}$.  The maximum differences of $\pm 0.1\%$ and $\pm 0.3 \%$ are assigned as the systematic uncertainty for the electron and muon samples, respectively.

The bottom hadrons' lifetimes ($B^+,~B^0,~B_s,~\rm{and}~\Lambda_b$) and
their decay products' impact parameters are strongly
correlated.  In order to estimate the uncertainty caused by bottom lifetime uncertainty, two additional Monte
Carlo samples were generated using BGENERATOR, a fast
$b\overline{b}$ Monte Carlo that approximates the NLO prediction by Ref.~\cite{MNR}.   All the bottom hadron lifetimes are
shifted by $\pm 1\sigma$ from their PDG values~\cite{pdg}.   The $F_{b}^{c\tau}$ shapes determined by these samples are then used in a re-fit of $f_{toward}$.  The estimate of the systematic uncertainty due to the bottom lifetime is
chosen to be the greatest differences from the standard fit.  The uncertainties estimated in the electron and muon samples are $\pm
0.3\%$ and $\pm 2.2\%$, respectively.

$B_s$, $B^+$, and $B^-$ have proper decay lengths of
$\sim470~\mu\rm{m}$, whereas $\Lambda_B$ has a proper decay length of
$387~\mu\rm{m}$.  Thus, the uncertainty in the fraction of bottom quarks
fragmenting to $\Lambda_b$ leads to the largest uncertainty of the $F_{b}^{c\tau}$ shape.  Using BGENERATOR, samples are generated
with the $\Lambda_b$ fragmentation fraction varied by $\pm 1 \sigma$ from the PDG values~\cite{pdg2002} .  The new
$F_b^{c\tau}$ shapes are used to re-fit $f_{toward}$, with the maximum
difference from the standard fit used as the estimate of the systematic
uncertainty, yielding a systematic uncertainty of $\pm 0.1\%$ and $\pm
0.2\%$ for the electron and muon samples.

Due to the limited number of residual conversions in the sample,  the
number of conversions pairing with $J/\psi$ for bottom decay ($n_{bconv}$) and with direct $J/\psi$ ($n_{dconv}$) cannot be fit independently.  Thus, the ratio between $n_{bconv}$ and $n_{dconv}$ is fixed to the fit ratio between
$J/\psi$ from bottom decay and directly produced $J/\psi$.  In order to
estimate the systematic uncertainty due to this  assumption, the fit of the
data is re-done with either $n_{bconv}$ or $n_{dconv}$ fixed to zero;
the difference between fits are used as an estimate of the systematic uncertainty, yielding a systematic uncertainty of $\pm 0.1\%$ for the electron sample.

The residual conversion shape ($F_{conv}$) is determined using data and simulation.
In data, the conversion radii of the found conversions indicate that a
large fraction of the conversion candidates have at
least 1 SVX$^\prime$ hit mis-assigned to the track.  The shape $F_{conv}$ is the sum of two
shapes: $F_{conv}^{GoodSVX}$, which describes the shape of residual conversion where SVX$^\prime$ hits are assumed to be correctly assigned, and $F_{conv}^{BadSVX}$, which describes the shape of residual conversion where
at least 1 SVX$^\prime$ hit is assumed to be incorrectly assigned.  $F_{conv}^{GoodSVX}$ and $F_{conv}^{BadSVX}$ are determined using 
Monte
Carlo described in section~\ref{conversions_shape}.  The value of $f_{conv}^{GoodSVX}$ is changed by $\pm 1 \sigma$ in
$F_{conv}$ in order to 
estimate the systematic uncertainty due to the $F_{conv}$ shape used.  The maximum difference of $\pm 0.2\%$ is assigned as a
conservative estimate of the systematic uncertainty due to the residual
conversion impact parameter shape used.

The direct impact parameter shapes ($F_{direct}^{d_0}$) are determined
by a fit to Monte Carlo samples in section~\ref{ipfit_back}.  The finite size of the Monte Carlo
samples lead to an uncertainty in the fit parameters of the shapes.  In
order to estimate the uncertainty in $f_{toward}$ due the
$F_{direct}^{d_0}$ shape uncertainty, each parameter is fixed to a value
$\pm 1 \sigma$ from the best fit value and the $F_{direct}^{d_0}$ shape is
re-fit.   The new shape is then used in the impact parameter--$c\tau$
fit.  The
largest negative and positive differences from the standard fit is conservatively assigned as
the systematic error, $^{+0.3}_{-0.4}\%$ for the electron sample and
$^{+7.4}_{-1.0}\%$ for the muon sample.

The direct and bottom $c\tau$ shapes are determined by a fit to the
data.  In the fit, the fraction of fake $J/\psi$ events ($f^{back}$) is fixed at the
predicted fraction.  In
order to estimate the effect on $f_{toward}$, the value of $f^{back}$ is
change by $\pm 1 \sigma$ and the $J/\psi$ mass signal region $c\tau$ fit
is re-done.  The resulting $F_{direct}^{c\tau}$ and $F_{b}^{c\tau}$
shapes are used in a re-fit of $f_{toward}$.  The greatest
difference from the standard fit is chosen to be a conservative estimate
of systematic uncertainty, yielding uncertainty estimates of $\pm 0.015 \%$ and
$\pm 0.01 \%$ for the electron and muon channels.

In this analysis, the number of events with a directly produced $J/\psi$
with a lepton from bottom decay ($n_{db}$) is assumed to be zero.  In
order to measure the effects of this assumption, a fit of $f_{toward}$
is performed where $n_{db}$ is a free parameter.  The difference in this fit from
the standard fit is assigned as the systematic uncertainty due to
$n_{db}$.  We assign a $\pm1.9\%$ uncertainty to $f_{toward}^{\mu}$ and
$\pm0.1\%$ to $f_{toward}^{e}$.

The individual systematic uncertainties are added in quadrature in order
to determine the combined systematic uncertainty.  The systematic
uncertainties for the electron and muon samples are $^{+0.5}_{-0.6}  \%$ and
$^{+8.0}_{-3.1} \%$, respectively.

\begin{table}
\begin{ruledtabular}
{
\renewcommand{\baselinestretch}{1.2}
\small\normalsize

\begin{tabular}{ccc}
Source & Electron & Muon \\ 
Sequential Rate& $\pm 0.1 \%$&$\pm 0.3\%$\\
B Lifetime& $\pm 0.3 \%$&$\pm 2.2 \%$  \\
Fragmentation Fractions& $\pm 0.1 \%$&$\pm 0.2 \%$  \\
$n_{bconv}/n_{dconv}$ ratio & $\pm 0.1\%$&  \\
Residual Conversion Shape & $\pm 0.2 \%$& \\
Direct Impact Parameter Shape & $^{+0.3}_{-0.4} \%$& $^{+7.4}_{-1.0}\%$\\
$f_{back}$ (for $F_{Direct}^{c\tau}$ and $F_{b}^{c\tau}$) & $\pm 0.02
\%$&$\pm 0.01 \%$ \\
$N_{db}$& $\pm0.1 \%$&$\pm1.9 \%$\\\hline
Total& $^{+0.5}_{-0.6} \%$&$^{+8.0}_{-3.1} \%$\\
\end{tabular}
\renewcommand{\baselinestretch}{1.0}
\small\normalsize
}
\end{ruledtabular}
\caption{Summary of the estimated values of the systematic uncertainty
for $f_{toward}$.}  
\label{table:systematics}
\end{table}
 
\subsubsection{Correction to b quark level}
At this time, no fragmenting NLO QCD calculation of bottom production at the Tevatron exists.  In order to compare to
next-to-leading order calculations, one must `correct' the experimental
measurement to the bottom quark level ($f_{toward}^{corr}$), using similar technique as in Ref.~\cite{cdf-97,cdf-96,cdf-00}.

The correction is:
\begin{equation}
C_{B\rightarrow b}=\frac{f_{toward,mc}^{{b\overline{b}}^{~90\%}}}
{f_{toward,mc}^{b\rightarrow J/\psi X; \overline{b}\rightarrow \ell Y}} 
\end{equation}

$f_{toward,mc}^{b\rightarrow J/\psi X; \overline{b}\rightarrow \ell Y}$ is
the $f_{toward}$ prediction by {\sc Pythia}, where the
$\Delta\phi$ is calculated between the $J/\psi$ and the additional lepton,
which both meet the selection criteria.  The quantity $f_{toward,mc}^{{b\overline{b}}^{~90\%}}$ is the fraction of bottom quarks
produced by {\sc Pythia} in the same hemisphere which pass the following criteria:
\begin{itemize}
\item{$p_T^{b_1}>p_T^{J/\psi}$ and $|y^{b_1}|<y^{J/\psi}$}
\item{$p_T^{b_2}>p_T^{\ell}$ and $|y^{b_2}|<y^{\ell}$}
\end{itemize}
$b_{1,2}$ can be either the bottom quark or antiquark.  No requirements
are made on the decay products of the bottom quarks.

\begin{figure}[htbp]
\includegraphics[width=8.6cm,]{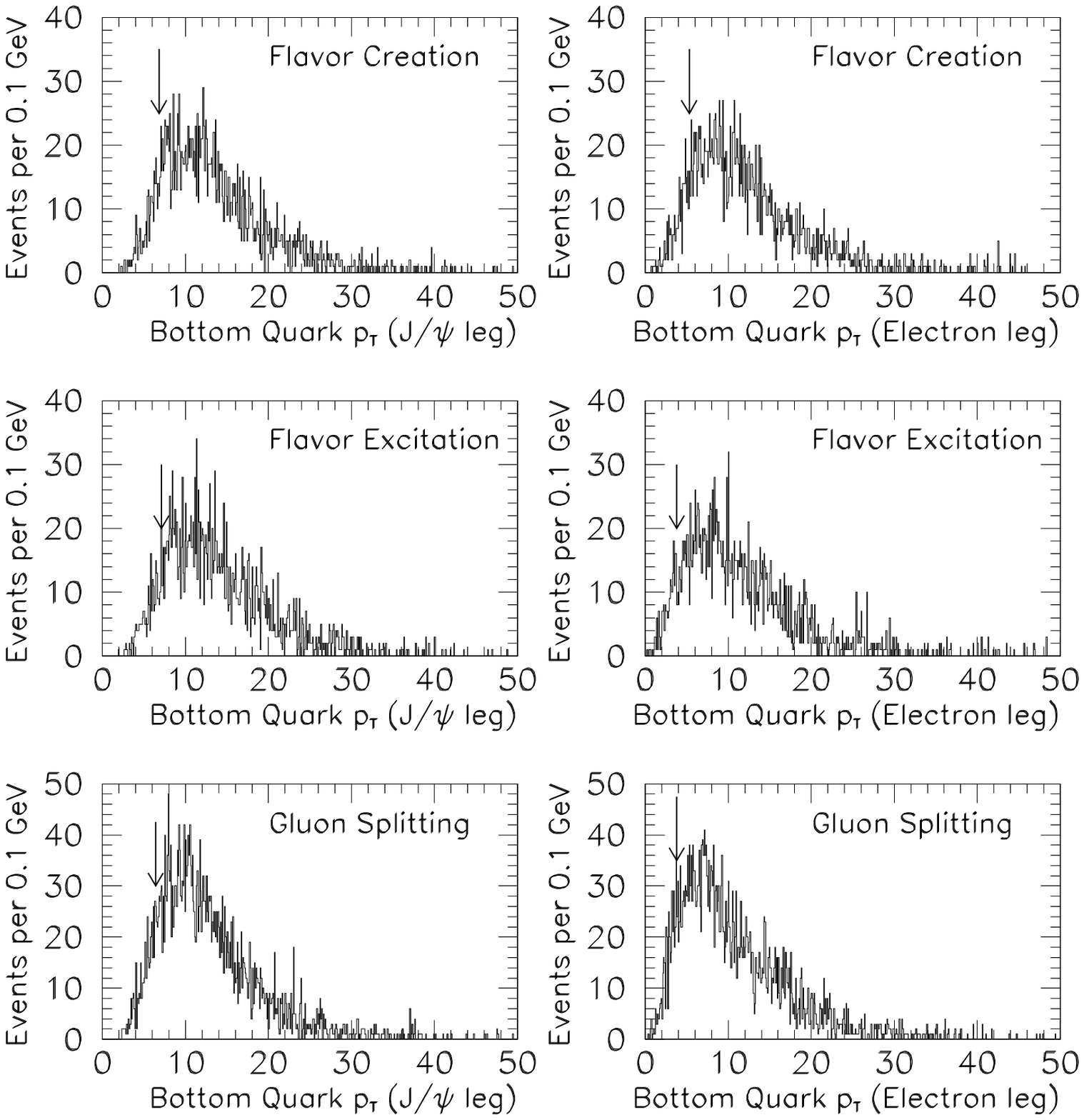}
\caption{$p_T$ of the bottom quarks in events that pass selection in
the electron {\sc Pythia} samples.  The arrows indicate the $90\%$ acceptance value.}
\label{fig:pteleeff}
\end{figure}

\begin{figure}[htbp]
\includegraphics[width=8.6cm]{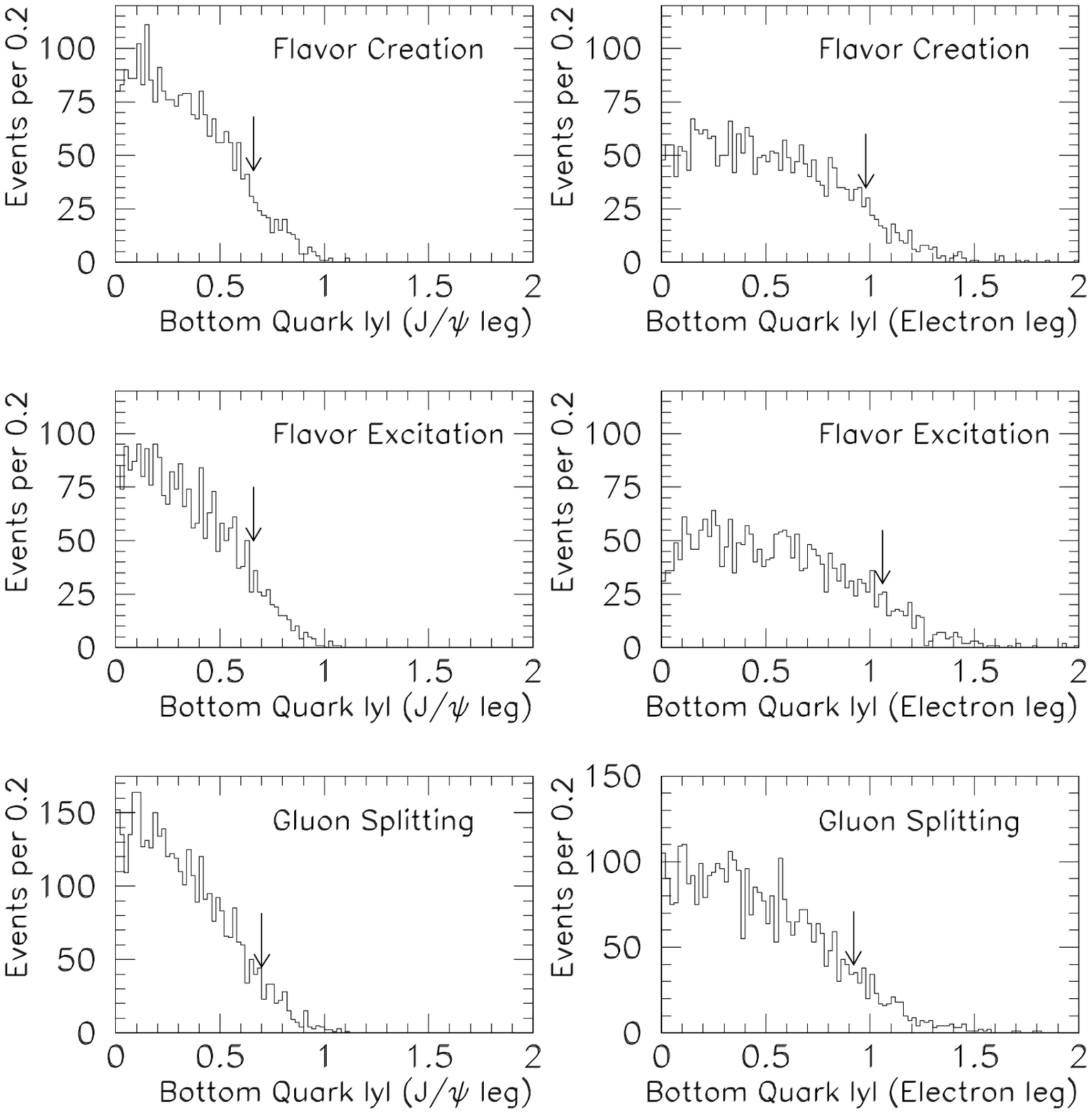}
\caption{$|y|$ of the bottom quarks in events that pass selection in
the electron {\sc Pythia} samples.  The arrows indicate the $90\%$ acceptance value.}
\label{fig:yeleeff}
\end{figure}

  The distributions of the $p_T$ and rapidity ($y$)
of bottom quarks in Monte Carlo events that pass $J/\psi$ and lepton selection
(shown in figure~\ref{fig:pteleeff}--\ref{fig:yeleeff}) are used to determine the
$p_T$ and rapidity regions for the calculation of the correction factor.
$p_T^{J/\psi}$ is defined to be the value of the bottom quark transverse
momentum in which $90\%$ of the
b quarks which decay to a $J/\psi$ have a higher momenta.  $y^{J/\psi}$ is defined to be the value
of  bottom quark's $|y|$ in which $90\%$ of the
b quarks which decay to a $J/\psi$ have a lower $|y|$.  $p_T^\ell$ and $y^\ell$ are defined in a
similar manner for the bottom quark that decayed into the additional lepton.
Table~\ref{table:eff90} shows the value determined in both the electron
and muon samples for three different production mechanisms.  The rapidities of all three mechanisms are very similar and are determined by
the detector geometry.  The $p_T$ values are different in
the three mechanisms.   Flavor creation produces two bottom quarks with
similar momenta, while gluon splitting and flavor excitation produce
quarks with dissimilar $p_T$.    The values of
$y^{J/\psi}$, $y^{\ell}$, $p_T^{J/\psi}$, and $p_T^{\ell}$ used in the
correction factor calculation is the average of the three production
mechanisms.   To estimate the size of the systematic uncertainty,
$C_{B\rightarrow b}$ is calculated for each production mechanism
separately.  The systematic uncertainty is estimated as the largest
difference between the individual and combined production mechanisms.

\begin{table}
\begin{ruledtabular}
\begin{tabular}{ccccc}
Sample & $p_T^{J/\psi}$& $y^{J/\psi}$ &$p_T^e$& $y^e$ \\
FC &6.8 GeV&0.66&5.3 GeV&0.98\\
FE &7.1 GeV&0.66&3.8 GeV&1.06\\
GS &6.4 GeV&0.70&3.8 GeV&0.92\\\hline
Ave&6.8 GeV&0.67&4.3 GeV&0.99\\
\end{tabular}
\vspace{0.5cm}
\begin{tabular}{ccccc}
Sample & $p_T^{J/\psi}$& $y^{J/\psi}$ &$p_T^\mu$& $y^\mu$ \\ 
FC &7.3 GeV&0.66&6.6 GeV&0.60\\
FE &7.0 GeV&0.66&5.8 GeV&0.66\\
GS &6.6 GeV&0.68&5.7 GeV&0.58\\\hline
Ave&7.0 GeV&0.67&6.0 GeV&0.61\\
\end{tabular}
\end{ruledtabular}
\caption{$90\%$ acceptance requirements on the bottom quarks decaying
to a $J/\psi$ or a lepton predicting by {\sc Pythia} Monte Carlo and a
detector simulation.  Top: Electron.  Bottom: Muon.}
\label{table:eff90}
\end{table}

Table~\ref{table:effcorr} shows the
$f_{toward,mc}^{{b\overline{b}}^{~90\%}}$, $f_{toward,mc}^{b\rightarrow
J/\psi X; \overline{b}\rightarrow \ell Y}$, and $C_{B\rightarrow b}$ for
the complete sample and the three separate production mechanisms.  The
values estimated are:
\begin{eqnarray}
C_{B\rightarrow b}^{e}&=&0.967\pm0.019(\rm{stat.})\pm0.088(\rm{syst.})\\
C_{B\rightarrow b}^{\mu}&=&0.968\pm0.026(\rm{stat.})\pm0.061(\rm{syst.})
\end{eqnarray}

\begin{table*}
\begin{ruledtabular}
\begin{tabular}{ccccc} 
 & FC & FE & GS & Combined  \\ 
$f_{toward,mc}^{{b\overline{b}}^{~90\%}}$ &$5.1\pm0.1\%$&$21.4\pm0.5\%$&$46.4\pm0.5\%$&$26.4\pm0.2\%$\\
$f_{toward,mc}^{b\rightarrow J/\psi X; \overline{b}\rightarrow \ell Y}$&$5.8\pm0.4\%$&$23.4\pm0.8\%$&$47.8\pm0.8\%$&$27.3\pm0.5\%$ \\\hline
$C_{B\rightarrow b}$ &$0.879\pm0.063$&$0.915\pm0.038$&$0.971\pm0.020$&$0.967\pm0.019$\\
\end{tabular}
\vspace{0.5cm}
\begin{tabular}{ccccc} 
 & FC & FE & GS & Combined  \\
$f_{toward,mc}^{{b\overline{b}}^{~90\%}}$ &$3.5\pm0.2\%$&$19.5\pm0.8\%$&$47.2\pm0.8\%$&$25.5\pm0.4\%$\\
$f_{toward,mc}^{b\rightarrow J/\psi X; \overline{b}\rightarrow \ell
Y}$&$3.4\pm0.5\%$&$20.4\pm1.3\%$&$49.3\pm1.2\%$&$26.3\pm0.7\%$ \\ \hline
$C_{B\rightarrow b}$ &$1.029\pm0.164$&$0.956\pm0.072$&$0.957\pm0.028$&$0.968\pm0.026$\\
\end{tabular}
\end{ruledtabular}
\caption{Correction factor between the experimental measurement and
the bottom quarks.  The errors quoted are statistical only.  Top: Electron sample.  Bottom: Muon sample.}
\label{table:effcorr}
\end{table*}
 
The measured toward fraction for the bottom quarks ($f_{toward}^{corr}$)
extracted using the correction factor ($C_{B\rightarrow b}$) is:

\begin{eqnarray}
f_{toward}^{corr,e}&=&18.6^{+6.3}_{-5.7}~ ^{+0.5}_{-0.6} \pm 1.7 \% \\
f_{toward}^{corr,\mu}&=&33.4^{+8.9}_{-7.9}~ ^{+7.7}_{-3.0} \pm 2.3\%
\end{eqnarray}
where the first error is the fit error, the second error is the
additional shape systematic uncertainties on $f_{toward}$, and the third error
is the uncertainty due to the correction to the bottom quark kinematics.

\subsection{Data--Theory Comparisons}

The measured toward fraction corrected to the quark level is compared
to the NLO QCD predictions~\cite{MNR}, using the same requirements as
for the
correction of the experimental measurements.  The NLO prediction ($f_{toward}^{NLO}$) is
made using $m_b=4.75~\rm{GeV}$, a renormalization/factorization scale
$\mu=\sqrt{m_b^2+(p_T^b+p_T^{\overline{b}})/2}$ and CTEQ5M~\cite{cteq} and
MRST99~\cite{mrst} parton distribution functions (PDFs).  To estimate the systematic uncertainty in the NLO
calculation, $m_b$ is varied from 4.5--5.0 GeV, and $\mu$ is varied from
0.5--2.0.  To study the effects of large initial state parton transverse
momenta ($k_T$), the NLO prediction is also made with $\langle k_T \rangle$ values of
0--4 GeV.  The $k_T$ effects are implemented in the same manner as in Ref.~\cite{kt}, where  a
$\langle k_T \rangle$ of 3--4 GeV per parton is predicted at the Tevatron.  $f_{toward}^{NLO}$ is predicted using MRST99 and
CTEQ5M PDFs, respectively, for the different input values of $\langle k_T \rangle$, $m_b$, and $\mu$.  
Table~\ref{table:combined} shows the summary of the predictions.
The NLO predictions do not depend strongly on the PDF selected.
The measured $f_{forward}^{corr}$ for both the electron and muon sample are consistent with the NLO prediction
for values of $\langle k_T \rangle$ between zero and 3~\rm{GeV}.  The renormalization/factorization scale uncertainties
in the NLO predictions and the statistical uncertainty in the measurement of $f_{toward}^{corr}$ prohibit a more precise 
determination of $\langle k_T \rangle$ from this analysis.

\begin{table}
\begin{ruledtabular}
\renewcommand{\baselinestretch}{1.2}
\small\normalsize
 \begin{tabular}{cc} 
\multicolumn{2}{c}{NLO MRST99(Electron)}\\ \hline
 $\langle k_T \rangle =0.0~\rm{GeV}$& $16.9\% \pm 0.2\% ~\rm{(stat.)}^{+3.6\%}_{-3.8\%} ~\rm{(sys.)}$\\ 
 $\langle k_T \rangle =1.0~\rm{GeV}$& $19.4\% \pm 0.3\% ~\rm{(stat.)}^{+4.8\%}_{-4.1\%} ~\rm{(sys.)}$\\ 
 $\langle k_T \rangle =2.0~\rm{GeV}$& $23.2\% \pm 0.4\% ~\rm{(stat.)}^{+5.8\%}_{-5.0\%} ~\rm{(sys.)}$\\ 
 $\langle k_T \rangle =3.0~\rm{GeV}$& $31.9\% \pm 0.6\% ~\rm{(stat.)}^{+5.3\%}_{-5.7\%} ~\rm{(sys.)}$\\ 
 $\langle k_T \rangle =4.0~\rm{GeV}$& $44.9\% \pm 0.7\% ~\rm{(stat.)}^{+5.4\%}_{-4.8\%} ~\rm{(sys.)}$\\ \hline
\multicolumn{2}{c}{NLO CTEQ5M(Electron)}\\ \hline
 $\langle k_T \rangle =0.0~\rm{GeV}$& $16.5\% \pm 0.2\% ~\rm{(stat.)}^{+3.7\%}_{-3.3\%} ~\rm{(sys.)}$\\ 
 $\langle k_T \rangle =1.0~\rm{GeV}$& $19.1\% \pm 0.3\% ~\rm{(stat.)}^{+4.9\%}_{-3.9\%} ~\rm{(sys.)}$\\ 
 $\langle k_T \rangle =2.0~\rm{GeV}$& $23.1\% \pm 0.4\% ~\rm{(stat.)}^{+5.2\%}_{-5.0\%} ~\rm{(sys.)}$\\ 
 $\langle k_T \rangle =3.0~\rm{GeV}$& $31.7\% \pm 0.6\% ~\rm{(stat.)}^{+6.0\%}_{-5.4\%} ~\rm{(sys.)}$\\ 
 $\langle k_T \rangle =4.0~\rm{GeV}$& $45.1\% \pm 0.7\% ~\rm{(stat.)}^{+5.5\%}_{-4.6\%} ~\rm{(sys.)}$\\ \hline
PYTHIA(Electron)& $26.4\% \pm 0.2\% ~\rm{(stat.)}$ \\ \hline
Data(Electron)& $18.6^{+6.3}_{-5.7}~ ^{+0.5}_{-0.6} \pm 1.7 \%$\\ 
\end{tabular}
\vspace{0.5cm}
 \begin{tabular}{cc}
\multicolumn{2}{c}{NLO MRST99(Muon)} \\ \hline
$\langle k_T \rangle =0.0~\rm{GeV}$& $16.7\% \pm 0.3\% ~\rm{(stat.)}^{+5.2\%}_{-3.6\%} ~\rm{(sys.)}$\\ 
$\langle k_T \rangle =1.0~\rm{GeV}$& $22.7\% \pm 0.6\% ~\rm{(stat.)}^{+6.1\%}_{-7.0\%} ~\rm{(sys.)}$\\ 
$\langle k_T \rangle =2.0~\rm{GeV}$& $24.7\% \pm 0.6\% ~\rm{(stat.)}^{+6.7\%}_{-6.5\%} ~\rm{(sys.)}$\\ 
$\langle k_T \rangle =3.0~\rm{GeV}$& $32.1\% \pm 0.6\% ~\rm{(stat.)}^{+6.4\%}_{-6.9\%} ~\rm{(sys.)}$\\ 
$\langle k_T \rangle =4.0~\rm{GeV}$& $44.9\% \pm 0.9\% ~\rm{(stat.)}^{+5.1\%}_{-6.6\%} ~\rm{(sys.)}$\\ \hline
\multicolumn{2}{c}{NLO CTEQ5M(Muon)} \\ \hline
$\langle k_T \rangle =0.0~\rm{GeV}$& $16.8\% \pm 0.3\% ~\rm{(stat.)}^{+4.1\%}_{-3.8\%} ~\rm{(sys.)}$\\ 
 $\langle k_T \rangle =1.0~\rm{GeV}$& $22.1\% \pm 0.5\% ~\rm{(stat.)}^{+5.6\%}_{-5.9\%} ~\rm{(sys.)}$\\ 
$\langle k_T \rangle =2.0~\rm{GeV}$& $23.8\% \pm 0.9\% ~\rm{(stat.)}^{+7.4\%}_{-5.0\%} ~\rm{(sys.)}$\\ 
$\langle k_T \rangle =3.0~\rm{GeV}$& $31.9\% \pm 0.9\% ~\rm{(stat.)}^{+5.7\%}_{-6.5\%} ~\rm{(sys.)}$\\ 
$\langle k_T \rangle =4.0~\rm{GeV}$& $44.4\% \pm 1.2\% ~\rm{(stat.)}^{+7.1\%}_{-5.0\%} ~\rm{(sys.)}$\\ \hline
PYTHIA(Muon) & $25.5\% \pm0.4\% ~\rm{(stat.)}$ \\ \hline
 Data(Muon)& $33.4^{+8.9}_{-7.9}~ ^{+7.7}_{-3.0} \pm 2.3\%$\\  
\end{tabular}
\renewcommand{\baselinestretch}{1.0}
\small\normalsize

\caption{Compilation of the corrected data results, the PYTHIA
predictions, and the NLO predictions of
$f_{toward}$   Top: Electron.
Bottom: Muon.}
\label{table:combined}
\end{ruledtabular}
\end{table}

\begin{figure*}[htbp]
       \centerline{
               \includegraphics[width=8.6cm]{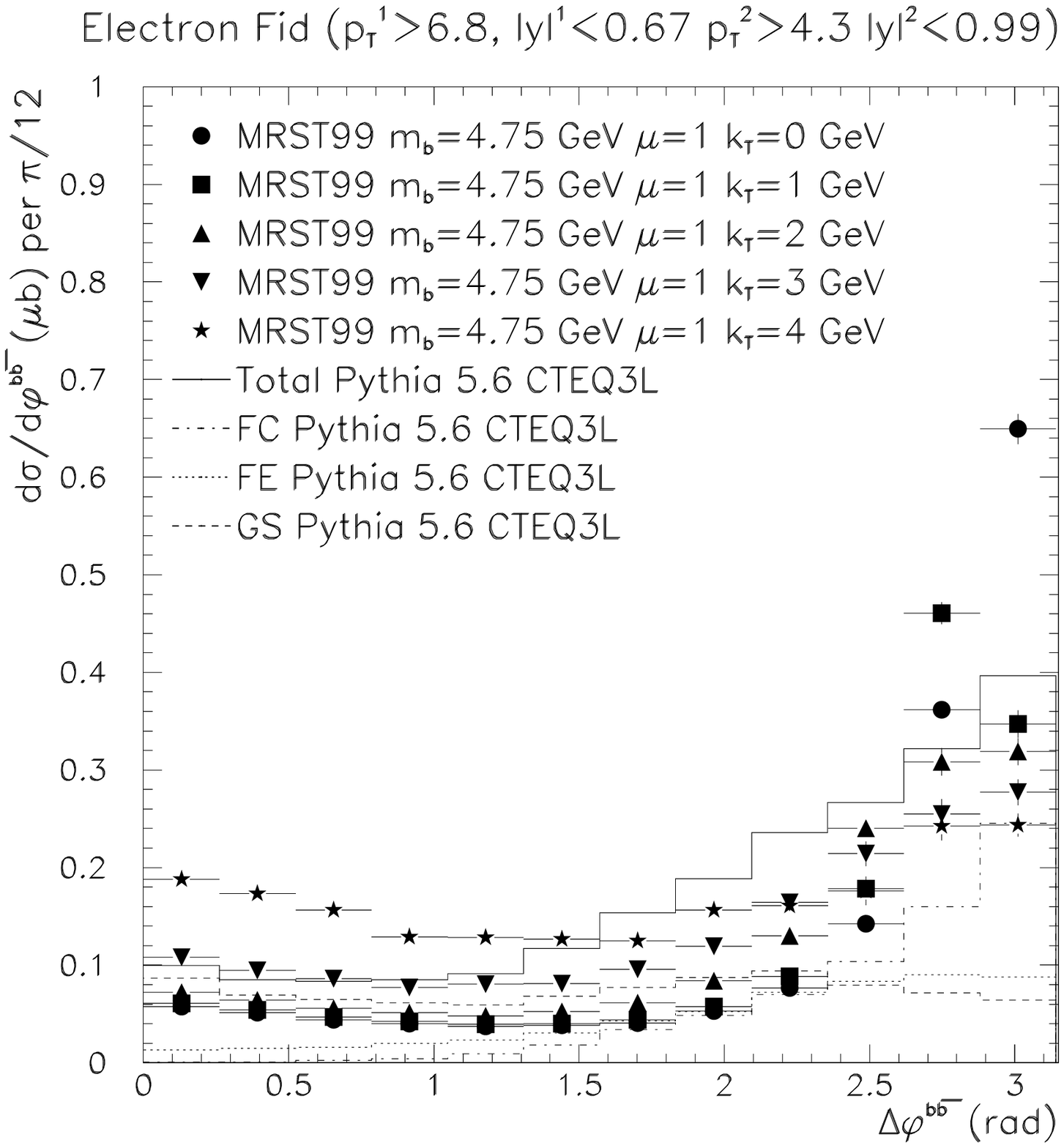}
	       \hspace{.1cm}
               \includegraphics[width=8.6cm]{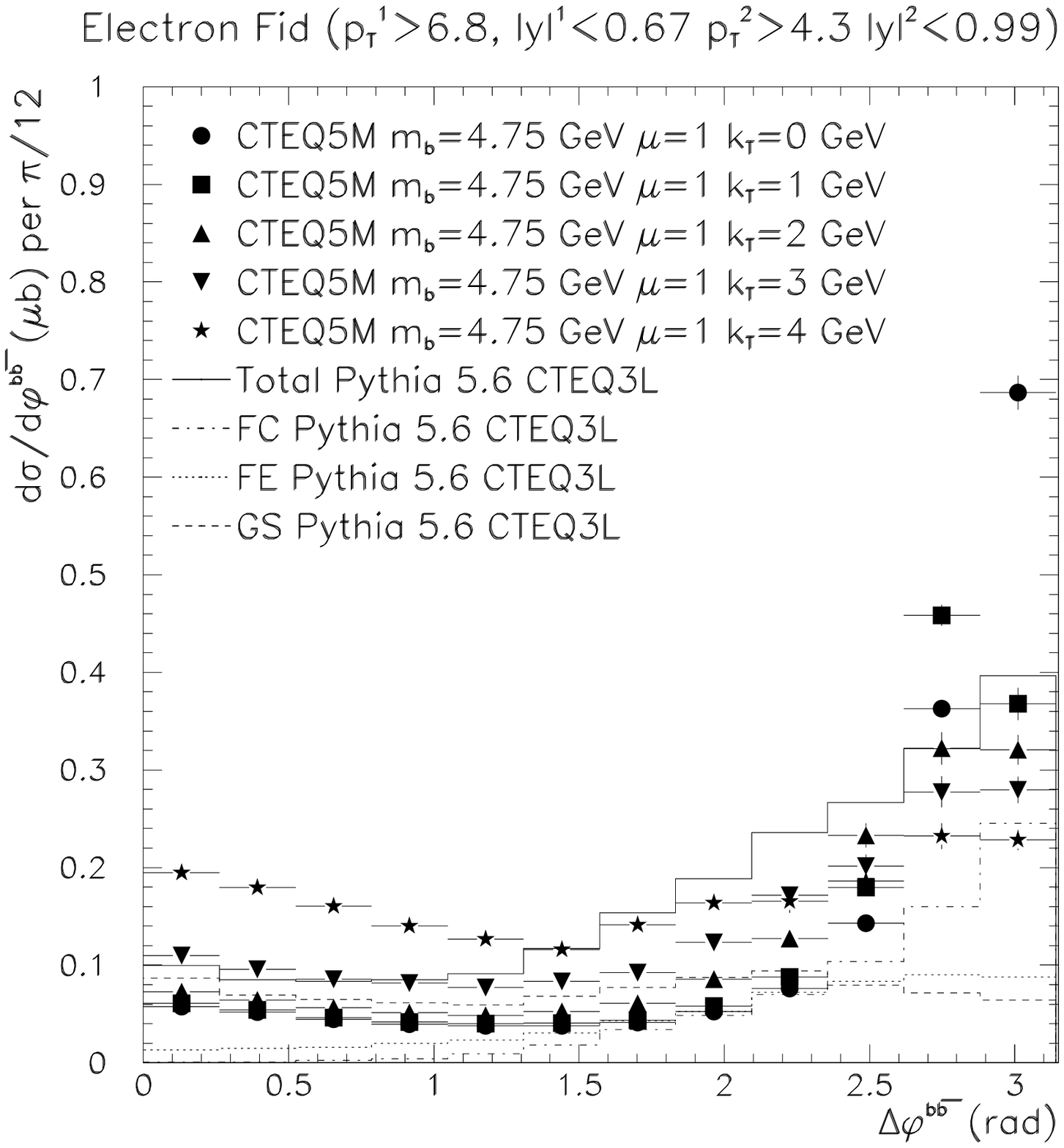}}
       \centerline{
               \includegraphics[width=8.6cm]{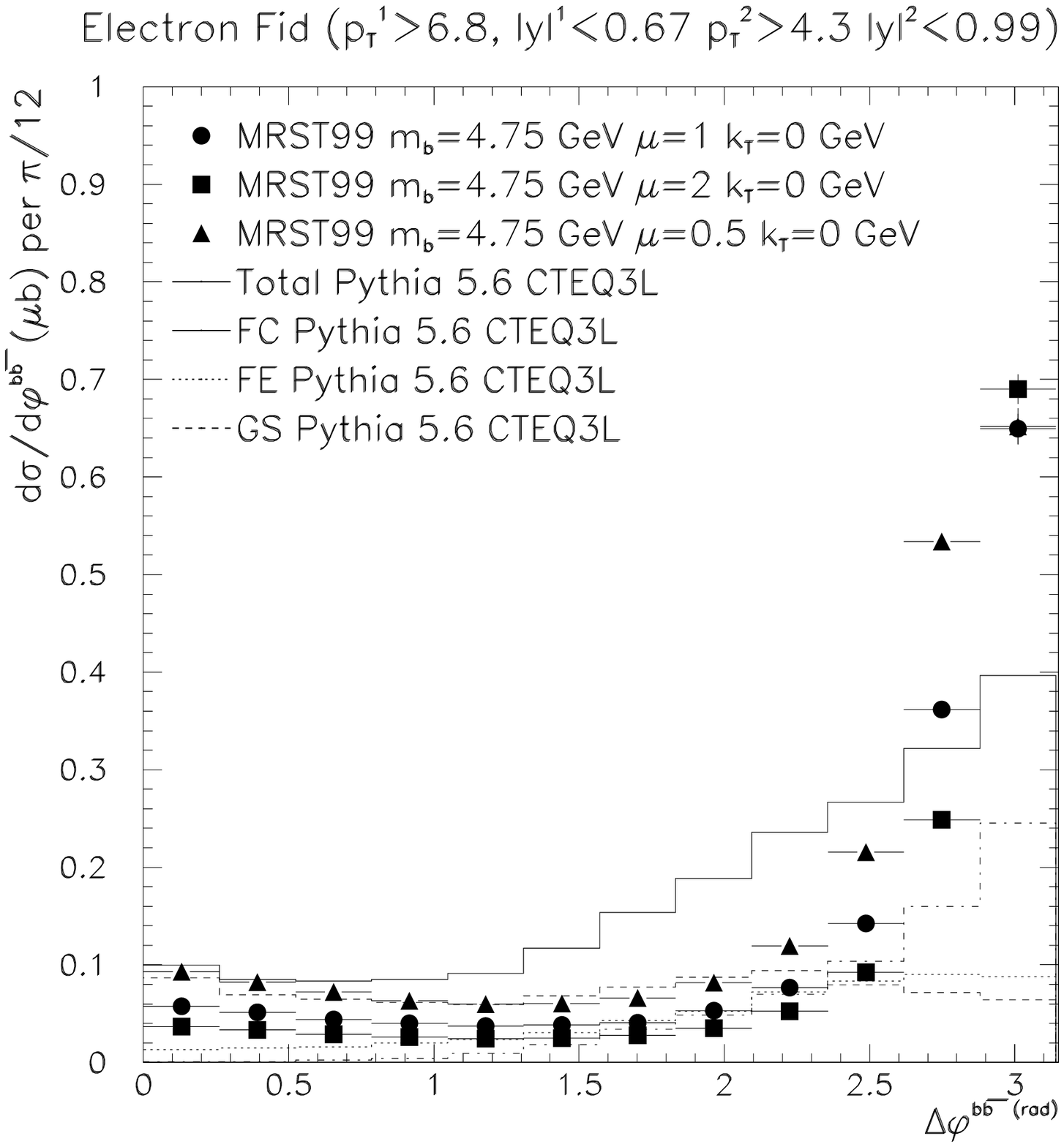}
	       \hspace{.1cm}
               \includegraphics[width=8.6cm]{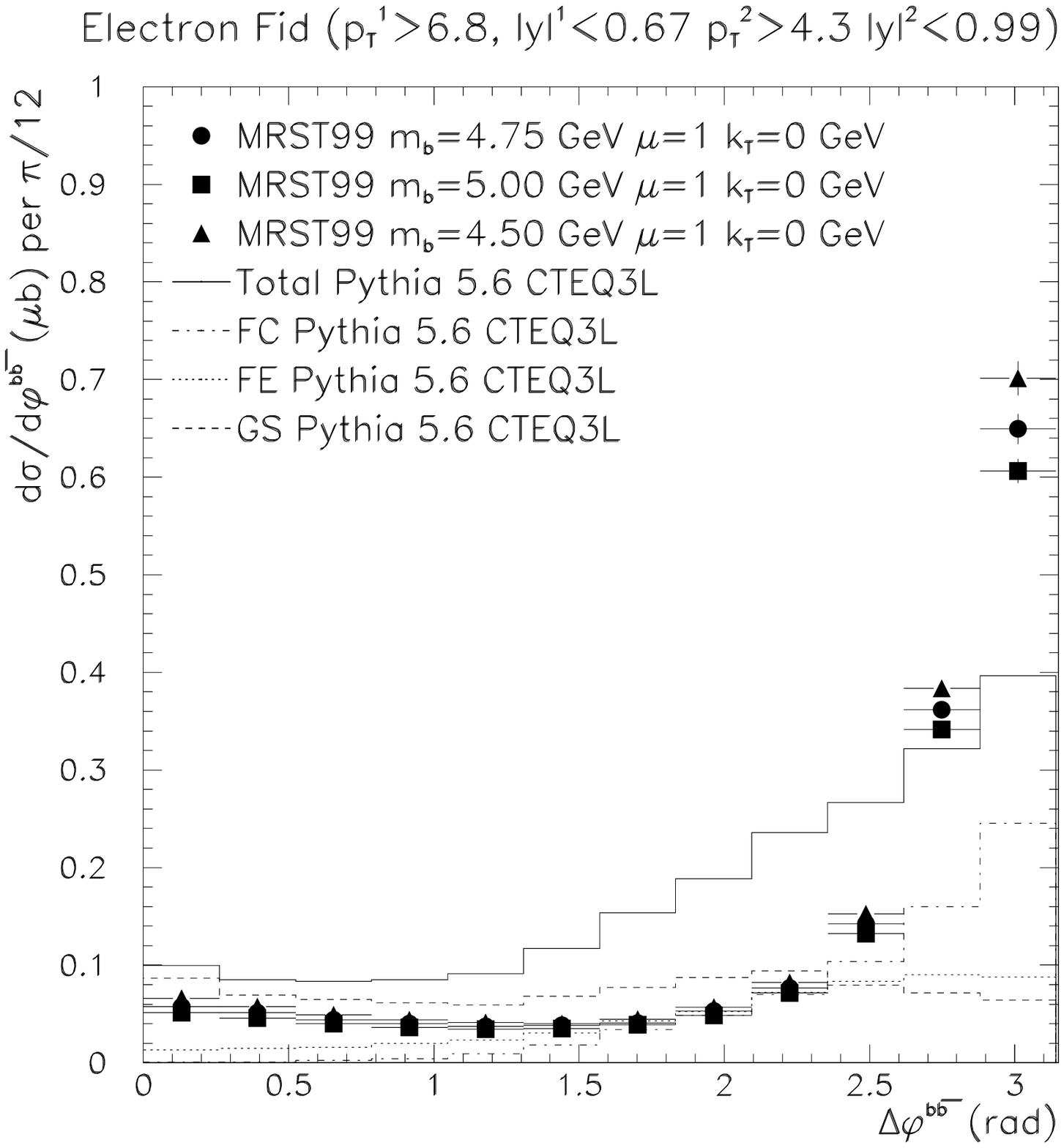}}

\caption{NLO prediction~\cite{MNR} of bottom spectra in the acceptance region.  The PYTHIA prediction is shown
as a reference.
Top Left: MRST99 PDF varying the additional $k_T$ smearing.
Top Right: CTEQ5M PDF varying the additional $k_T$ smearing.
Bottom Left: MRST99 PDF varying the renormalization scale $\mu$. Bottom
Right: MRST99 PDF varying the bottom quark mass $m_B$.}
\label{fig:dphi_theory}
\end{figure*}

Figure~\ref{fig:dphi_theory} illustrates the
effects of varying the PDF, $\langle k_T \rangle$, $m_b$, and
renormalization/factorization scale on
the NLO predictions in the electron acceptance region.  Varying $m_b$ mass does not affect the predicted
shape, but instead only affects the total cross section predicted.  The
two different PDFs studied yield very similar shape and total cross
section predictions.  Only scale
and $\langle k_T \rangle$ variations yield appreciably different shape
and total cross section
predictions.  Varying the renormalization/factorization scale changes
the total cross section as expected; lowering the scale increases the
total cross section.  In addition, varying the scale changes the
predicted rate at large $\Delta \phi^{b\overline{b}}$ ($>2.9$ radians)
relative to the rest of the distribution, while the shape of $<2.9$
radian region varies little. Varying the scale changes the relative rates of
$p\overline{p} \rightarrow b\overline{b}$ to  $p\overline{p} \rightarrow
b\overline{b}g$ in the NLO prediction.  Varying the $\langle k_T
\rangle$ on the other hand, changes the predicted $\Delta
\phi^{b\overline{b}}$ in a more continuous manner.  With the increased
number of $J/\psi +\ell$ expected in Run II, a differential azimuthal
cross section measurement with 6--12 bins in $\Delta\phi$ should be able
to separate scale uncertainty and $k_T$ smearing effects.

\section{Conclusions} 

We have presented two new measurements of the $\Delta\phi$ distribution for bottom anti-bottom pairs produced through QCD interactions 
at the Tevatron.    These measurements are specifically targeted to measure the $\Delta\phi$ distribution down to arbitrarily 
small opening angles for $b\overline{b}$ pairs produced with low transverse momentum, where previous measurements have lacked sensitivity.  The small $b\overline{b}$ opening angle region is 
of interest because in this region, the higher-order $b\overline{b}$ production mechanisms dominate over flavor creation.   
The data presented here are consistent with other measurements and cannot be described solely by flavor creation. 
Both measurements indicate that a significant fraction of the $b\overline{b}$ pairs (roughly 25\%) are 
produced with $\Delta\phi < 90^{\circ}$, in agreement with the conclusion from previous analyses~\cite{cdf-96,field} that flavor excitation 
and gluon splitting play a significant role in $b\overline{b}$ production at the Tevatron.  The results of these measurements 
are consistent with the parton shower Monte Carlo models of {\sc Pythia} and {\sc Herwig} as implemented in Ref.~\cite{field} and with
NLO QCD predictions.  Neither non-perturbitive~\cite{nonpert} nor supersymmetric~\cite{squark} production mechanisms are needed in order to describe the measured $\Delta\phi$ spectra. 

\begin{acknowledgments}
We thank the Fermilab staff and the technical staffs of the participating institutions for their vital contributions.  This work was supported by the U.S. Department of Energy and National Science Foundation; the Italian Istituto Nazionale di Fisica Nucleare; the Ministry of Education, Culture, Sports, Science and Technology of Japan; the Natural Sciences and Engineering Research Council of Canada; the National Science Council of the Republic of China; the Swiss National Science Foundation; the A.P. Sloan Foundation; the Bundesministerium fuer Bildung und Forschung, Germany; the Korean Science and Engineering Foundation and the Korean Research Foundation; the Particle Physics and Astronomy Research Council and the Royal Society, UK; the Russian Foundation for Basic Research; the Comision Interministerial de Ciencia y Tecnologia, Spain; work supported in part by the European Community's Human Potential Programme under contract HPRN-CT-20002, Probe for New Physics; and this work was supported by Research Fund of Istanbul University Project No. 1755/21122001.
\end{acknowledgments}

\appendix
\section{Monte Carlo Parameters for the Secondary Vertex Tag Correlation Analysis}
\label{sec:bvtx-mc}

The specific Monte Carlo generator settings used for the secondary vertex tag correlation analysis are specified below.  For a more explicit discussion of the Monte Carlo, consult Ref.~\cite{lannon-thesis}.

\subsection{{\sc Pythia}}
Version 6.203 of {\sc Pythia}~\cite{pythia1,pythia2} was used for this analysis.  Flavor creation events are generated with the process MSEL = 5, while flavor excitation and gluon splitting use MSEL = 1.  The parameter PARP(67), which holds the value that is multiplied by the $Q^2$ of the hard scatter to determine the maximum virtuality of the initial-state shower, was used to control the amount of initial state radiation in the {\sc Pythia} samples.  Three different values of PARP(67) were used and for each setting, other {\sc Pythia} parameters were manipulated to give the best match to the CDF data.  The naming convention for the three ISR samples is given in Table~\ref{table:isr}.  This tuning was done with the CTEQ5L~\cite{cteq5l} parton distribution functions and different {\sc Pythia} parameters may be required to achieve the same tuning with a different set of parton distribution functions.  Table~\ref{table:pythia-par} gives the values of the parameters used for this tuning.  Parameters not mentioned in Table~\ref{table:pythia-par} were left at the default values for this version of {\sc Pythia}.

\begin{table}
\begin{ruledtabular}
\begin{tabular}{ll}
Sample Name & ISR Setting \\
\hline
High  & PARP(67) = 4.0 \\
Medium  & PARP(67) = 3.0 \\
Low  & PARP(67) = 1.0 \\
\end{tabular}
\caption{\label{table:isr}The naming convention for the three ISR samples for {\sc Pythia}.}  
\end{ruledtabular}
\end{table}

\begin{table*}
\begin{ruledtabular}
\begin{tabular}{|c|p{9cm}|c|c|}
Parameter&Meaning &PARP(67)=3.0,4.0&PARP(67)=1.0\\
\hline
MSTP(81) & Multiple-parton interaction switch & \multicolumn{2}{c|}{ 1 (Multiple Parton Interactions ON)} \\
\hline
MSTP(82) & Model of multiple parton interactions & \multicolumn{2}{c|}{3 (Varying impact parameter assuming} \\
 & & \multicolumn{2}{c|}{a single Gaussian matter distribution)} \\
\hline
PARP(82)& $p_T$ turn-off when using single Gaussian model of multiple interactions & 1.7& 1.6 \\
\hline
PARP(85) & Probability that a multiple parton interaction produces two gluons with color connections to the ``nearest neighbors'' & \multicolumn{2}{c|}{1.0}\\
\hline
PARP(86) & Probability that an MPI produces two gluons either as described above or as a closed gluon loop. The rest of the MPIs produce quark-antiquark pairs & \multicolumn{2}{c|}{1.0} \\
\hline
PARP(89) & Determines the reference energy E0 & \multicolumn{2}{c|}{1800.} \\
\end{tabular}
\caption{\label{table:pythia-par}The table above shows the { \sc Pythia} setting used to tune the underlying event to data for the CTEQ5L parton distribution set and three different initial-state radiation settings.}  
\end{ruledtabular}
\end{table*}
          
\subsection{{\sc Herwig}}

Version 6.400 of {\sc Herwig}~\cite{herwig} was used for this analysis.  The flavor creation and flavor excitation samples were generated with IPROC = 1705, and the gluon splitting sample was made using IPROC = 1500.  Only one {\sc Herwig} sample was generated for each production mechanism, again using the CTEQ5L parton distribution functions.  $\Lambda_{QCD}$ was set to 192 MeV and the CLPOW parameter was set to 1.25 to match the observed frequency of $B$ baryons at CDF.  All other {\sc Herwig} parameters were left at their default values for this version.

\section{${\boldmath J/\psi}$--lepton Correlation Monte Carlo Event Samples}
\label{monte carlo}

Various simulated data sets have been necessary for the measurement of $f_{toward}$ in the $J/\psi$--lepton data sample.   The follow section will detail the method of generating the simulated events used in the measurement.

For the bottom impact parameter description, {\sc Pythia}~\cite{pythia1,pythia2} with the CTEQ3L~\cite{cteq} parton distribution
functions is used.   The bottom quarks are hadronized using the Bowler
fragmentation function~\cite{lund} and using the LUND
string fragmentation model.  The resulting
bottom hadrons are decayed using the CLEO decay model~\cite{cleo}.  The
events are then passed through a detector simulation~\cite{qfl} and the
trigger simulation.  The same selection criteria is
applied to the $J/\psi$ candidates in Monte Carlo as in data.

For the bottom decay impact parameter shape for muons, the muons are required to be
fiducial in both the CMU and the CMP muon subsystems with a SVX$^\prime$ track and a $p_T > 3~\rm{GeV}$.

  For the bottom decay shapes for electrons, the electrons are required to
be in the CEM fiducial region with a SVX$^\prime$ quality track
with a $p_T > 2~\rm{GeV}$.  The efficiency of the electron identification
criteria is simulated in the same manner as Ref.~\cite{bcprd}.  The CPR,
the CES, and the CTC $dE/dx$ selection criteria do not depend on the isolation of the
electron, due to the fine segmentation of the CPR, the CES and the CTC.
Therefore, the efficiencies as a function of $p_T$ of the CPR and the CES selection
derived by Ref.~\cite{cescpr} using conversions can be used.   The CTC
$dE/dx$ efficiency as a function of $p$ is defined by the selection
criteria.  The rate of signal events being removed as conversions and the $E_{had}/E_{EM}$ and $E/p$ efficiencies
depend on the isolation of the track.  Therefore, the values simulated in
the Monte Carlo have to be used to determine the efficiency.

For residual conversion studies and impact parameter shape determination, the necessary simulated sample was generated in the following manner.  A sample of $\pi^0$ is simulated in
the detector.  A soft electron conversion candidate is required to have a found track with SVX$^\prime$ information in the electron fiducial region with a $p_T>2~\rm{GeV}$.  The
efficiency of electron identification requirements is simulated in the
same manner as for the bottom impact parameter shape sample.  

\begin{figure}[htbp]
\includegraphics[width=8.6cm]{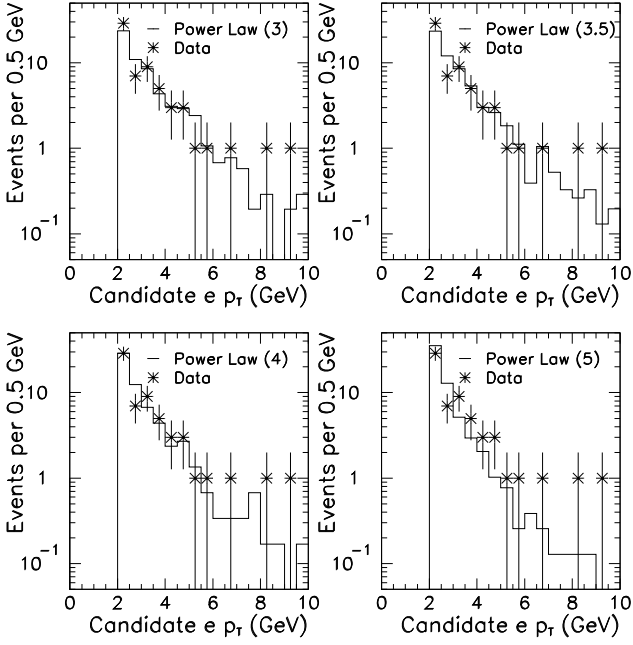}
\caption{The $p_T$ spectra of the SLT conversion candidates.  Top Left: 3rd order power law.  Top Right:
3.5 order power law.  Bottom  Left: 4th order power law.  Bottom
Right: 5th order power law.}
\label{fig:powercnv}
\end{figure}

The simulated $\pi^0$ are generated with a power law spectra for $p_T$ and a flat
$\eta$ distribution.  The order of the power law is varied in order to
match the $p_T$ spectra of the found SLT conversion candidates in data.
The shape of the found pair candidates' $p_T$ is used as a cross check
of the power law description of the conversions.
Figure~\ref{fig:powercnv} shows the Monte Carlo spectra normalized to
the data for a power law of 3, 3.5, 4, and 5.  The 3.5
order power law describes the data well and is used for the calculation of
the efficiency.  The 3rd and 4th order power law is used as a estimate
of the systematic uncertainty. 

\begin{figure}[htbp]
 \includegraphics[width=8.6cm]{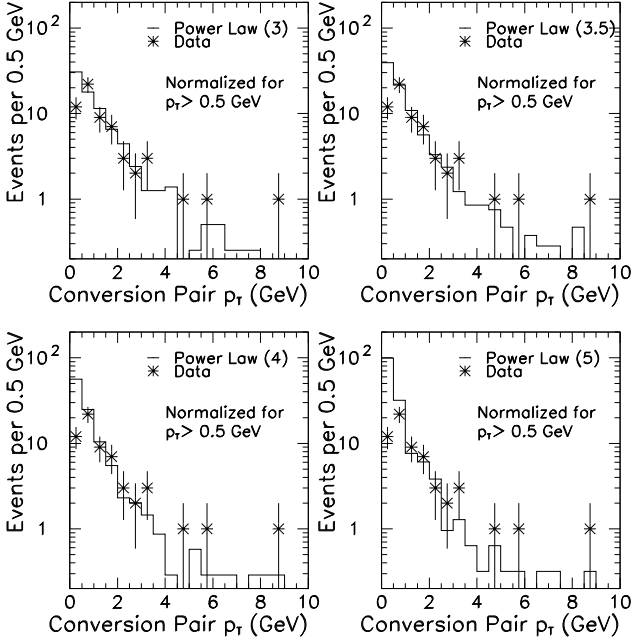}
\caption{The $p_T$ spectra of the pair candidates.  The Monte Carlo is normalized to the data
with $p_T>0.5~\rm{GeV}$.  Top Left: 3rd order power law.  Top Right:
3.5 order power law.  Bottom  Left: 4th order power law.  Bottom
Right: 5th order power law.}
\label{fig:powerpair}
\end{figure}

The $p_T$ of the pair candidates is shown in
figure~\ref{fig:powerpair}.   The 3.5 order power law describes the
shape of events with $p_T>0.5~\rm{GeV}$ (where the tracking is
assumed to be fully efficient).  Half the difference between the 3rd and 4th order
power law spectra is used as the systematic uncertainty of the estimate.

The impact parameter--c$\tau$ shape of the $B_c$ background was determined using a Monte Carlo sample.    The $B_c$ mesons are generated
according to the NLO fragmentation model from Ref.~\cite{bcmodel} with a flat
rapidity spectra.  The particles are decayed using the semi-leptonic
decay model of Ref.~\cite{igsw} and passed through a detector and trigger
simulation.  The selection criteria used is identical to
section~\ref{ipfit}.

  Single bottom quarks are generated according to the next-to-leading order QCD
predictions by Ref.~\cite{nde} and fragmented using the Peterson
fragmentation model~\cite{peterson}.  The resulting
bottom hadrons are decayed using the CLEO decay model~\cite{cleo},
requiring a $J/\psi \rightarrow \mu^+ \mu^-$ decay.  The
events are then passed through a detector simulation~\cite{qfl} and the
trigger simulation.  For both the calculation of electron
and muon fake rates, the $J/\psi$ is required to pass the same selection
criteria as data.  The sample is normalized to the number of $J/\psi$
events from bottom decay fit in data in section~\ref{jpsifit}.

\section{${\boldmath J/\psi}$--lepton Correlation Log-Likelihood Function}
\label{fit}
  The fit parameters will be in lower case, the constraints will be upper
case, and the errors on the constraints (if applicable) are denoted $\Delta
\rm{(Constraint)}$.  The superscripts indicate the additional lepton
used ($e$ for electrons, $\mu$ for muons) and the $\Delta\phi$ region (t
for toward, a for away).   For example, $N_{B_c}^{e,t}$ is the number
of $B_c \rightarrow J/\psi e X$ background events estimated in the
electron sample in the toward $\Delta\phi$ region.

\subsection{Data}
The inputs to the fit on an event-by-event basis are the $J/\psi$ candidate's
$c\tau$  and the additional lepton candidate's impact parameter.  In the following sections, $x$ will denote
the impact parameter, and $y$ will denote the $c\tau$.  The numbers of
candidates in the $J/\psi$ mass sideband and signal regions in both
$\Delta\phi$ regions are used as a constraint in the likelihood. In the
electron sample, the numbers of found conversions in $J/\psi$ mass
sideband and signal regions in both $\Delta\phi$ regions also used as
a constraint.  Conversion
constraints are discussed later in their respective sections.  

\begin{equation}
P(n_{signal},N_{signal})=
\frac{(n_{signal})^{N_{signal}}}{N_{signal}!}e^{-n_{signal}}
\end{equation}  
with the appropriate $n_{signal}$ and $N_{signal}$ for the given sample and
$\Delta\phi$ region.  $n_{signal}$ is not a fit parameter, but is a
function of the other fit parameters, shown in section~\ref{subsec:sums}.
 
\subsection{${\boldmath b \rightarrow J/\psi X;\overline{b} \rightarrow \ell X'}$
signal}
The shapes which are used for the $b \rightarrow J/\psi X;\overline{b} \rightarrow
\ell X'$ signal are described by the fit functions in
sections~\ref{jpsifit} and~\ref{ipfit}.  The impact
parameter and $c\tau$ are assumed to be uncorrelated. Therefore, the
shape which describes the signal is the product of the impact parameter
shape ($F_b^{d_0}(x)$) and the $c\tau$ shape
($F_b^{c\tau}(y)$) for bottom decay.  The parameters that are used in
$F_b^{d_0}(x)$ are different for the electron and muon fits.

  The number of $b\overline{b}$ events
fit is $n_{b\overline{b}}$ with the superscripts given by sample and
$\Delta\phi$ region.  For example, the number of $b\overline{b}$ events
fit in the toward $\Delta\phi$ region in the electron sample is
$n_{b\overline{b}}^{e,t}$.  The $b\overline{b}$ contribution of the
shape component of the likelihood is given by
$\frac{n_{b\overline{b}}}{n_{signal}} \cdot F_b^{d_0}(x) \cdot
F_b^{c\tau}(y)$ with the appropriate superscripts for the additional
lepton type and $\Delta\phi$ region.

\subsection{Unconstrained, uncorrelated backgrounds}
The impact parameter--$c\tau$ shapes of the three sources of uncorrelated
backgrounds without constraints, considered in this analysis, are
constructed using the functions derived in
sections~\ref{jpsifit} and~\ref{ipfit_back}.  The fit parameters
for these three backgrounds are:
\begin{itemize}
\item{$n_{dd}$: the number of events with the $J/\psi$ candidate and
with the
additional lepton candidate both directly produced} 
\item{$n_{bd}$: the number of events with the $J/\psi$ candidate from bottom
decay and with the
additional lepton candidate produced directly} 
\item{$n_{db}$: the number of events with a directly produced $J/\psi$
candidate and with an
additional lepton candidate from bottom decay}
\end{itemize} 
where the superscripts indicate of the sample and $\Delta\phi$ region.
The number of events with a directly produced $J/\psi$ candidate and an
additional lepton candidate from bottom decay is assumed to be small and
$n_{db}$ is fixed to zero.  This parameter is released and fit for as an
estimate of systematic uncertainty due to this assumption.

The shape component of the likelihood for these three backgrounds is
assembled in the same manner as the $b\overline{b}$ signal.

\subsection{Residual conversion background}
\label{sec:resconv}
The total number of predicted residual conversions is $R_{conv} \cdot
N_{conv}$, where $R_{conv}$ is the ratio between the number of residual versus found
conversions and $N_{conv}$ is the number of found conversions in the
sample.  The number of found conversions removed from the two
$\Delta\phi$ regions with the $J/\psi$ candidate in the mass signal
region is $N_{conv}^{t}=6$ and $N_{conv}^{t}=9$, respectively.  In
section~\ref{conversions}, $R_{conv}$ is estimated to be
$1.00\pm0.37$, using
data and Monte Carlo techniques.  Residual conversions are assumed to
pair with all three sources of uncorrelated $J/\psi$ candidates: fake
$J/\psi$ , directly produced $J/\psi$, and bottom decay $J/\psi$. The same value of
the fit parameter $r_{conv}$ is used for all sources of $J/\psi$
candidates that pair with the residual conversions.  The value of
$r_{conv}$ is constrained as a Gaussian probability in the likelihood.  
\begin{equation}
G(r_{conv}-R_{conv},\Delta R_{conv})=\frac{1}{\sqrt{2}\Delta
R_{conv}}e^{-\frac{1}{2}\left(\frac{r_{conv}-R_{conv}}{\Delta R_{conv}}\right)^2}
\end{equation}

The fit parameters that set the scale for the number of residual conversions
events with the $J/\psi$ candidate from bottom decay and direct
production are $n_{bconv}$ and $n_{dconv}$.  The parameters represent
the number of found conversions with the $J/\psi$ candidate from the
given source.  The number of residual conversions fit from these two
sources are $r_{conv}\cdot n_{bconv}$ and $r_{conv}\cdot n_{dconv}$.  The number of
residual conversions already included in the sideband shape component is
$r_{side}n_{side}^{e}f_{conv}^{d_0}$, where
$f_{conv}^{d_0}=\frac{r_{conv}\cdot
n_{convside}}{N_{sideband}}$ is the fit fraction of $J/\psi$ mass sideband events
where the electron is a residual conversion.  

Due to the relatively small number of residual conversions, fitting all
three pairing of $J/\psi$ candidate types with conversions is not
possible.  In order to constrain this component of the fit farther, the ratio of
between $n_{bconv}$ and $n_{dconv}$ is assumed to be the same as the
ratio between $J/\psi$ mesons from bottom decay and $J/\psi$ mesons
produced directly (at the primary vertex).  The fraction of $J/\psi$
mesons from bottom decay is fit to be $16.6\% \pm 0.2\%$ in section~\ref{jpsifit}, yielding the
relationship $n_{bconv}^{t/a}=0.2\cdot n_{dconv}^{t/a}$.  As an estimate of
the systematic uncertainty, $n_{bconv}^{t/a}$ and $n_{dconv}^{t/a}$ are fixed
to zero in separate fits in order to probe the full range of ratio  $n_{bconv}^{t/a}:n_{dconv}^{t/a}$.

The number of found conversions in the two $\Delta\phi$
regions is used as a constraint on the fit of the residual conversions.
The number of found conversions fit is the number of residual
conversions fit divided by the ratio of residual versus found
conversions: 
\begin{equation}
n_{conv}\equiv
n_{bconv}+n_{dconv}+\frac{r_{side}n_{side}^{e}n_{convside}}{N_{sideband}}
\end{equation}
The constraint using the number of found conversions is the Poisson
probability of finding $N_{conv}$ conversion candidates with a mean value of 
number of found conversion fit.
\begin{equation}
P(n_{conv},N_{conv})=
\frac{(n_{conv})^{N_{conv}}}{N_{conv}!}e^{-(n_{conv})}
\end{equation}  

\subsection{Fake ${\boldmath J/\psi}$ backgrounds}

The fake $J/\psi$ impact parameter--$c\tau$ background shape
($F_{sideband}^\mu$) is determined in
section~\ref{sideband} from a fit to the data for the muon sample.  The
predicted number of events for this background is the ratio between the
number of fake $J/\psi$ events in the $J/\psi$ mass signal and sideband
region ($R_{side}$) times the number of events seen in data with the $J/\psi$
candidate in the mass sideband regions ($N_{side}$) for the given sample and
$\Delta\phi$ region.

  In section~\ref{jpsisel}, the ratio is
determined to be $R_{side}=0.501\pm0.044$ from a fit of the total
$J/\psi$ data sample.  The same value for the fit parameter $r_{side}$
is used in both $\Delta\phi$ regions in the sample, but can be different in the
electron and muon samples.  The fit value of $r_{side}$ is constrained
using a Gaussian factor in the likelihood function.  
\begin{equation}
G(r_{side}-R_{side},\Delta R_{side})=\frac{1}{\sqrt{2}\Delta
R_{side}}e^{-\frac{1}{2}\left(\frac{r_{side}-R_{side}}{\Delta R_{side}}\right)^2}
\end{equation}

 The corresponding
fit parameter $n_{side}$, for the given sample and $\Delta\phi$
region, is constrained 
using the Poisson probability of measuring $N_{side}$ events for a
sample with an average of $n_{side}$ events.
\begin{equation}
P(n_{side},N_{side})=
\frac{(n_{side})^{N_{side}}}{N_{side}!}e^{-n_{side}}
\end{equation}  

The contribution of the shape component of the likelihood is
$\frac{r_{side}n_{side}}{n_{signal}} \cdot F_{side}(x,y)$ for the given sample and
$\Delta\phi$ region.

The fake $J/\psi$ background component in the electron sample is treated
differently due to the presence of residual conversions in the
background.  The fake $J/\psi$ shape is fit at
the same time as the $J/\psi$ signal region.   The $f_{conv}$ component
of the fake $J/\psi$ impact parameter--$c\tau$ shape is a composite of
two variables which are constrained.  $f_{conv}=\frac{r_{conv}\cdot
n_{convside}}{N_{sideband}}$ where $r_{conv}$, $n_{convside}$, and
$N_{sideband}$ are the fit ratio of residual to found conversion, fit
number of found conversions, and
$N_{sideband}=N_{side}^{e,t}+N_{side}^{\mu,a}$.   The parameter $n_{convside}$ is
constrained by the number of conversions found in the sideband using the Poisson probability:

\begin{equation}
P(n_{convside},N_{convside})=
\frac{(n_{convside})^{N_{convside}}}{N_{convside}!}e^{-n_{convside}}
\end{equation}  

\subsection{${\boldmath B_c \rightarrow J/\psi \ell X}$ backgrounds}
 $B_c \rightarrow
J/\psi \ell X$ background is predicted to only populate the toward region
in $\Delta\phi$.  The expected number of $B_c$ events is 
constrained in the likelihood as a Gaussian probability factor.
\begin{equation}
G(n_{B_c}-N_{B_c},\Delta N_{B_c})=\frac{1}{\sqrt{2}\Delta
N_{B_c}}e^{-\frac{1}{2}\left(\frac{n_{B_c}-N_{B_c}}{\Delta N_{B_c}}\right)^2}
\end{equation}
where $\Delta N_{B_c}$ is the positive-sided error of
$N_{B_c}$ if $(n_{B_c}-N_{B_c})\ge0.0$, and the negative-sided error otherwise.
 
\subsection{${\boldmath b \rightarrow J/\psi \ell_{fake} X}$ backgrounds}
As in the $B_c$ background, $b \rightarrow J/\psi \ell_{fake} X$
background events are only expected to populate the toward region in
$\Delta\phi$, and therefore the background is only fit for in the toward region
in the two samples.   The expected number is $b \rightarrow J/\psi \ell_{fake} X$ events is used as a 
constraint to the likelihood in the same way as $B_c$.
\begin{eqnarray*}
G(n_{B_{fake}}-N_{B_{fake}},\Delta N_{B_{fake}})&=&\frac{1}{\sqrt{2}\Delta
N_{B_{fake}}}\\
&\times& e^{-\frac{1}{2}\left(\frac{n_{B_{fake}}-N_{B_{fake}}}{\Delta N_{B_{fake}}}\right)^2}
\end{eqnarray*}

\subsection{$n_{signal}$ sums}
\label{subsec:sums}

The number of events fit in the $J/\psi$ mass signal region is a
function of fit parameters described previously in this section.  Listed
below are the functions for number of events fit for the two samples and
$\Delta\phi$ regions.
\begin{widetext}
\begin{eqnarray}
n_{signal}^{\mu,t}&=&n_{b\overline{b}}^{\mu,t} +n_{bd}^{\mu,t}
+n_{db}^{\mu,t} +n_{dd}^{\mu,t}+ r_{side}^{\mu} \cdot n_{side}^{\mu,t}+n_{B_c}^{\mu,t} + n_{B_{fake}}^{\mu,t}\\
n_{signal}^{\mu,a}&=&n_{b\overline{b}}^{\mu,a} +n_{bd}^{\mu,a}
+n_{db}^{\mu,a} +n_{dd}^{\mu,a}+r_{side}^{\mu} \cdot n_{side}^{\mu,a}\\
n_{signal}^{e,t}&=&n_{b\overline{b}}^{e,t} +n_{bd}^{e,t}+n_{db}^{e,t} +n_{dd}^{e,t}+ r_{side}^{e} \cdot n_{side}^{e,t}+n_{B_c}^{e,t}+ n_{B_{fake}}^{e,t}+r_{conv} \cdot n_{bconv}^t+r_{conv}\cdot n_{dconv}^t \\
n_{signal}^{e,a}&=&n_{b\overline{b}}^{e,a} +n_{bd}^{e,a}+n_{db}^{e,a} +n_{dd}^{e,a}+ r_{side}^{e} \cdot n_{side}^{e,a}+r_{conv} \cdot n_{bconv}^t +r_{conv}\cdot n_{dconv}^t 
\end{eqnarray}
\end{widetext}

\subsection{Impact parameter--${\boldmath c\tau}$ shape component}
The complete functions for the shape components of the fit are listed
below for the two samples and $\Delta\phi$ regions.  As a reminder, $x$ is
the additional lepton candidate's impact parameter and y is the $J/\psi$
candidate's $c\tau$.
\begin{widetext}
\begin{eqnarray}
F_{shape}^{\mu,t}(x,y)&=&\frac{1}{n_{signal}^{\mu,t}}\Biggl[ n_{b\overline{b}}^{\mu,t} 
	          F_b^{c\tau,\mu}(y)
	          F_b^{d_0,\mu}(x)+n_{bd}^{\mu,t} F_{b}^{c\tau,\mu}(y)
	          F_{direct}^{d_0,\mu}(x)+n_{db}^{\mu,t} F_{direct}^{c\tau,\mu}(y)
	          F_{b}^{d_0,\mu}(x)\\
		 &&+n_{dd}^{\mu,t} F_{direct}^{c\tau,\mu}(y)
	          F_{direct}^{d_0,\mu}(x)+n_{B_{fake}}^{\mu,t}F_{B_{fake}}^{\mu}(x,y)+n_{B_c}^{\mu,t}F_{B_c}^{\mu}(x,y)+r_{side}^{\mu}n_{side}^{\mu,t}F_{side}^{\mu}(x,y)\Biggr]\\
F_{shape}^{\mu,a}(x,y)&=&\frac{1}{n_{signal}^{\mu,a}}\Biggl[ n_{b\overline{b}}^{\mu,a} 
	          F_b^{c\tau,\mu}(y)
	          F_b^{d_0,\mu}(x)+n_{bd}^{\mu,a} F_{b}^{c\tau,\mu}(y)
	          F_{direct}^{d_0,\mu}(x) 
	         +n_{db}^{\mu,a} F_{direct}^{c\tau,\mu}(y)
	          F_{b}^{d_0,\mu}(x)\\
&&+n_{dd}^{\mu,a} F_{direct}^{c\tau,\mu}(y)
	          F_{direct}^{d_0,\mu}(x)+r_{side}^{\mu}n_{side}^{\mu,a}F_{side}^{\mu}(x,y)\Biggr]\\
F_{shape}^{e,t}(x,y)&=&\frac{1}{n_{signal}^{e,t}}\Biggl[ n_{b\overline{b}}^{e,t} 
	          F_b^{c\tau,e}(y)
	          F_b^{d_0,e}(x)+n_{bd}^{e,t} F_{b}^{c\tau,e}(y)
	          F_{direct}^{d_0,e}(x)+n_{db}^{e,t} F_{direct}^{c\tau,e}(y)
	          F_{b}^{d_0,e}(x)\\
&&+n_{dd}^{e,t} F_{direct}^{c\tau,e}(y)
	          F_{direct}^{d_0,e}(x)
+n_{B_{fake}}^{e,t}F_{B_{fake}}^{e}(x,y)+n_{B_c}^{e,t}F_{B_c}^{e}(x,y)\\
&&+r_{side}^{e}n_{side}^{e,t}F_{side}^{e}(x,y)+r_{conv}n_{bconv}^{t}F_{conv}(x)F_{b}^{c\tau,e}(y)+r_{conv}n_{dconv}^{t}F_{conv}(x)F_{direct}^{c\tau,e}(y)
\Biggr]\\
F_{shape}^{e,a}(x,y)&=&\frac{1}{n_{signal}^{e,a}}\Biggl[ n_{b\overline{b}}^{e,a} 
	          F_b^{c\tau,e}(y)
	          F_b^{d_0,e}(x)+n_{bd}^{e,a} F_{b}^{c\tau,e}(y)
	          F_{direct}^{d_0,e}(x) 
	         +n_{db}^{e,a} F_{direct}^{c\tau,e}(y)
	          F_{b}^{d_0,e}(x)\\
&&+n_{dd}^{e,a} F_{direct}^{c\tau,e}(y)
	          F_{direct}^{d_0,e}(x)+r_{side}^{e}n_{side}^{e,a}F_{side}^{e}(x,y)+r_{conv}n_{bconv}^{a}F_{conv}(x)F_{b}^{c\tau,e}(y)\\
&&+r_{conv}n_{dconv}^{a}F_{conv}(x)F_{direct}^{c\tau,e}(y)
\Biggr]
\end{eqnarray}
\end{widetext}

\subsection{Bin constraints component}
The constraints which are specific to a given region in $\Delta\phi$ and sample
are listed below:

\begin{widetext}
\begin{eqnarray}
C_{bin}^{\mu,t}&=&P(n_{signal}^{\mu,t},N_{signal}^{\mu,t}) \cdot
P(n_{side}^{\mu,t},N_{side}^{\mu,t}) \cdot
G(n_{B_c}^{\mu,t}-N_{B_c}^{\mu,t},\Delta N_{B_c}^{\mu,t}) \cdot
G(n_{B_{fake}}^{\mu,t}-N_{B_{fake}}^{\mu,t},\Delta
N_{B_{fake}}^{\mu,t})\\
C_{bin}^{\mu,a}&=&P(n_{signal}^{\mu,a},N_{signal}^{\mu,a}) \cdot
P(n_{side}^{\mu,a},N_{side}^{\mu,a})\\
C_{bin}^{e,t}&=&P(n_{signal}^{e,t},N_{signal}^{e,t}) \cdot
P(n_{side}^{e,t},N_{side}^{e,t}) \cdot
G(n_{B_c}^{e,t}-N_{B_c}^{e,t},\Delta N_{B_c}^{e,t})\\
&\times& 
G(n_{B_{fake}}^{e,t}-N_{B_{fake}}^{e,t},\Delta
N_{B_{fake}}^{e,t}) \cdot P(n_{conv}^{t},N_{conv}^{t})\\
C_{bin}^{e,a}&=&P(n_{signal}^{e,a},N_{signal}^{e,a}) \cdot
P(n_{side}^{e,a},N_{side}^{e,a}) \cdot P(n_{conv}^{a},N_{conv}^{a})\\
\end{eqnarray}
\end{widetext}

\subsection{Global constraints component}
The global constraints are the simplest component of the likelihood.
The functions of the global constraints are listed below:
\begin{eqnarray}
C_{global}^{\mu}&=&G(r_{side}^{\mu}-R_{side},\Delta R_{side})\\
C_{global}^{e}&=&G(r_{side}^{e}-R_{side},\Delta R_{side})\\
&\times&
G(r_{conv}-R_{conv},\Delta R_{conv}) 
\end{eqnarray}

\subsection{Log-likelihood function}
Finally, the log-likelihood can be assembled from the functions
developed in the previous sections.  The likelihood function for the
muon sample is:
\begin{eqnarray*}
\mathcal{L}&=& C_{global}^{\mu} \prod\limits^{t,a}_i\left[ C_{bin}^{\mu,i}\prod\limits^{N_{signal}^{\mu,i}}_j \left(F_{shape}^{\mu,i}(x_{i,j}^\mu,y_{i,j}^\mu)\right)\right]
\end{eqnarray*}
where  $x_{i,j}^\mu$ and
$y_{i,j}^\mu$ are the impact parameter of the additional muon
candidate and the $c\tau$ of the $J/\psi$ candidate for $j^{th}$ event
in the $i^{th}$ $\Delta\phi$ region in the $J/\psi$ signal region.

The likelihood function for the electron sample is similar to the muon
likelihood.  The electron likelihood includes conversion terms and the
fit of the $J/\psi$ mass sideband region.

\begin{eqnarray*}
\mathcal{L}&=& C_{global}^{e} \prod\limits^{t,a}_i\left[
C_{bin}^{e,i}\prod\limits^{N_{signal}^{e,i}}_j
\left(F_{shape}^{e,i}(x_{i,j}^e,y_{i,j}^e)\right)\right]\\
&\times& 
P(n_{convside},N_{convside}) \cdot \prod\limits^{N_{sideband}}_k \left(
F_{conv}(x_{k}^e,y_{k}^e) \right)
\end{eqnarray*}
where  $x_{i,j}^e$ and
$y_{i,j}^e$ are the impact parameter of the additional electron
candidate and the $c\tau$ of the $J/\psi$ candidate for $j^{th}$ event
in the $i^{th}$ $\Delta\phi$ region in the $J/\psi$ signal region, and
$x_{k}^e$ and $y_{k}^e$ are the impact parameter of the additional electron
candidate and the $c\tau$ of the $J/\psi$ candidate for $k^{th}$ event
 in the $J/\psi$ sideband region.

\section{Tests of ${\boldmath J/\psi}$--lepton Correlation Log-Likelihood Fit}
\label{toymc}
The impact parameter--$c\tau$ likelihood fit is tested using a large set of
toy Monte Carlo samples.   First, the input means for the various fit
components are chosen to be
similar to the results in data.   The constrained terms are chosen to
be consistent with the constraint. These inputs are Poisson fluctuated to determine the composition of each sample.  Each event is
assigned an impact parameter and $c\tau$ according to the shape function
used to describe that type of event .

Next, the fit constraints not yet varied ($R_{side}$, $R_{conv}$, $N_{B_{fake}}$,
$N_{B_c}$, $N_{conv}^{t}$, $N_{conv}^{a}$, and $N_{convside}$) are fluctuated
using the appropriate statistic.  The fluctuated
constraints are then used in the fit of the toy Monte Carlo sample.
  
A total of 1000 samples are generated and fit for both the electron and
muon samples.  The fit values are not forced to be non-negative.  The pull is calculated for each
fit value relative to the non-fluctuated input quantities.  The pull is
equal to width of the $\frac{n-\mu}{\sigma_n}$ distribution where n is
the fit value, $\sigma_n$ is the fit error returned, and $\mu$ is the
average value of the parameter input.  The bias, which is the mean
of the $\frac{n-\mu}{\sigma_n}$ distribution, is also measured.
Finally, the average difference between the fitted value and input
parameter is calculated, $\overline{(x-\overline{x})}$.
 
The pulls, biases, and average differences for all variables are
acceptable for both test samples.  All pulls are
within $\pm6\%$ of 1 and all biases are within $\pm0.12\sigma$ of
0.  Allowing the likelihood to have negative components yields fit results with
meaningful fit values and errors.   

\begin{figure*}[htbp]
               \includegraphics[width=8.6 cm]{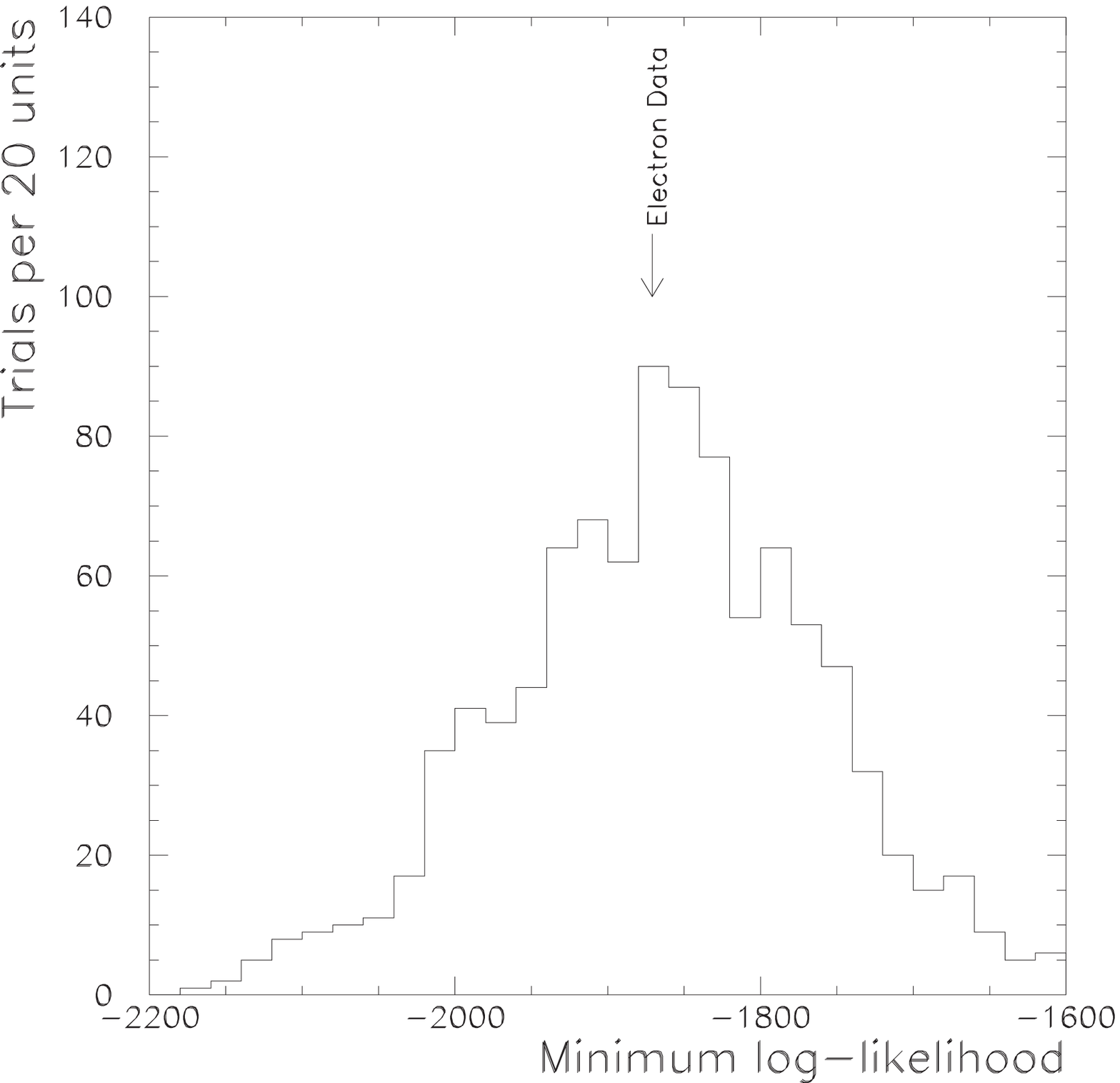}
               \includegraphics[width=8.6 cm]{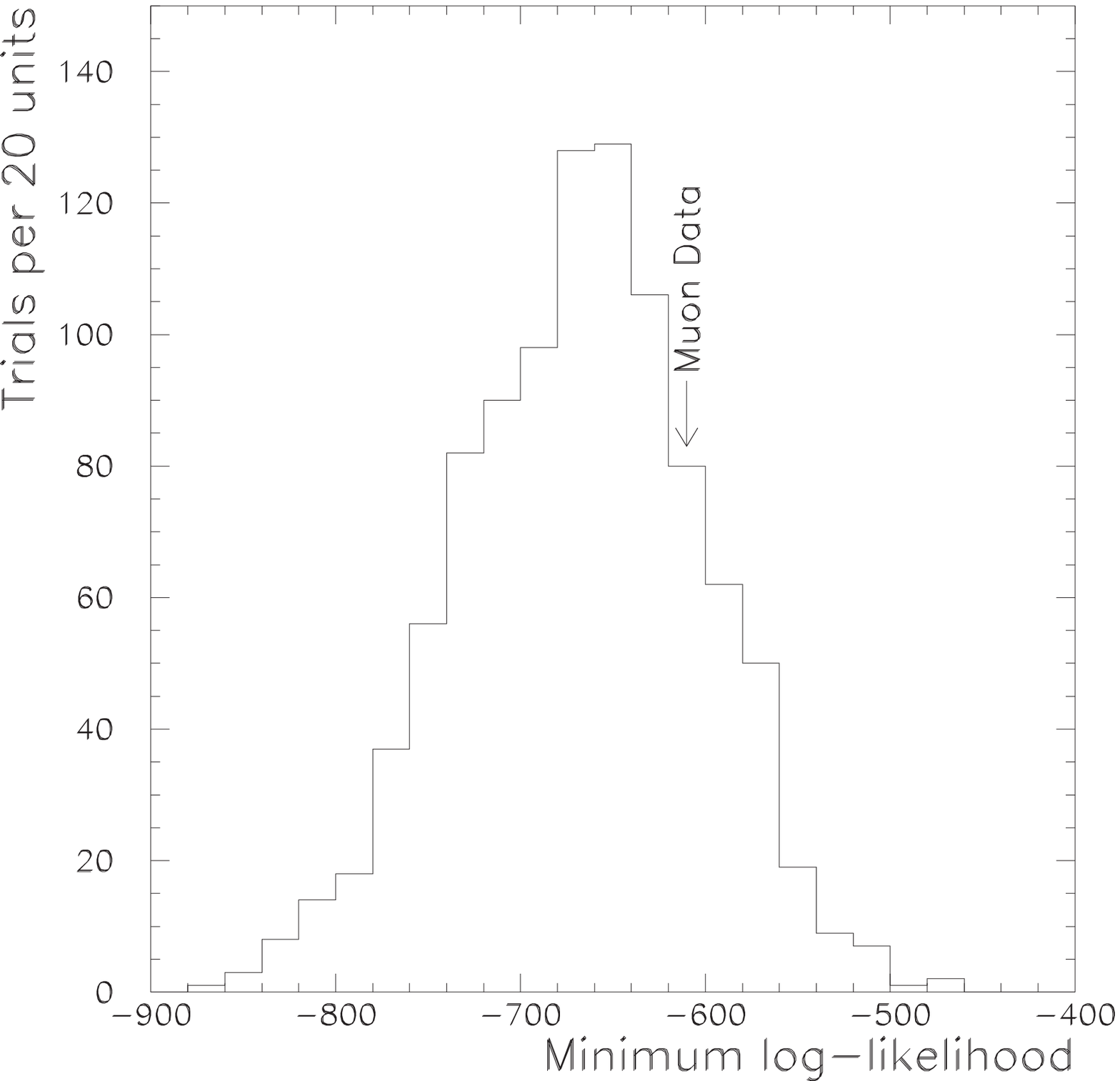}
\caption{The minimum log-likelihood distributions of the toy Monte Carlo  Left: Electrons Right: Muons.}
\label{fig:pullslike}
\end{figure*}

Figure~\ref{fig:pullslike} shows the minimum log-likelihood distributions of
both samples.  We find that $19.6\%$ of the muon toy Monte Carlo samples and
$49.8\%$ of the electron toy Monte Carlo samples have a higher minimum
log-likelihood than the data.  The minimum log-likelihood distributions
along with the biases, pulls, and average differences give confidence
that the likelihood is working properly.

The muon toy Monte Carlo samples have an input mean of
$f_{towards}^{input}=34.5\%$ and a fit mean of
$f_{towards}^{fit}=34.5\pm 0.4\%$.  The width of the fit $f_{toward}$
distribution is $10.9\pm0.3\%$, which is consistent with the error seen in
data of $^{+9.2}_{-8.2}\%$.   The electron toy Monte Carlo samples have an input mean of $f_{towards}^{input}=19.2\%$ and a fit mean
of  $f_{towards}^{fit}=18.6\pm 0.2\%$.  The width of the fit $f_{toward}$ distribution is  $6.0\pm0.1$, which is consistent with the error seen in data of 
$^{+6.5}_{-5.8}\%$.

\bibliography{b_corr}

\end{document}